\documentclass[hyper]{JHEP3}
\usepackage{setspace}
\usepackage{graphicx}
\usepackage{cite}
\def\beq{\begin{equation}}
\def\eeq{\end{equation}}
\def\beqn{\begin{eqnarray}}
\def\eeqn{\end{eqnarray}}

\newcommand{\sq}{\tilde{q}}
\newcommand{\gl}{\tilde{g}}
\newcommand{\mgl}{m_{\tilde{g}}}
\newcommand{\msuL}{m_{\tilde{u}_L}}
\newcommand{\suL}{\tilde{u}_L}
\newcommand{\suR}{\tilde{u}_R}
\newcommand{\msuR}{m_{\tilde{u}_R}}
\newcommand{\msq}{m_{\tilde{q}}}
\newcommand{\rtar}{\rightarrow}
\newcommand{\mttwo}{M_{T2}}

\newcommand{\mttwmin}{M_{T2}({\rm min})}
\newcommand{\mttwsub}{M_{T2}({\rm sub})}
\newcommand{\mttwmod}{M_{T2}^{\rm mod}({\rm min})}

\def\gsim  {\hspace{0.3em}\raisebox{0.4ex}{$>$}\hspace{-0.75em}\raisebox{-.7ex}{$\sim$}\hspace{0.3em}}
\def\lsim  {\hspace{0.3em}\raisebox{0.4ex}{$<$}\hspace{-0.75em}\raisebox{-.7ex}{$\sim$}\hspace{0.3em}}

\title{Controlling ISR in sparticle mass reconstruction}

\author{Mihoko M.\ Nojiri$^{1,2,3}$, Kazuki Sakurai$^{4,5}$ \\
${}^1$Theory Group, KEK, 1-1 Oho, Tsukuba, Ibaraki 305-0801, Japan\\
${}^2$The Graduate University for Advanced Studies (SOKENDAI),1-1 Oho,
Tsukuba, Ibaraki 305-0801, Japan\\
$^3$ Institute for the Physics and Mathematics of the Universe, The University of Tokyo, Chiba 277-8568, Japan\\
$^4$Cavendish Laboratory, J.J.\ Thomson Avenue, Cambridge CB3 0HE, UK\\
$^5$Department of Applied Mathematics and Theoretical Physics\\ 
Wilberforce Road, Cambridge CB3 0WA, UK\\
\\
       E-mail: \email {nojiri@post.kek.jp,sakurai@hep.phy.cam.ac.uk}
        }

\received{\today}               
\accepted{\today}               

\preprint{ KEK TH-1388 \\ Cavendish-HEP-10/13 \\DAMTP-2010-57  }

\abstract{
Use of inclusive $\mttwo$ distribution for sparticle mass determination 
is discussed. We define new parameters $\mttwmin$ and $\mttwmod$, 
which are a 
kind of minimum of sub-systerm $\mttwo$ values. Their endpoints are 
less affected by initial state radiations. 
We demonstrate that both $\msq$ and $\mgl$
can be extracted from the endpoints of the distributions in the wide region
of parameter space expected in CMSSM. 
We also present a comparison with $M_{TGen}$ distributions.
}

\keywords{Hadronic Colliders, Supersymmetry Phenomenology, Beyond Standard Model}

\begin{document}

\section{Introduction}
The quest  of physics beyond the standard model   
is now reaching new phase with the LHC experiments\cite{atlas, cms} 
and various new dark matter (DM) searches\cite{cdms2, xenon100}. 
In models with conserved parity for stability of a dark matter, 
new coloured particles may be produced in pairs at the LHC 
and each of them decays into particles involving the stable dark matter.  
This gives missing transverse momentum to the events, which would be useful
in separating the signal from QCD backgrounds. 
Especially, in supersymmetric (SUSY) models squarks and gluino are produced and 
decay into lighter particles involving the lightest SUSY particle (LSP) 
which is usually identified as the dark matter.
The signature is events with jets (+leptons) + missing transverse momentum. 

Because  the dark matter  cannot be detected in the LHC detectors, 
masses of SUSY particles cannot be reconstructed as resonances. 
Various techniques to determine 
the masses of SUSY particles have been studied (See Ref.\cite{review-lester} for a recent review.). 
Important kinematical variables are 
peak position of $M_{\rm eff}$ distribution, 
the endpoints of invariant mass distributions of jets and leptons\cite{Hinchliffe:1996iu, Bachacou:1999zb, Hinchliffe:1999zc, 
Allanach:2000kt, Gjelsten:2004ki, Gjelsten:2005aw, Miller:2005zp, 
Costanzo:2009mq, Burns:2009zi, Matchev:2009iw}, 
the endpoint of $M_{T2}$ distributions\cite{
lester:1999tx, Barr:2003rg, Lester:2007fq, Cho:2007qv, Gripaios:2007is,
Barr:2007hy, Cho:2007dh, shimizu, takeuchi, Cheng:2008hk, hiramatsu,
Barr:2009jv, Matchev:2009fh, Konar:2009wn, Konar:2009qr}.  
Among those, most clean channel is probably the 
invariant mass distributions of  $jll$ channel although the branching ratio of this mode is rather small. 

In the early stage of the LHC, it is more useful to study the distribution with 
jet + $E_{Tmiss}$ channel\cite{shimizu, takeuchi}. 
The number of signal events is rather small in  the discovery phase, 
therefore if we require multiple leptons in the final state, 
we would not be able to obtain an event sample large enough to do analysis.   
For this problem, quantities that are less sensitive to the detail of the decay pattern are useful. 

In previous papers\cite{shimizu, takeuchi}, 
we have defined  an inclusive $M_{T2}$, 
which is calculated for any SUSY events at the LHC using hermisphere method. 
Hemisphere method is the way to define two sets of particles in the event, 
and each set (called hemisphere) contains  particles coming from the same parent particle decay 
in the limit that parent sparticles are highly boosted into back-to-back\cite{hemi1, hemi2}. 
We have shown an interesting correlation 
between the endpoint of the inclusive $M_{T2}$ distribution and mass 
of the heavier of squark and gluino.  
We have also demonstrated that a subsystem $M_{T2}$, $\mttwsub$,  
is useful to determine the gluino mass for $\mgl<\msq$ case\cite{takeuchi}.

One of the important issues for mass reconstruction 
is existence of initial state radiation (ISR)\cite{
hiramatsu, Plehn:2005cq, Alwall:2008qv, Papaefstathiou:2009hp, Papaefstathiou:2010ru}. 
When we produce heavy 
particles at the LHC, there would be a hard ISR jet 
which could have $p_T$ about the same order 
of the mass of the produced particle.  
The ISR jets may be involved among the jets for 
the invariant mass or $M_{T2}$ calculations, then 
the distributions are smeared. 
In Ref.\cite{hiramatsu}, it is shown that the effect of leading ISR can be removed by minimizing 
``$\mttwo$ after removing a jet"  for several choices of a removing jet. 
Smearing  due to ISR is significantly reduced
by this procedure for a limited example $\gl\gl$ production process followed by 
$\tilde{g}\rightarrow \tilde{\chi}^0_1 jj$. 


In this paper we define a $\mttwmin$ to reconstruct 
the endpoint of $\mttwo$ distribution under the effect of ISR for general 
squark and gluino production processes.  
The definition is  also consistent with the previously defined $\mttwsub$ 
to obtain $\mgl$ when $\mgl < \msq$. 
We  extend our studies so that we can 
cover $\msq \le \mgl$ case as well, so that $\mgl$ is reconstructed by the appropriate choice of quantities. For this purpose, 
we define a $\mttwmod$ 
for the events that two highest  $p_T$ jets coming from two 
body decays with large mass differences  between parent and daughter particles 
and there are two jets  which are prominent over the other jets. 
As its nature, 
the $\mttwmod$ works for $\msq \le \mgl$.  
We provide parton level and jet  level studies and demonstrate that these quantities 
are actually important.

This paper is organized as follows.  In section 2, we introduce basic parameters such 
as $\mttwo$ and $\mttwsub$, which has been defined in the 
previous papers by using hemispehre algorithm, and summarize previous analyses.  
We also define 
new parameters  $\mttwmin$, and $\mttwmod$.  
These parameters are defined 
to  reduce ISR effects but for different squark and gluino decay patterns. 
Section 3 is for parton level study of hemisphere analysis, which is efficient for inclusive analysis. 
Section 4 describes the parton level distribution of 
$\mttwmin$ and $\mttwmod$ with and without ISR. 
The needs of  $\mttwmod$ 
will be explained in this section. 
In section 5, we discuss  jet level $\mttwo$, $\mttwmin$ and $\mttwmod$ distributions. 
The measured endpoints depend on the dominant decay pattern of sparticles. 
We found  the decay pattern can be determined from simpler quantities 
such as $p_T$ of the two highest $p_T$ jets and hermisphere mass distributions. 
The inclusive $\mttwo$ study can be applied in the early stage of the LHC. 
The expected distribution at $\sqrt{S}=7$\,TeV and $\int {\cal L} dt=1\,{\rm fb^{-1}}$ 
is shown in section 6.  Section 7 is devoted for discussions and conclusions.

\section{Inclusive $M_{T2}$ distribution, ISR, and mixed production}
\label{sec:isr-mixed}

\begin{figure}[!t]
\begin{center}
\includegraphics[width=4.6cm,angle=90]{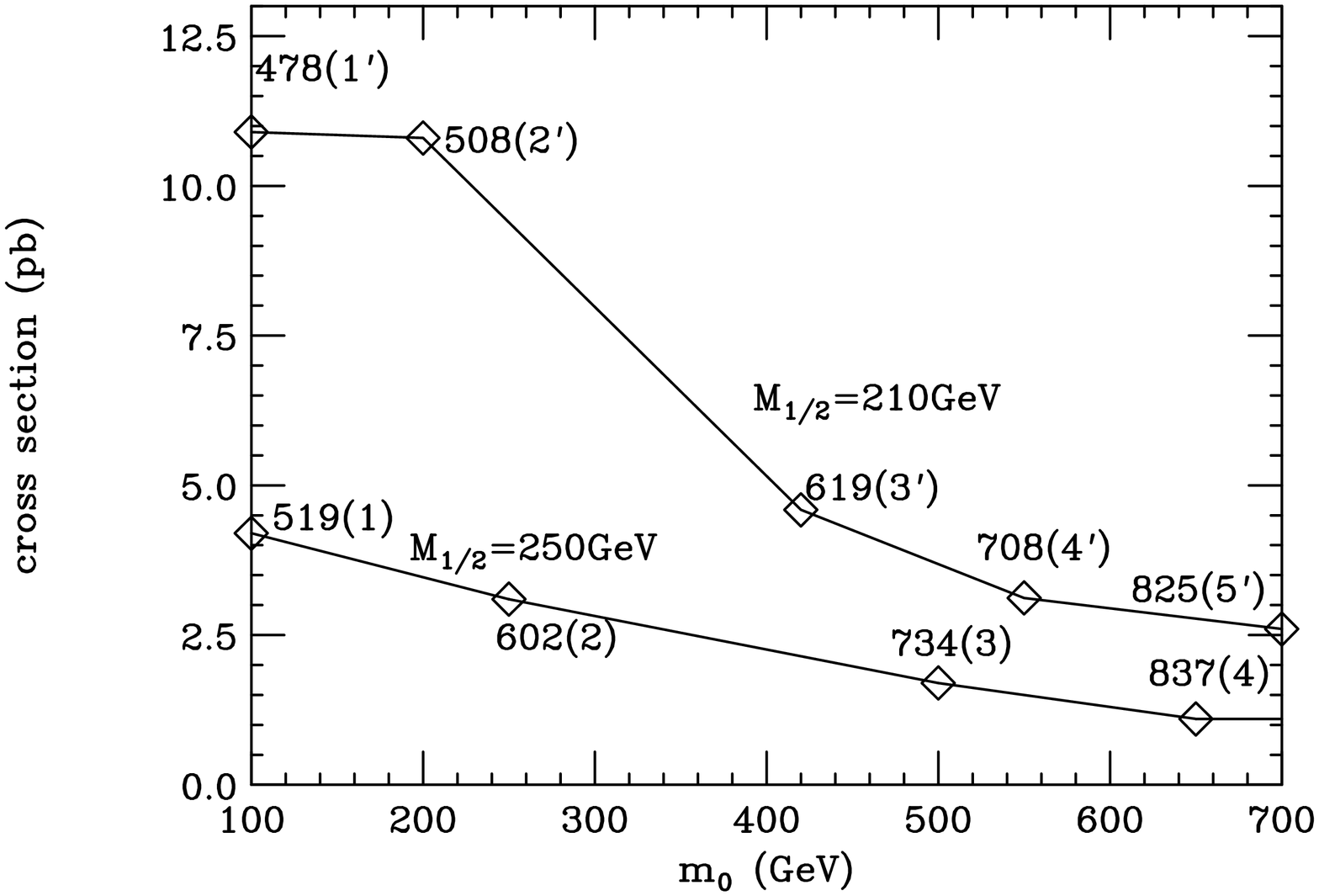} \hskip 0.5cm 
\includegraphics[width=4.6cm,angle=90]{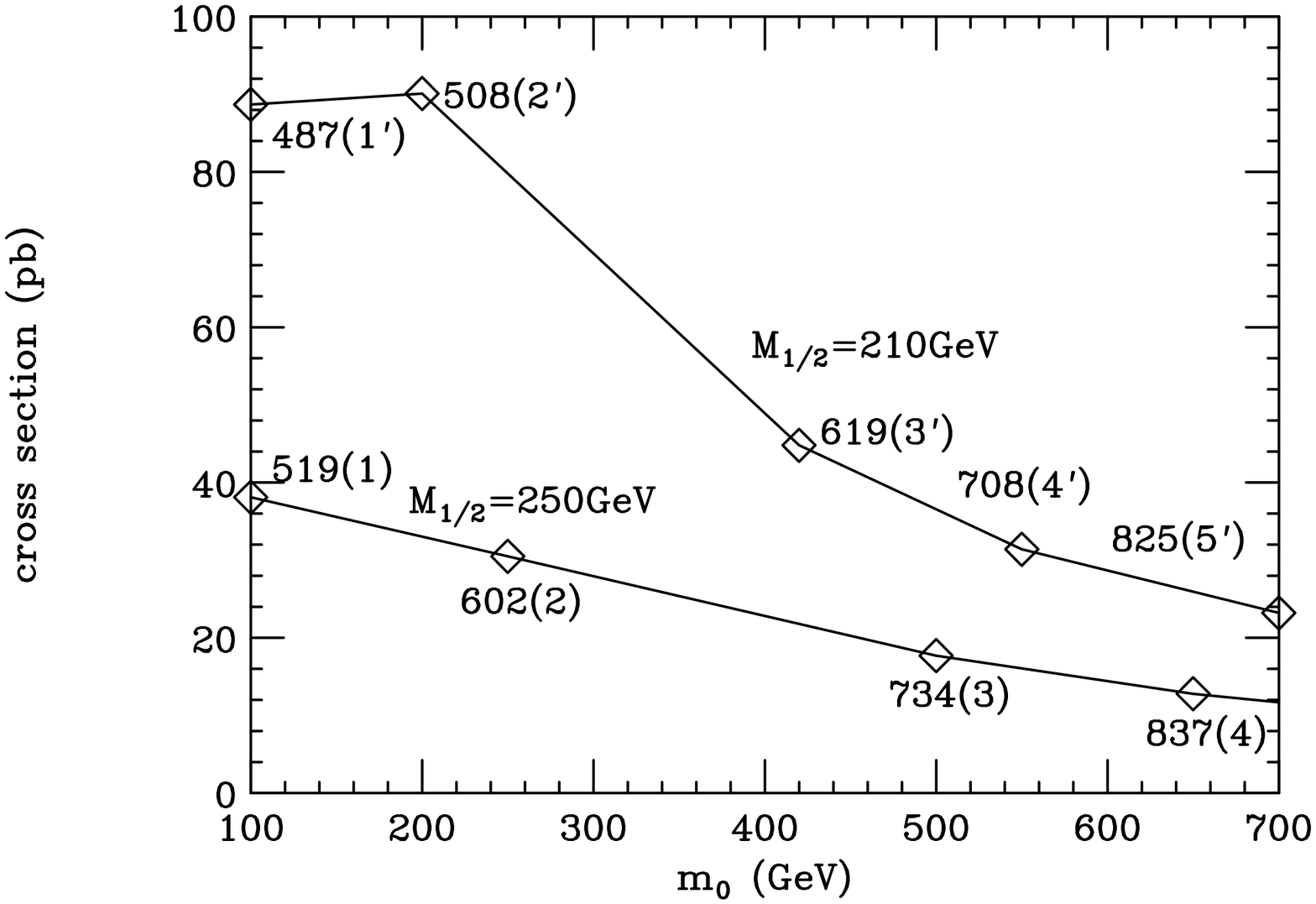}
\caption{
Total SUSY production cross sections for the model points in Table\,1 
as a function of 
$m_0$ for fixed gaugino masses  at 7 TeV (left) and 14 TeV (right).
The CMSSM points are produced by ISASUSY\cite{Paige:2003mg}, 
and the cross sections are estimated by HERWIG\cite{herwig1,herwig2,herwig3}.  
The $\tilde u_L$ masses and names of the model points are also shown in the figure.}
\label{crosssection}
\end{center}
 \end{figure}

The discovery of supersymmetric particles  at  the CERN LHC has been studied in depth\cite{
Baer:1995nq, Baer:1995va, Abdullin:1998pm, Baer:1998sz, Abdullin:1998nv,
Allanach:2000ii, Baer:2003wx, Izaguirre:2010nj, Baer:2010tk, cmsnote
}.  
In recent articles\cite{Baer:2010tk, cmsnote}, the discovery region at $\sqrt{s}=7$\,TeV and $\int {\cal L}dt= 1\,$fb$^{-1}$  in CMSSM has been studied.
One can read from Fig.\,13 in Ref.\cite{cmsnote} that 
the region $m_0<850\,(750)$\,GeV at $M_{1/2}<210\,(250)$\,GeV can be explored, respectively. 
Here  $m_0$ and $M_{1/2}$ are 
universal scalar and gauino masses.   
In Fig.\,\ref{crosssection}, we show the total  
SUSY production cross section as a function of $m_0$  
for $M_{1/2}= 210$ (250) GeV at $\sqrt{S}=7$ TeV (left) and 14 TeV (right).  
The model points appeared in Fig.\,\ref{crosssection} are defined in Table\,\ref{masstable}.
The cross section at each model point is calculated by HERWIG\cite{herwig1, herwig2, herwig3}. 
The cross section at the  boundary of the 
discovery region for  $\sqrt{S}=7$\,TeV with $\int {\cal L} dt = 1$\,fb$^{-1}$ is 
roughly $\sigma_{SUSY}=1$\,pb. 
After the 7\,TeV run, 
the collision at 14\,TeV is scheduled.
The cross sections at 14\,TeV is about 10 times larger than those at 7\,TeV. 

By studying production cross sections of sparticles at the LHC, 
patterns for sparticle masses can be investigated. 
The production cross section is sensitive to the mass scale of produced particles, 
as can be seen in Fig.\,\ref{crosssection}, and it is an important piece of information 
on the mass scale.  
Increasing mass by 100\,GeV leads 50\,\% reduction of the cross section at 
$m_{\tilde{g}} \sim m_{\tilde{u}_L}\sim 500$\,GeV.  
The uncertainty of the squark and gluino production cross section would be less than 
10\,\%\cite{Beenakker:1996ch}. 

\begin{table}
\begin{center}
\begin{tabular}{|c|c|c|c|c|c|}
\hline
&Point 1'& 2'& 3'&4'& 5'\cr
\hline 
$m_0$(GeV)           & 100& 200&420 & 550& 700\cr
$M_{1/2} $   & 210& 210& 210&210 & 210 \cr
$\mgl$        & 522& 528& 541& 550& 558 \cr
$\msuL$     & 478&  508& 619 &708 & 825\cr
 $\msuR$     & 466 &  496& 611& 702&820\cr
$\sigma_{\rm SUSY}({\rm 14TeV})$ in pb &88.7 & 90.1&44.8 & 31.4  & 23.2 \cr 
$\sigma_{\rm SUSY}({\rm 7TeV})$ in pb &10.9 & 10.8&4.59 & 3.12  & 2.60 \cr 
$\sigma({\gl\sq })  /\sigma_{\rm SUSY}$ at 14TeV &0.52 &0.46&0.49&0.48& 0.45\cr
$\sigma({\gl\gl }) /\sigma_{\rm SUSY}$ at 14TeV   &0.14 &0.14&0.21&0.28&0.37\cr
$\sigma({\sq\sq }) /\sigma_{\rm SUSY}$ at 14TeV &0.21 &0.18&0.15&0.12&0.08\cr
$Br(\suL\rtar\gl)$   &0  &0  & 0.26 & 0.48&  0.56  \cr 
$Br(\suR\rtar\gl)$   &0  &0  & 0.60 & 0.82& 0.87 \cr 
\hline
\end{tabular}
\vskip 1cm
\begin{tabular}{|c|c|c|c|c|c|} 
\hline
&Point 1 & 2& 3& 4& 5\cr
\hline 
$m_0$           & 100& 250& 500 & 650& 750\cr
$M_{1/2} $    & 250& 250& 250&250& 250 \cr
$\mgl$         & 612& 620& 636& 646& 651 \cr
$\msuL$      & 560& 602& 734 &837 & 913\cr
$\msuR$     & 544& 589& 724& 828&906\cr
$\sigma_{\rm SUSY}({\rm 14TeV})$ in pb &38.1 & 30.5 &17.7 & 12.8 & 10.6  \cr 
$\sigma_{\rm SUSY}({\rm 7TeV})$   in pb & 4.2& 3.1& 1.7 & 1.2 &  1.1\cr 
$\sigma({\gl\sq }) /\sigma_{\rm SUSY}$ at 14TeV  & 0.47&0.48 &0.46& 0.43&0.39\cr
$\sigma({\gl\gl })  /\sigma_{\rm SUSY}$ at 14TeV &  0.11 & 0.12 & 0.18 & 0.22 &0.25 \cr
$\sigma({\sq\sq }) /\sigma_{\rm SUSY}$ at 14TeV & 0.22&0.21 &0.16&0.12&0.0 9\cr
$Br(\suL\rtar\gl)$   &0  &0  & 0.249 & 0.469&  0.611  \cr 
$Br(\suR\rtar\gl)$   &0  &0  & 0.582 & 0.813& 0.891 \cr 
\hline
\end{tabular}
\end{center}
\caption{Masses and important branching ratios in some  CMSSM points. Points  1 $-$ 5 for fixed gaugino mass $M_{1/2}=250$\,GeV, Points 1' $-$ 5' for 
$M_{1/2}=210$\,GeV. Mass spectrum are calculated by ISASUSY\cite{Paige:2003mg}  
and cross sections are calculated by HERWIG\cite{herwig1,herwig2,herwig3}.  }\label{masstable}
\end{table}

One can determine the mass spectrum directly from 
kinematical distributions of sparticle decay products. 
If both squark and gluino masses can be 
determined, one can further study their interaction 
by comparing the measured production cross section 
with the theoretical value, because the dependence 
on the other parameters are rather small\cite{Datta:1,Kane:1}. 

In SUSY models with conserved R parity,  two LSPs  escape from detection for each 
SUSY event,  therefore  masses of sparticles cannot be measured as resonances.
(In this paper the LSP is assumed to be the lightest neutralino, $\tilde \chi_1^0$.) 
Endpoints  of  invariant mass 
distributions of jets and leptons 
 distribution can be used to determine the masses.
For example in $\tilde{q}\rightarrow \tilde{\chi}^0_2\rightarrow \tilde{l}
\rightarrow \chi^0_1$ produce a $jll$ final state and endpoints of the 
invaiant mass distribution $m_{jll}$, $m_{jl}(min)$, $m_{jl}(max)$, $m_{ll}$ 
can be used to determine the all sparticles masses involved in the decay. 
However, branching ratios of  favorable modes such as $jll$ 
are generally small, and they may not be 
useful in the early stage of the experiments. 
Therefore  it is important to find inclusive quantities sensitive to  squark and gluino masses so that 
one can use the most of the signal events at the LHC. 

Typical energy scale of  sparticle production  processes may be 
estimated from the  ``effective mass",  
\begin{equation}
M_{\rm eff} \equiv \sum_i |p_{Ti}| + E_T^{\rm miss} \,, 
\end{equation}
where in this paper the sum is taken for the jets with $p_T> 50$\,GeV and $\eta<$2.5. 
A peak value of the $M_{\rm eff}$ distribution is correlated with squark and gluino masses, 
although the relation is rather qualitative\cite{Tovey:meff}.

More recently, we have proposed an ``inclusive $M_{T2}$" 
which is calculated from jet (and lepton) momenta 
and the missing transverse momentum\cite{shimizu, takeuchi}.
The definition of $M_{T2}$ is the following\cite{lester:1999tx, Barr:2003rg}:
\begin{equation}
M_{T2} \equiv \min_{p^{\chi}_{T1}+p^{\chi}_{T2}=p_T^{\rm miss}} \left[\max\left(M_T(p_{v_1},p_1^{\chi} ),M_T(p_{v_2},p_2^{\chi} )\right) \right] .
\end{equation}
Here  $v_1$ and $v_2$ are two selected visible systems in an event,
and $p^{\chi}_i$ is  a test LSP momentum with a test mass $m_{\chi}$.
If  $v_1$ and $v_2$ arise from two on-shell particles $A_1$ and $A_2$ as $A_i\rightarrow v_i \chi^0_1$, 
respectively, the $M_{T2}$  is bounded from above as 
\begin{equation}
M_{T2} \le \max( m_{A_1},m_{A_2}) ~~~ {\rm for} ~  m_{\chi}=m_{\chi^0_1}
\label{mt2eq}\end{equation}
by its construction. 
Therefore by measuring the 
endpoint of $M_{T2}$ distribution, one can measure the 
mass of the heavier of parent particles. 
Moreover, the LSP mass can be determined by measuring a
kink position of the $M_{T2}^{\rm max}(m_\chi)$ as a function of 
the test mass $m_{\chi}$\cite{Cho:2007qv, Barr:2007hy, Cho:2007dh}.
Throughout this paper, we fix $m_{\chi} = m_{\tilde \chi_1^0}$ for simplicity.

To  determine masses of squark or gluino from the $\mttwo$ endpoint, 
decay products of sparticles should be correctly identified to $v_i$. 
Unless the event has relatively simple topology such 
as $pp\rightarrow $$ \tilde{q}\tilde{q}\rightarrow (j\tilde \chi^0_1) (j \tilde \chi^0_1)$
or $pp \rightarrow \tilde{g}\tilde{g}$$\rightarrow$  $(jj \chi^0_1) (jj \chi^0_1)$, 
some of the jets are mis-grouped into wrong $v_i$.
It has been proposed in\cite{hemi1,hemi2} hemisphere algorithm is useful to define $v_i$. 
It is defined as follows; 
\begin{enumerate}
\item Take two jets as the first seeds of the two hemisphere $H_1$ and $H_2$, 
$J_1\in H_1$ and $J_2\in H_2$: $J_1$ is the highest $p_T$ jet,
and  $J_2$ is the jet ($i$) whose
$p_{Ti} \times \Delta R(p(J_1),p(i) )$ is the largest in the event. 
\item Associate the other jets to the one of the hemispheres  based on 
a distance measure $d$, so that  $j_k$ belongs to $H_i (i=1,2)$ 
if $d(i,k) <d(j,k) $.
Here  $d$ is defined as
\begin{equation}
d(p_{\rm hemi}^{(i)},p_k) = (E_{\rm hemi}^{(i)}-|p_{\rm hemi}^{(i)}|\cos\theta_{ik})\frac{E^{(i)}_{\rm hemi}}{(E^{(i)}_{\rm hemi}+E_k)^2} .
\end{equation}  
\item  Take the sum of the jet momenta that belong to $H_i$ 
 and regard it as a new seed. 
Repeat the processes 2 and 3, while keeping the first seeds $J_1$ and $J_2$ in the different hemispheres, till the assignment converges.  
\end{enumerate}
After defining the hemisphere, we define the inclusive $M_{T2}$ 
using $p_{v_i}= \sum_{k\in H_i} p_k$. 
In this paper we only use jets with $p_T>$50\,GeV and $|\eta|<$2.5. 

Alternatively, it was proposed to use $M_{TGen}$ 
which is the minimum of the $M_{T2}$s for all  choices of jet combinations\cite{Lester:2007fq}
\begin{equation}
M_{TGen} \equiv \min_{1 \le \alpha \le 2^{n-1}-1} M_{T2}(p_{H_i^\alpha},p_{H_j^\alpha}, p_T^{\rm miss}, m_{\chi})
\end{equation}
where $\alpha$  denotes a possible combination to split jets into two groups $H_1^\alpha$  and $H_2^\alpha$,
and minimization is taken over all possible $2^{n-1}-1$ combinations. 
Furthermore, $p_{H_i^\alpha} = \sum_{k \in H_i^\alpha } p_k$ and $n$ is the number of visible objects. 
If all jets and leptons are decay products of $A_1$ and $A_2$, $M_{TGen} \le \max\{m_{A_1}, m_{A_2}\}$.

At hadron collision, there are also particles coming from ISR 
in addition to those from sparticle decays. 
Transverse momentum of the ISR jet
can be as large as those from squark and gluiino decays. 
The ISR jets in the visible systems $v_1$ and $v_2$ smear the $M_{T2}$ endpoint significantly. 

In Ref.\cite{hiramatsu} a new definition of $M_{T2}$ that reduce ISR effect is defined, 
 and mass determination based on the quantity 
is demonstrated  for a $pp\rightarrow \gl\gl X$ process
followed by a gluino decay $\gl \to jj \chi^0_1$, where $X$ is a ISR partons. 
The ISR improved $\mttwo$, $\mttwmin$, is defined as follows; 
\begin{enumerate}
\item  Calculate $M_{T2}(i)$ which is a $M_{T2}$ calculated with 
a given  grouping procedure but without involving the  $i$-th jet in $p_T$ order.
In this paper, we take hemisphere algorithm after removing the $i$-th 
jet for the grouping. 
\item  Take the minimum of the $M_{T2}(i)$ over the certain range of $i$, 
\begin{equation}
M_{T2}({\rm min}) \equiv \min_i M_{T2}(i) .
\label{eq:mindef}
\end{equation}
\end{enumerate}
In Ref.\cite{hiramatsu}, the minimization is taken up to the fifth jet
and $\mttwo$ is calculated by 4 jets because a gluino forced to 
decay into $jj\tilde{\chi}^0_1$.
The contamination above the endpoint is significantly reduced 
for $\mttwmin$ distribution.  
In the paper, however, only the limited process was studied, and 
the  techniques should be extended to the other  SUSY processes, 
especially  gluino-squark co-production
because it is the dominant production process 
for a wide region of the parameter space (See Table\,\ref{masstable}.).

A merit to use another definition of the $M_{T2}$ for gluino-squak co-production,  $M_{T2}({\rm sub})$, 
has been discussed in Ref.\cite{takeuchi}. 
It is defined as the $M_{T2}$ but the highest $p_T$ jet is not included, namely, 
\begin{equation} 
M_{T2}({\rm sub}) \equiv M_{T2}(1) .
\end{equation}
In the parameter region where $\msq > \mgl$, 
the subtraction of the highest $p_T$ jet
tends to reduce the full-system of $\tilde q$-$\tilde g$ into the subsytem of  $\tilde g$-$\tilde g$
 or $\tilde g$-$\tilde \chi_i$ ($\tilde \chi_i$ denotes charginos or neutralinos.) 
 effectively because squarks either decay into $\tilde g$ or $\tilde \chi_i$ with high probability. 
 In this case, we have
\begin{equation}
M_{T2}({\rm sub})  \le m_{\tilde g} .
\label{eq:subdef}
\end{equation}
It is shown that the endpoint of the $\mttwsub$ correlates with gluino mass very well. 
By definition, $M_{T2}({\rm min})< M_{T2} ({\rm sub})$,  therefore 
the $M_{T2}({\rm min})$ distribution may improve the sensitivity to the gluino mass
by reducing the tail of the distribution.

The parameter region with $\msq<\mgl$ is not considered in Refs.\cite{shimizu, takeuchi, hiramatsu}.   
For such a parameter region, 
gluino decays into $j \tilde q$, 
and the squark further decays into $j \tilde{\chi}_i$.  
The jet from the squark two body decay 
has significant energy of the order of  $|p_{Tj}| \sim m_{\tilde{q}}/2$ 
for CMSSM   like mass spectrum  with $\msq \gg m_{\tilde \chi_i}$.
The $p_T$ of the jet from the two body decay tends to be 
 harder than that of ISR jet,  therefore 
including the highest $p_{T}$ jet for the minimization 
for  $M_{T2}({\rm min})$ may  remove a jet from the squark decay with high 
probability, making the distribution near the endpoint rather flat. 
Therefore,  we introduce a variable $M_{T2}^{\rm mod} {\rm (min)}$ that is defined as
\begin{equation}
M_{T2}^{\rm mod} {\rm (min)} \equiv \min_{i \ne 1, 2}  M_{T2}(i) 
\label{eq:moddef}
\end{equation}
for $\msq<\mgl$. 
The quantities defined in Eqs.\,(\ref{eq:mindef}), (\ref{eq:subdef}) and (\ref{eq:moddef}) can
be straightforwardly extended to the ones using $M_{TGen}$ variables.
  
Table.\,\ref{tab:pre_sum} summarises the results of inclusive $M_{T2}$ studies which have been done so far.
The aim of this paper is to provide more comprehensive study of mass 
determination using the quantities defined above,
in a wide range of CMSSM parameter space.

\begin{table}[!t]
\hspace{2.5cm}
\begin{tabular}{|c|l|c|c|c|c|c|}
\hline
 & \multicolumn{3}{|c|}{$m_{\tilde q} > m_{\tilde g}$}  & $m_{\tilde g} < m_{\tilde q}$  \\ \hline
 Production & ~~~$\tilde g \tilde g$   & $\tilde q \tilde g$ & Inclusive  & Any \\ \hline
$M_{T2}^{\rm max}$ & 
$> m_{\tilde g}$\cite{hiramatsu} & $\simeq m_{\tilde q}$ & $\simeq m_{\tilde q}$\cite{shimizu,takeuchi} & ? \\ \hline
$M_{T2}^{\rm max}$(sub) & 
$\simeq m_{\tilde g}$  & $\simeq m_{\tilde g}$ & $\simeq m_{\tilde g}$\cite{takeuchi} & ? \\ \hline
$M_{T2}^{\rm max}$(min) & 
$\simeq m_{\tilde g}$\cite{hiramatsu} & ? & ? & ? \\ \hline
\end{tabular}
\label{tab:pre_sum} 
\caption{Summary table of previous studies on inclusive $M_{T2}$.}
\end{table}
%

\section{Validation of hemisphere algorithm}

One of the difficulties to apply  $M_{T2}$ distirubiton for general SUSY processes is 
in defining appropriate two visible systems, $v_1$ and $v_2$.
If $v_1$ and $v_2$ are originated from decays $A_1$ and $A_2$, $A_i \to v_i \tilde \chi_1^0$,
the $M_{T2}$ (We call it $M_{T2}$(true).) is bounded above by $\max \{ m_{A_1}, m_{A_2} \}$.
When there are several jets and leptons in an event,  
it is generally difficult to find such a ``correct" assignment.

Finding the correct assignment may not be necessary to reconstruct the endpoint.
Instead, we may use a jet assignment algorithm that possesses the following properties:
\begin{description}
\item[(i)]  $M_{T2}$ calculated from the algorithm does not exceed $\max \{ m_{A_1}, m_{A_2} \}$ significantly,\,  
\item[(ii)]  A large number of events contribute to the endpoint region.  
\end{description}

\begin{figure}[t]
\begin{center}
\includegraphics[width=15cm] {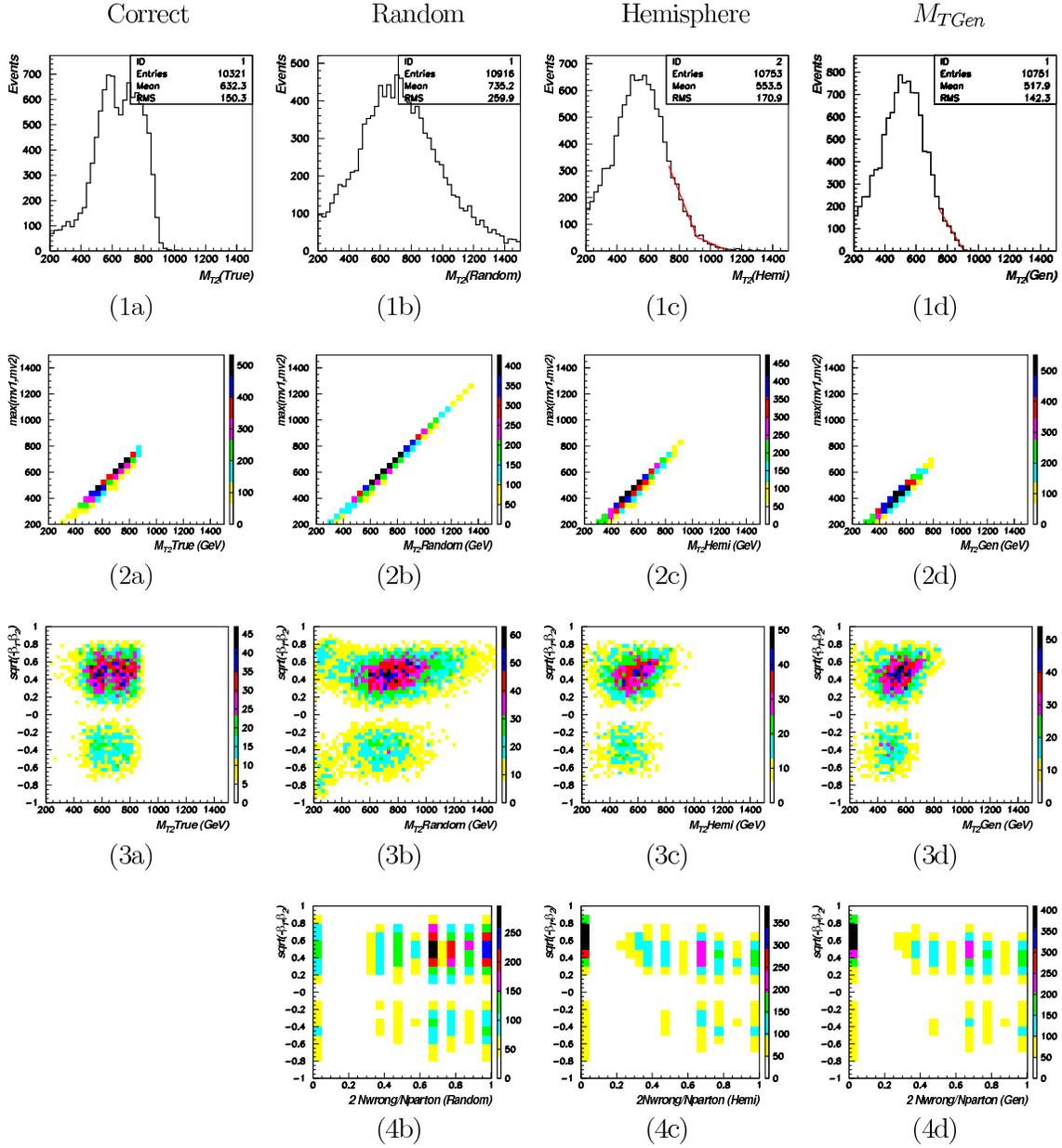}
\caption{
(1a)$-$(1d): $M_{T2}$ distributions.~~
(2a)$-$(2d): Distributions on a
($M_{T2}(m_{\tilde \chi_1^0})$, ${\rm sign}(-\beta^2_{12}) \sqrt{|\beta^2_{12}|}$) plane.~~
(3a)$-$(3d): Distributions on a 
($M_{T2}(m_{\tilde \chi_1^0})$, ${\rm sign}(-\beta^2_{12}) \sqrt{|\beta^2_{12}|}$) plane.~~
(4b)$-$(4d): Distributions on a
($2N_{\rm wrong}/N_{\rm parton}$, ${\rm sign}(-\beta^2_{12}) \sqrt{|\beta^2_{12}|}$) plane.~~
The correct, random, $M_{TGen}$ and hemisphere assignments are adopted
in figures ($i$a), ($i$b), ($i$c) and ($i$d), respectively, where $i=1-4.$~~  
We generate $6\times 10^4$ events at Point 5 with 7 TeV proton centre of mass energy. 
}
\label{fig:sec3figs}
\end{center}
\end{figure}

Fig.\,\ref{fig:sec3figs} (1a) shows the $M_{T2}$(true) distribution in parton level.
As can be seen, the $M_{T2}$(true) distribution has a sharp endpoint at 
$\max\{ m_{\tilde q}, m_{\tilde g}\}=m_{\tilde q} \simeq910$\,GeV.
We generate 60000 events  at Point 5 with 7\,TeV proton centre of mass energy,
and ISR is not included.
We apply the following ``minimal cuts" to reduce the standard model background\footnote{
Recent studies adopt much higher cuts to reduce backgrounds, 
but they do not  reduce the signal significantly.}
\begin{description}
\item[ 1.]  $E_{T}^{\rm miss}> \max(100\,{\rm GeV}, 0.2M_{\rm eff}$),\,
\item[ 2.]  $M_{\rm eff} > 500$\,GeV,\,
\item[ 3.]  $N^{\rm jets(parton)}_{p_T>100\,{\rm GeV}} \ge 1$,
\end{description}
in addition, we also define 
\begin{description}
\item[ 4.]   $N^{\rm jets(parton)}_{p_T>50\,{\rm GeV}} \ge 4$~~({\rm ``4~jet~cut"}),~~~{\rm or}
~~~$N^{\rm jets(parton)}_{p_T>200\,{\rm GeV}} \ge 2$~~({\rm ``2~jet~cut"}), 
\end{description}
for section \ref{sec:jet}.

If $v_i$ is not defined appropriately,
$M_{T2}$ can exceed $\max \{ m_{A_1}, m_{A_2} \}$.
Fig.\,\ref{fig:sec3figs} (1b) shows the $M_{T2}$ distribution 
where partons are randomly assigned into $v_1$ or $v_2$
so that each system has at least one constituents.    
We can see that the distribution has a large tail,
and the endpoint structure is not seen.

To see what kinds of events and assignments generate large $M_{T2}$ values,
we show distributions with the random assignment on 
($M_{T2}$, $\max(m_{v1}, m_{v2})$) 
and
($M_{T2}$, ${\rm sign}(-\beta^2_{12}) \sqrt{|\beta^2_{12}|}$) 
planes in Fig.\,\ref{fig:sec3figs} (2b) and (3b), respectively. 
Here, $\beta^2_{12} \equiv \overrightarrow \beta_1 \cdot \overrightarrow \beta_2$
is an inner product between velocity vectors of $A_1$ and $A_2$ in the lab frame.
From figure (2b), we can see that the $M_{T2}$ is linearly dependent  on the heavier of 
$m_{v1}$ and $m_{v2}$. 
Thus, the assignments that provide large $M_{T2}$
also provide large $m_{v_i}$.
In addition, from figure (3b) we can see that
the endpoint of the $M_{T2}$ distribution increases as increasing
${\rm sign}(-\beta^2_{12}) \sqrt{|\beta^2_{12}|} $ from 0 to 0.8.
The tail of the $M_{T2}$ distribution is mainly caused by the events
with large ${\rm sign}(-\beta^2_{12}) \sqrt{|\beta^2_{12}|}$.
In such events, $A_1$ and $A_2$ are highly boosted into back-to-back.

\begin{figure}[t]
\begin{center}
\vspace{-5mm}
\includegraphics[width=7cm] {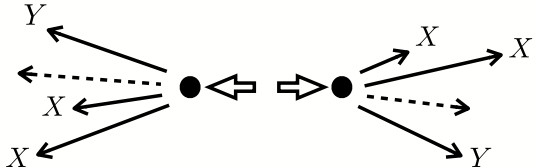}
\caption{A schematic picture of an assignment 
that provides a dangerously large $M_{T2}$ value in an event with $\sqrt{\hat s} \gg M_1+M_2$.
Dashed arrows represent the missing particles. 
Visible objects are grouped into $X$ or $Y$.}
\label{fig:configuration}
\end{center}
\end{figure}

Fig.\,\ref{fig:configuration} is a schematic picture of a wrong assignment that provides large $M_{T2}$.
In the picture two initial particles are highly boosted into back-to-back, and daughter particles are
incorrectly assigned into groups, $X$ and $Y$.
As can be seen, if two daughters, $i$ and $j$, have different origin,
the back-to-back boost makes their angle $\theta_{ij}$ and momentum magnitudes large.
This results a large invariant mass of the visible systems.
 
For the events with large back-to-back boosts
and two daughter particle momenta from the same sparticle makes  small angle
compared to ones from different origin.
The idea of hemisphere algorithm is based on this observation,
although the angle $\theta_{ij}$ is modified into more elaborate distant measure, $d$.
Interestingly, even if the algorithm fails to find the correct assignment,
$m_{v_i}$ would not be too large because members in the same group have relatively small $d$.
In the algorithm, the highest $p_T$ jet is assigned into a different group from 
the one contains the jet with the largest $p_{T_i} \Delta R(p(J_1),p(i))$.
This also prevents from making $m_{v_i}$ too large.

Fig.\,\ref{fig:sec3figs} (1c), (2c) and (3c) correspond to (1b), (2b) and (3b), respectively
but hemisphere algorithm is used.
From figures ($i$c) ($i=1,2,3$), we can see that the tails in figures ($i$b) are removed significantly, 
and the endpoint structure at $\max \{ m_{A_1}, m_{A_2} \}$ is recovered.
Fig.\,\ref{fig:sec3figs} (4b) and (4c) shows distributions on 
($2N_{\rm wrong}/N_{\rm parton}$, ${\rm sign}(-\beta^2_{12}) \sqrt{|\beta^2_{12}|}$) planes 
for the random and hemisphere algorithm, respectively.
Here, $N_{\rm wrong}$ is the number of wrong assignments.
As expected,
in events with large ${\rm sign}(-\beta^2_{12}) \sqrt{|\beta^2_{12}|}$ ($\gsim 0.4$),
the algorithm successfully selects the correct assignment with high probabilities.
Compared to (4b), the improvement is significant.
Note that large back-to-back boosts require $\sqrt{\hat s} \gg m_{A_1}+m_{A_2}$,
where $\sqrt{\hat s}$ is centre of mass energy of colliding partons.
An event sample used in Fig.\,\ref{fig:sec3figs} is at Point 5 with $\sqrt{S} = 7$\,TeV,
which has the largest sparticle mass scale and the smallest centre of mass energy of protons
in our samples shown in Table.\,\ref{masstable}.
The efficiencies of the algorithm would be better for the other samples, or at 
higher $\sqrt{S}$

Another known algorithm is $M_{TGen}$ method.
By its construction, this variable is smaller than or equal to $M_{T2}$(true) in event by event basis.
Fig\,\ref{fig:sec3figs} ($i$d) ($i=1-4$) correspond to ($i$c) but $M_{TGen}$ method is adopted.
From figure (4d), we can see that $M_{TGen}$ can select the correct assignment
in events with large ${\rm sign}(-\beta^2_{12}) \sqrt{|\beta^2_{12}|}$ ($\gsim 0.4$)
as hemisphere algorithm.
This is because in such events all wrong assignments can provide larger $m_{v_i}$ 
than that from the correct assignment because of the back-to-back boost.

Compared to hemisphere algorithm, $M_{TGen}$ distribution does not have a tail
beyond $\max \{ m_{A_1}, m_{A_2} \}$ in parton level without ISR.
We fit distribution (1c) to two linear functions, $f(M_{T2})=a_{\pm} (M_{T2}-M_{T2}^{\rm edge})+b$
for ${\rm sign}(M_{T2}-M_{T2}^{\rm edge})=\pm 1$, 
and (1d) to a linear function, $f(M_{TGen}) = c - d M_{TGen}$.
The fitted endpoints is $912.2\pm 9.6\,$GeV for (1c).
On the other hand, $M_{TGen}^{\rm edge}=c/d$ are $906.8\pm 85.2$ for (1d).
The main source of the large error in the $M_{TGen}^{\rm edge}$ is poor statistics 
near the endpoint region.
To improve the error we can fit the distribution away from the endpoint.
However in this case, the central value become lower than the input squark mass. 
Both results are consistent with $\max \{ m_{\tilde q}, m_{\tilde g} \}=m_{\tilde q}\simeq910$\,GeV.

As can be seen, the slope of the $M_{TGen}$ distribution is sharper than that of the $M_{T2}$ distribution
with hemisphere algorithm. 
This feature relies on the property that the $M_{TGen}$ can not exceed the mass of a sparticle.
We will show that this property is not held for the $M_{TGen}$(min) and the $M_{TGen}^{\rm mod}$(min).
In the rest of this paper, we adopt hemisphere algorithm unless it is explicitly stated,
because both algorithms have similar performance 
but $M_{TGen}$ calculation is more computational intensive.
In addition, as we will show later, hemisphere algorithm can be also applied 
to find out decay topology in the events.

\section{Parton level $M_{T2}(\rm min)$ and $M_{T2}^{{\rm mod}}({\rm min})$ distributions with and without ISR }
\label{sec:parton}

In section \ref{sec:isr-mixed},
we defined $M_{T2}$(min) and $M_{T2}^{\rm mod}({\rm min})$ variables
for gluino mass measurement.
The idea of these variables is that the events above the $M_{T2}$(true) endpoint 
can be removed by subtracting the hardest ISR jet.
Although those variables are designed for the events with a hard ISR jet,
 we first show parton level $M_{T2}$(min) and $M_{T2}^{\rm mod}({\rm min})$
distributions without ISR.
SUSY processes are the mixture of the events with and without hard ISR jets. 
We therefore check that parton level 
$M_{T2}$(min) and $M_{T2}^{\rm mod}({\rm min})$ 
distributions do not ``collapse" near the 
endpoint by removing a parton from sparticle decays.

\begin{figure}[t!]
\begin{center}
\includegraphics[width=6cm] {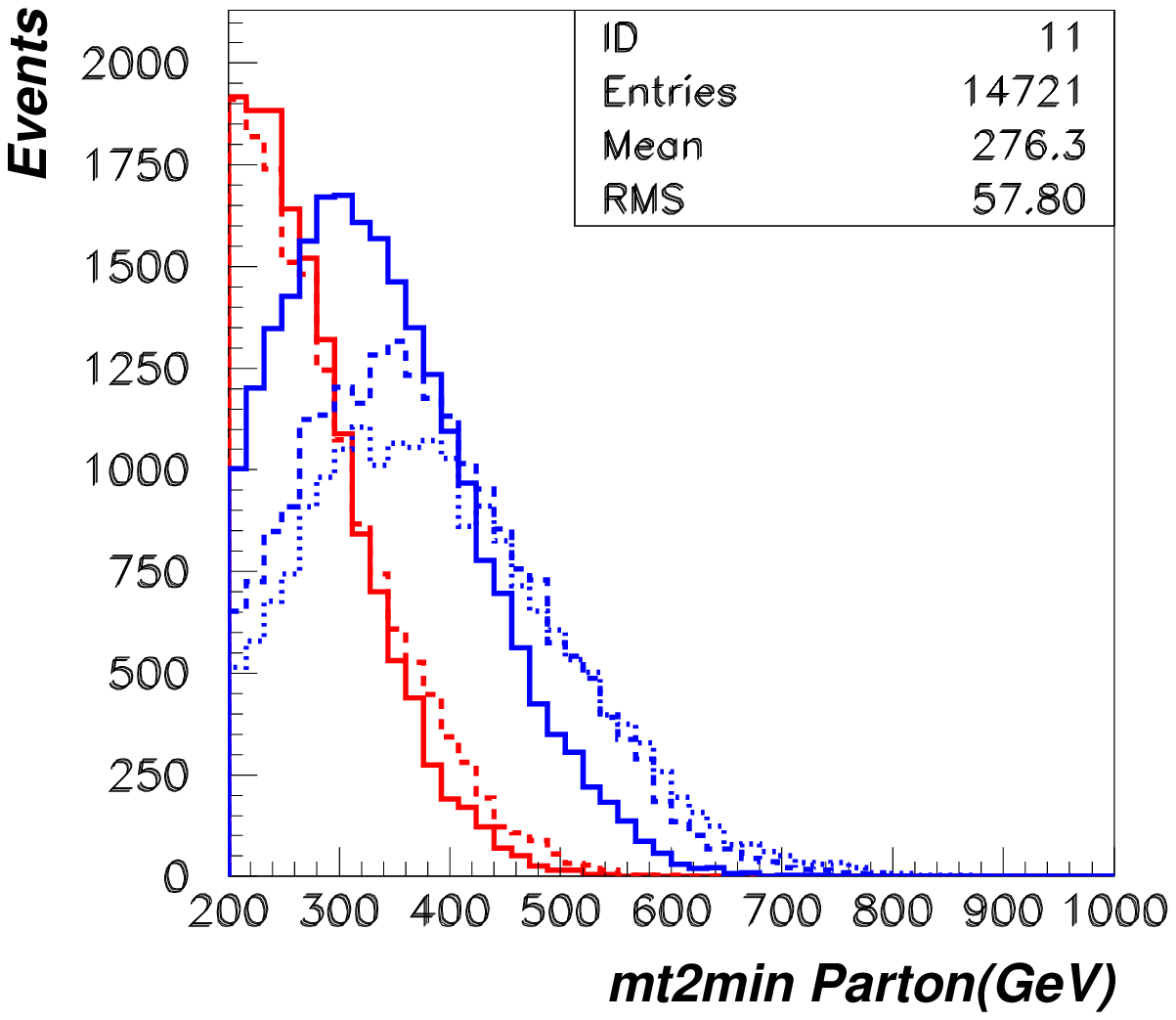}
\hspace{5mm}
\includegraphics[width=6cm] {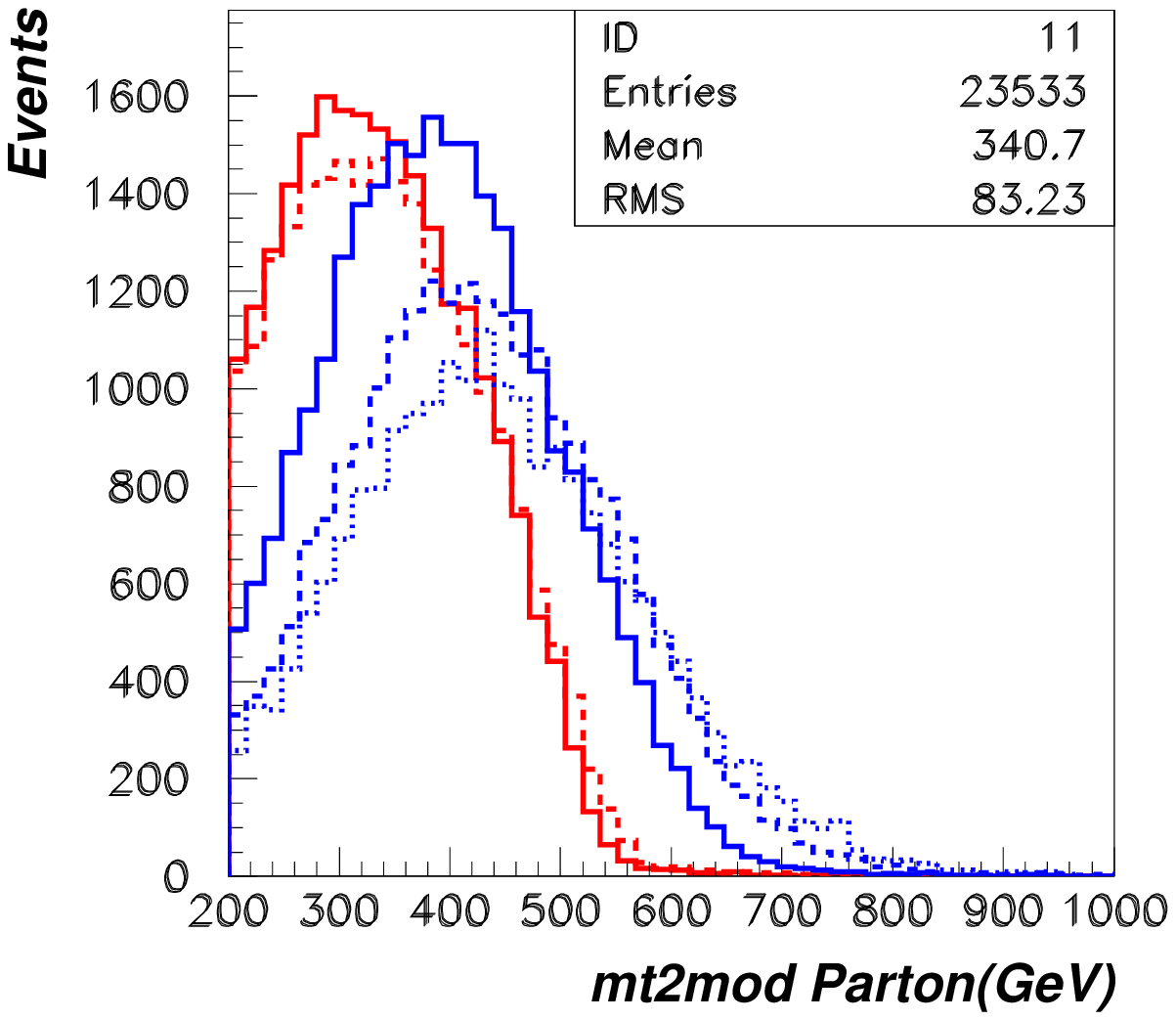}
\caption{Left; $m_{T2}({\rm min})$ distributions in parton level. 
Right; $m_{T2}^{\rm mod}({\rm min})$ distributions in parton level.
Red-solid, -dashed, Blue-solid, -dashed and -dotted distributions correspond to
Points 1, 2, 3, 4 and 5, respectively. 
We generate 60000 events at 14\,TeV.
}
\label{fig:mt2minmodp}
\end{center}
\end{figure}

\begin{figure}[t!]
\begin{center}
\hspace{-1cm}
\includegraphics[width=10cm] {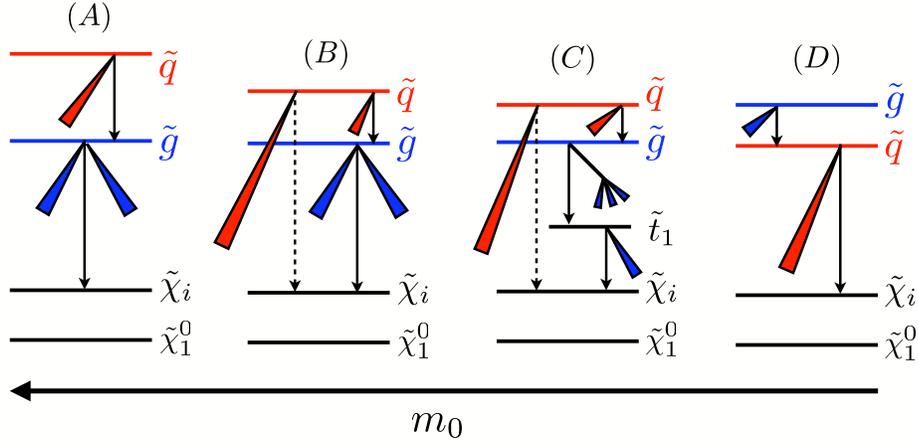}
\caption{Schematic picture depicting decay patterns for various mass spectra.
A, B, C and D correspond to $m_{\tilde q} \gg m_{\tilde g}$, $m_{\tilde q} > (m_{\tilde t_1}+m_t) > m_{\tilde g}$,
$m_{\tilde q} >  m_{\tilde g} > (m_{\tilde t_1}+m_t)$ and $m_{\tilde g} >  m_{\tilde q}$, respectively.
Red and blue triangles represent jets from squark and gluino decays, respectively.
Dashed arrows represent the decays are not main modes. 
} 
\label{fig:decay_figs}
\end{center}
\end{figure}

Fig.\,\ref{fig:mt2minmodp} (left) shows 
$M_{T2}({\rm min})$ distributions,
where red-solid and -dashed distributions correspond to Points 1 and 2 ($m_{\tilde g} > m_{\tilde q}$),
and blue-solid, -dashed and -dotted ones correspond to Points 3, 4 and 5 ($m_{\tilde q} > m_{\tilde g}$),
respectively. Here 
we generate 60000 events at 14\,TeV, and 
the minimal cuts defined in the previous section is applied.
For Points 3, 4 and 5, the endpoints of the distributions roughly agree with 
 the gluino masses
636, 646 and 651\,GeV, respectively.
On the other hand, for Points 1 and 2, the endpoints of the distributions are 
about $100-200$\,GeV smaller than the input gluino masses.

The difference in the distributions at Points 1, 2 and those at Points 3, 4, 5
comes from the ordering on the gluino masses.
Fig.\,\ref{fig:decay_figs} shows various mass spectra and corresponding decay patterns.
Points 5 and 3, 4 correspond to types $A$ and $B$ in Fig.\,\ref{fig:decay_figs}, respectively. 
A squark dominantly  decays to $j \tilde g$, and a gluino decays to $jj \tilde \chi_i$
in the region. 
Although there is a large mass hierarchy between gluino and weak gauginos,
$p_T$ of jets from the gluino decay are relatively mild because of the three body decay.     
Consequently, the events tend to have multiple jets with modest $p_T$. 

On the other hand, mass spectra at Points 1 and 2 correspond to type $D$ in Fig.\,\ref{fig:decay_figs}.  
A squark decays to two body final state $j \tilde \chi_i$ producing a high $p_T$ jet
for the mass spectrum.
Because a gluino decays to $j \tilde q$, SUSY events usually contain two high $p_T$ jets
from the squark decays.   
If we subtract one of the high $p_T$ jets from the event,
 the remaining system loose too much energy and calculated $M_{T2}$
can not reach at the gluino mass.

A variable $M_{T2}^{\rm mod}$(min)  is designed 
to reconstruct the gluino mass in type $D$ spectrum.
In the definition, we do not subtract the two highest $p_T$ jets.
Fig.\,\ref{fig:mt2minmodp} (right) shows 
$M_{T2}^{\rm mod}({\rm min})$ distributions in parton level,
where the colour and line schemes are the same as the LHS figure.
The endpoints for Points 1 and 2 are consistent with
 the input gluino masses
612 and 629\,GeV, respectively.  
On the other hand, for Points 3, 4 and 5, the endpoints are about 100\,GeV larger than the gluino masses.
The events above the input gluino mass are mostly $\tilde g$-$\tilde q$ production events.
This suggests that $M_{T2}$(min) and $M_{T2}^{\rm mod}$(min) should be used
for appropriate mass spectrum and decay pattern.
Namely we need to know the mass ordering of gluino and squark in the gluino measurement.
We will discuss this point in the next section.

\begin{figure}[t!]
\begin{center}
\includegraphics[width=6cm] {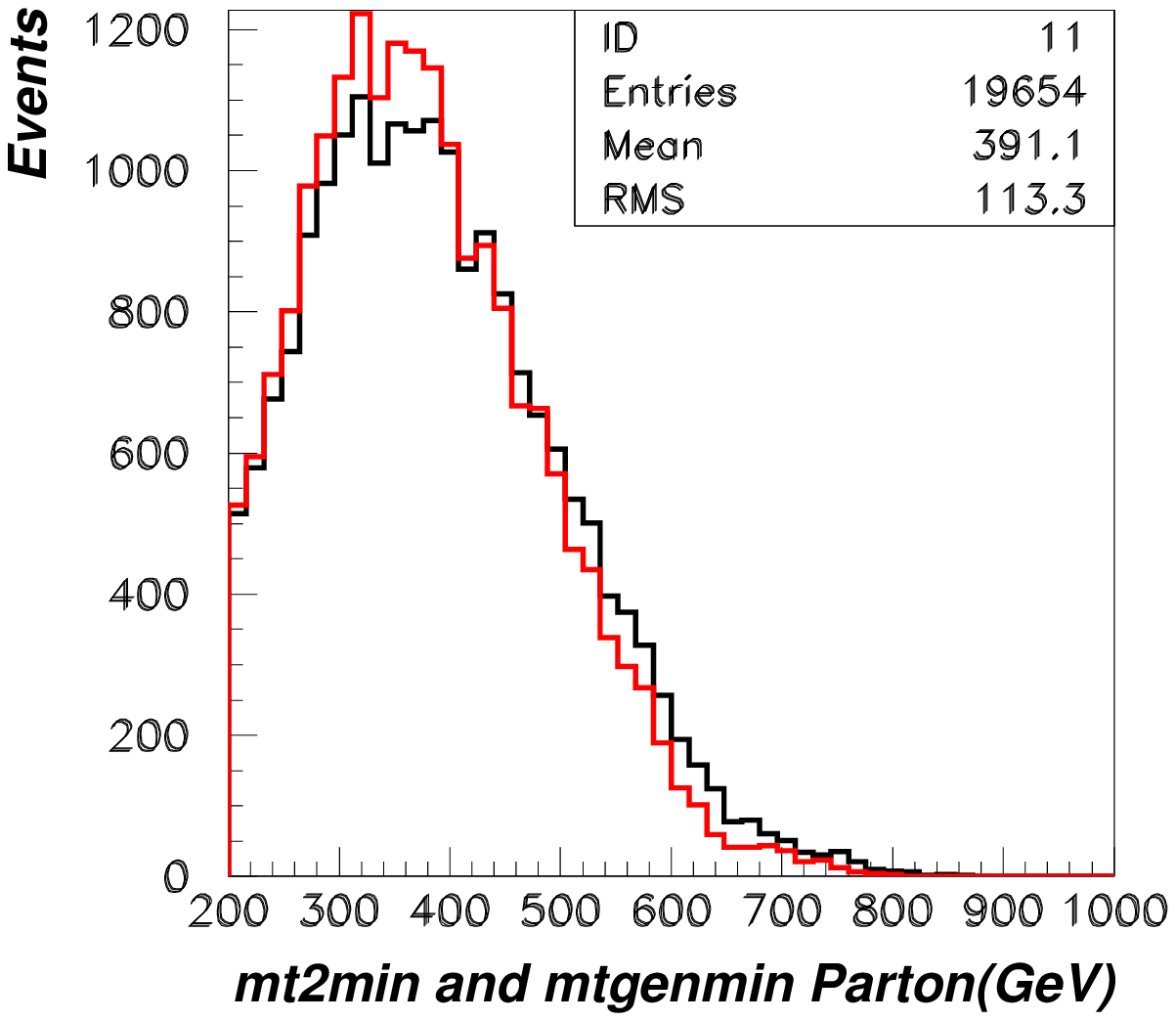}
\hspace{5mm}
\includegraphics[width=6cm] {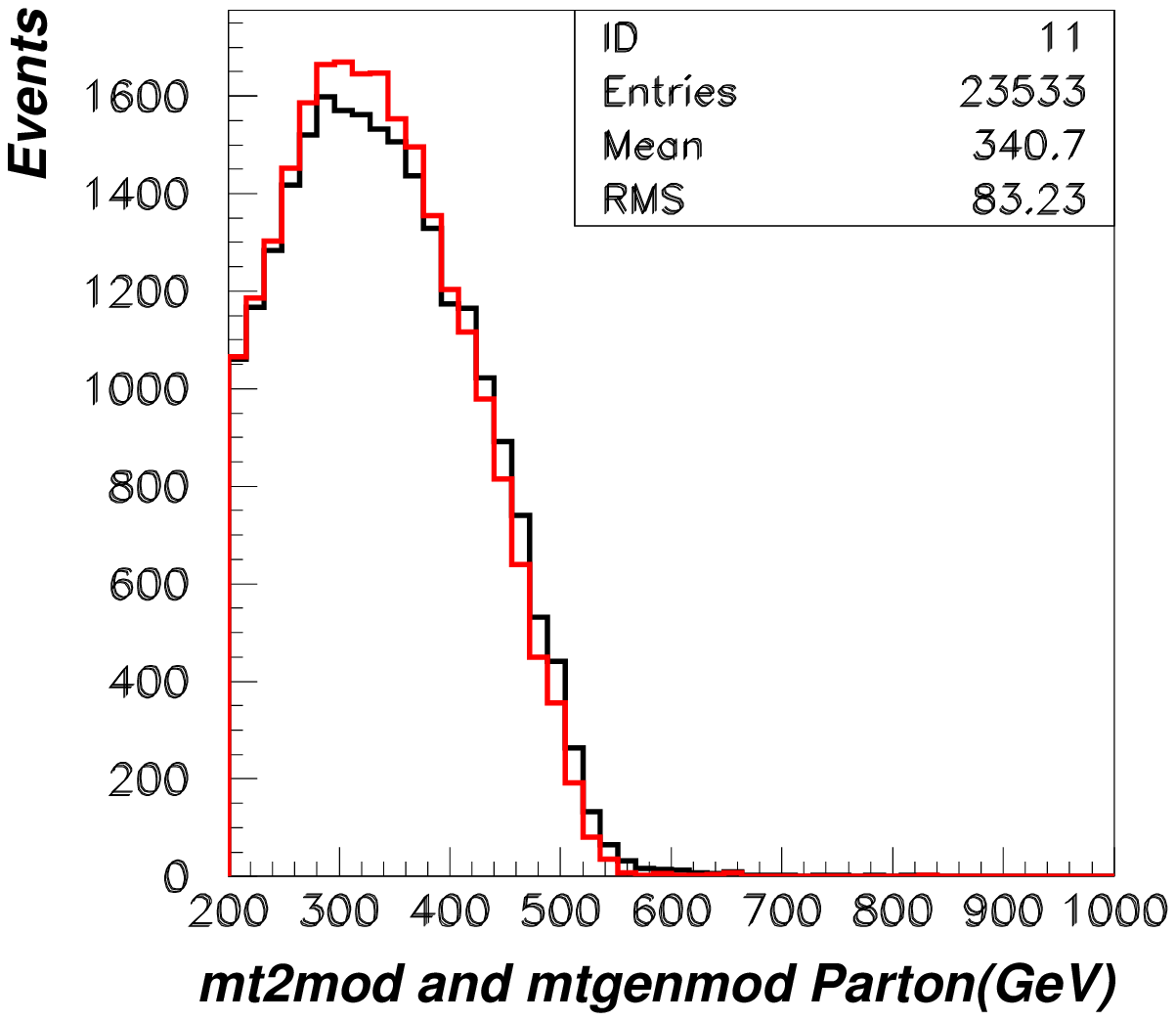}
\caption{Left; parton level $m_{T2}({\rm min})$ (black) and $m_{TGen}({\rm min})$ (red) distributions 
at Point 5. 
Right; parton level $m_{T2}^{\rm mod}({\rm min})$ (black) and $m_{TGen}^{\rm mod}({\rm min})$ (red) 
distributions at Point 1.}
\label{fig:mt2mingen}
\end{center}
\end{figure}

$M_{TGen}^{\rm mod}$(min) and $M_{TGen}^{\rm mod}$(min) variables can be defined 
analogously to $M_{T2}^{\rm mod}$(min) and $M_{T2}^{\rm mod}$(min), respectively.
The left figure in Fig.\,\ref{fig:mt2mingen} shows
comparisons between $M_{T2}$(min) (black) and $M_{TGen}$(min) (red) distributions at Point 5.
Both the distributions have endpoints near the input gluino mass but also tails which 
mainly come from $\tilde q$-$\tilde g$ production events.
Similarly, the right figure in Fig.\,\ref{fig:mt2mingen} shows
$M_{T2}^{\rm mod}$(min) (black) and $M_{TGen}^{\rm mod}$(min) (red) distributions at Point 1.
The endpoints are given by the input squark mass.
At Point 1 ($\mgl > \msq$), $\tilde g$-$\tilde g$ production events can produce $M_{T2}^{\rm mod}$(min)
and $M_{TGen}^{\rm mod}$(min) larger than the squark mass.
However the cross section of this production process is subdominant and the tails are tiny.
To reconstruct the gluino mass, inclusion of ISR is necessary as we will show in the next section.   
We may slightly improve the distributions by adopting $M_{TGen}^{(\rm mod)}$(min) instead of
$M_{T2}^{(\rm mod)}$(min), although it would be computational intensive especially for the events
with a large number of jets.

We now discuss effect of hard initial state radiations associated with 
hard processes at LHC. 
It was explicitly shown in Ref.\cite{hiramatsu}  that
\begin{enumerate} 
\item  The hardest ISR parton in $\tilde{g}$ and $\tilde{q}$ production processes could be harder than 
partons coming from $\gl$ three body decay $\gl \rightarrow j j  \tilde \chi^0_1$. 
The ISR  causes endpoint smearing of  the inclusive $M_{T2}$ distributions.
Significant events  appears beyond the $M_{T2}$(true) endpoint  
 for $\gl$-$\gl$ production when the $M_{T2}$ is calculated from the four highest $p_T$ jets. 
 
\item  The $\mttwmin$ distribution is less affected by ISR, because the leading 
ISR effect can be removed. It is explicitly shown that 
 $M_{T2}^{\rm max}({\rm min}) \simeq \mgl$ for the process $pp\rightarrow \gl \gl$   
 when a gluino is forced to decay through three body decay mode  $\gl \rightarrow  jj \tilde \chi_1^0$.  
\end{enumerate}

\begin{table}
\begin{center}
\begin{tabular}{|c||c|c|c|c|}
\hline         
 & Point 1& Point 3 & Point 5  & Point 3' \cr
 \hline
$\sigma(\gl\gl j)$(inclusive) /$\sigma(\gl\gl)$ (exclusive)  & 1.48\ & 1.61& 1.72& 1.42 \cr
$\sigma(\gl\sq j)$(inclusive) /$\sigma(\gl\sq)$(exclusive)&
 0.69& 0.77& 0.84& 0.69  \cr
\hline 
\end{tabular}
\caption{Ratio between the number of events with and without a hard ISR parton. See text for the detail} 
\label{inexratio}
\end{center}
\end{table}

The ratio of events with a hard ISR to those without any hard ISR is summarized in 
Table\,\ref{inexratio}. 
Here we show the ratio of the number of events,  
$N$(inclusive  $\gl\gl j$ or $\gl\sq j$ sample)/$N$(exclusive 
$\gl\gl$ or $\gl\sq$ sample)  calculated by Madgraph/Madevent\cite{madgraph}
with Pythia parton shower\cite{phythia}. 
Here,  the ``exclusive sample"  is generated from $\gl$-$\gl$ or $\gl$-$\sq$ 
process with parton shower, but if parton shower is resolved, namely
if a cluster of the partons  is  isolated from the initial state with more than a certain $k_T$  distance, 
the event  is rejected.  
 On the other hand the ``inclusive sample" means 
the events generated from $pp \rightarrow \gl\gl j$ or $\gl\sq j$ matrix elements with full parton  showers, where $j$ denotes  gluon or quark. 
Again, if the event does not have a resolved parton cluster, the event 
is rejected. 
Altogether,  there is no overlap  between exclusive and inclusive samples.  
The parton level distribution of the \lq\lq matched sample" is therefore the sum of the events of the inclusive sample and exclusive sample.
 The $k_T$ cut off scale is chosen so that there are no discontinuity in the total distribution.\footnote{
The $k_T$ cut depends on the shower algorithm. 
We use pythia $p_T$ ordered shower.  The $k_T$ cut is 60\,GeV and 
the pythia shower scale is 100\,GeV.}  
As it has already discussed in Ref.\cite{hiramatsu}, 
the fraction of the inclusive events in $\gl\gl j$ matched samples is always higher than that in
$\gl\sq j$ matched samples. 
This is because the difference in colour factors between $\gl$ and $\sq$.

\begin{figure}
\begin{center}
\includegraphics[width=6cm]{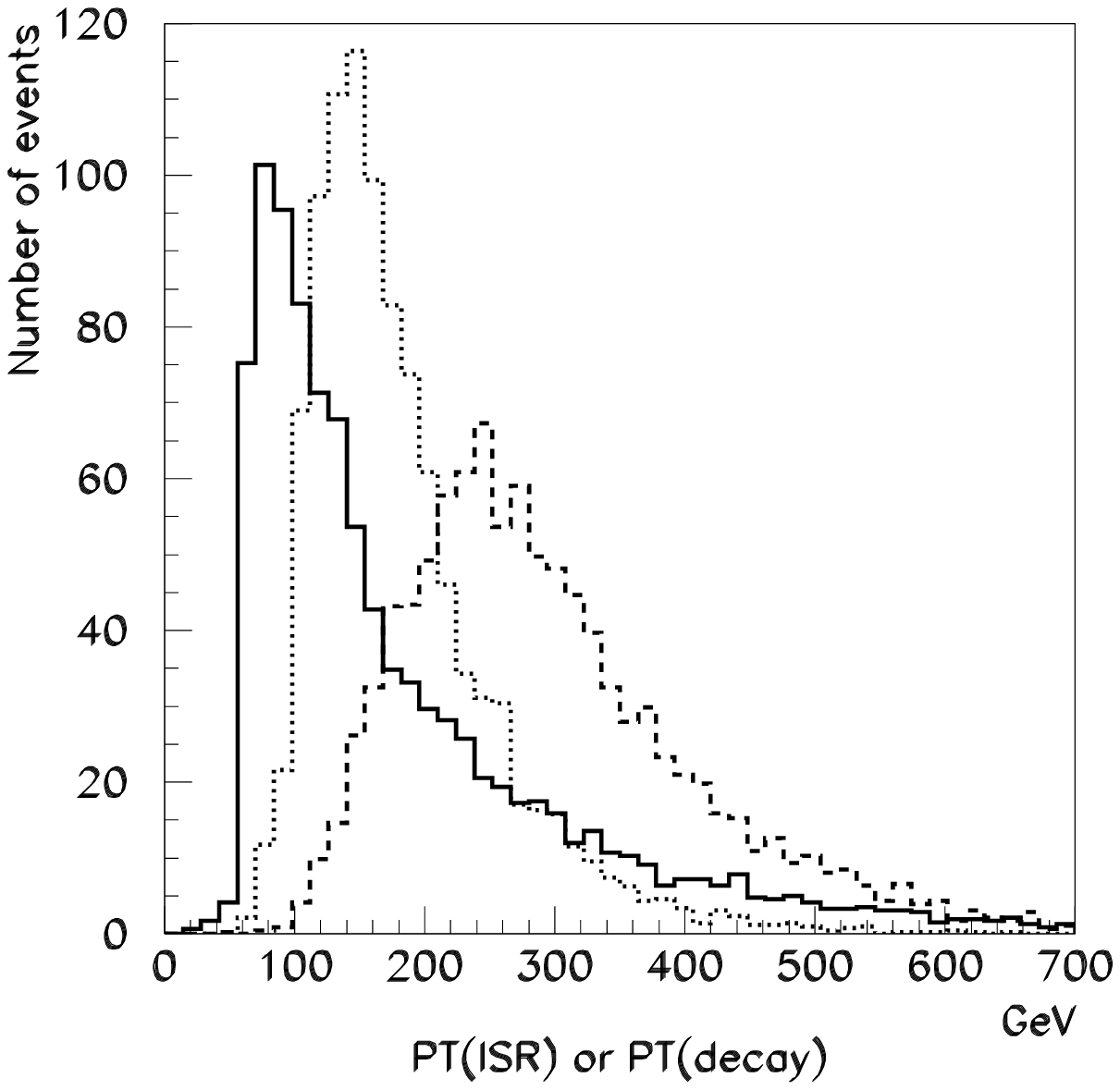}
\includegraphics[width=6cm]{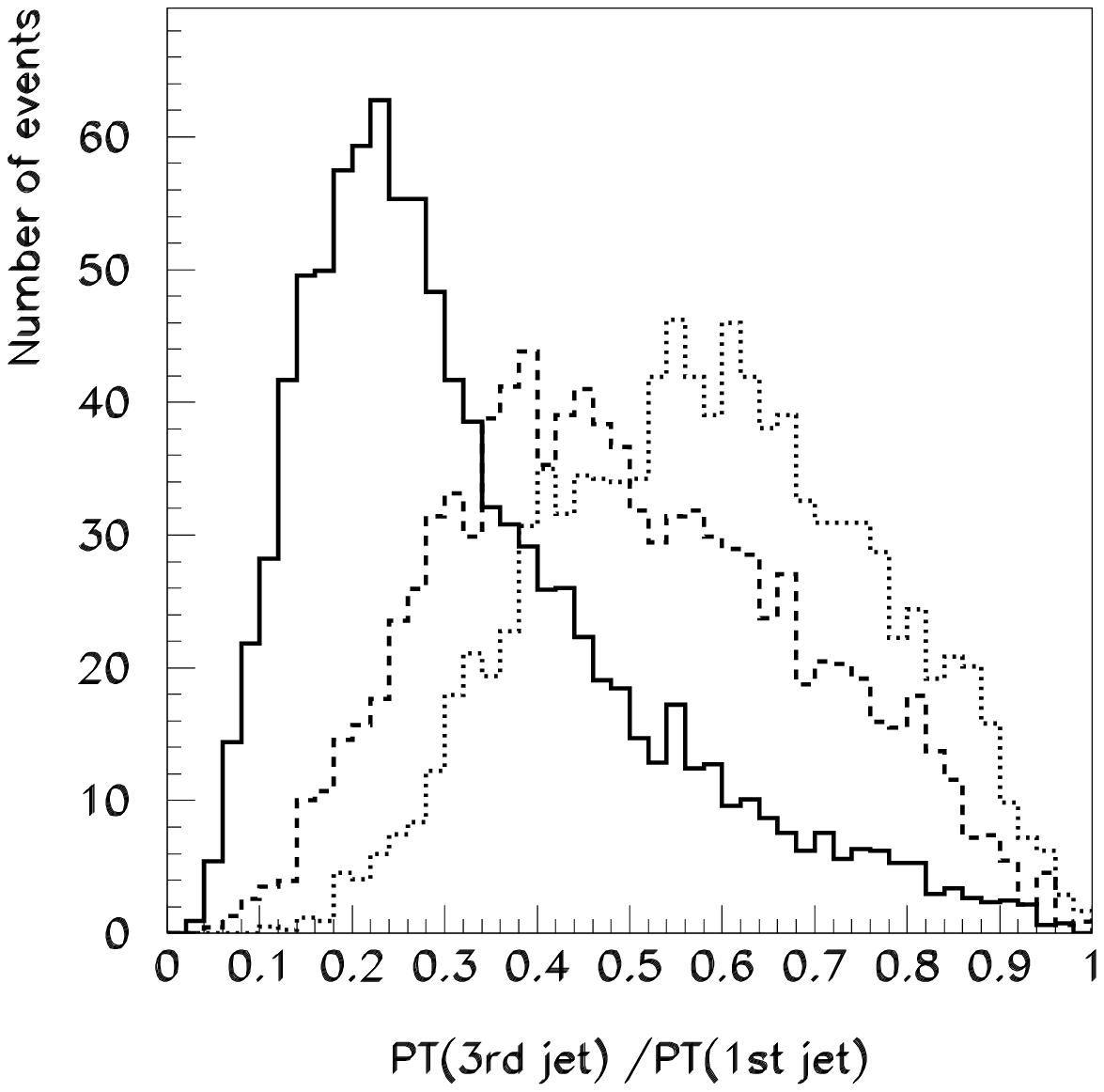}
\end{center}
\caption{Left: the $p_T$ distribution of ISR for $\gl\gl j$ process at Point 5 (solid), the highest 
$p_T$ decay products  at Point 5(thick dotted) and Point 3' (thin dotted). 
Right: ratio of the $p_T$ of the third and first $p_T$ partons  from gluino decays 
at  Point 1 (solid) Point 5 (thick dashed) and Point 3'(thin dashed) for the same process. The distributions are scaled from  20000 $\gl\gl j$ matched events to 1000 events at each point.  }\label{working}
\end{figure}

In Fig.\,\ref{working} (left) we show the $p_T$ distribution of the ISR  parton at Point 5 (a solid line).  
The $p_T$ is 200\,GeV in average. The $p_T$ distributions of the ISR are roughly the same 
for all model points in this paper. 
On the other hand,  the $p_T$ distributions of the partons from squark/gluino decay  
are rather model parameter dependent. 

In the same figure we show the $p_T$ distributions of the highest $p_T$ partons 
from $\gl$ decays for the $\gl\gl j$ matched sample at Point 5 (thick dotted) and at Point 3' (thin dashed). 
The gluino dominantly decays through three body final state $\tilde g \to jj \tilde \chi_i$ at Point 5.
At Point 3' the mass spectrum corresponds to type $C$ in Fig.\,\ref{fig:decay_figs},
where $m_{\tilde g} > (m_{\tilde t_1}+m_t)$.
In this type of spectrum, a gluino dominantly decays into two-body final
state $\tilde g \to \tilde{t}_1 t$,
and the top and stop further decay to lighter particles.
Thus, the highest $p_T$ jet is relatively soft compared to that of Point 5 and
the scale of $p_T$ is close to that of ISR.
On the other hand, the distribution of the highest $p_T$ jet at Point 1 is similar to that at Point 5.

Not only the distribution of the highest $p_T$ jet, 
but also the distribution of the other jets could be different. 
In Fig.\,\ref{working} (right), we plot the distributions of $p_{T3}/p_{T1}$ 
at Points 1 (solid), 5 (thick dashed) and 3' (thin dashed). 
At Point 1, the highest $p_T$ jet takes significant part of the total 
energy of the event because $\tilde{q} \rightarrow q \tilde \chi_i$ modes dominate. 
While at Point 5, four partons from the gluino three body decays have roughly the same order of $p_T$. 
At Point 3', the tendency is even stronger.

\begin{figure} 
\begin{center}
\includegraphics[width=4.7cm]{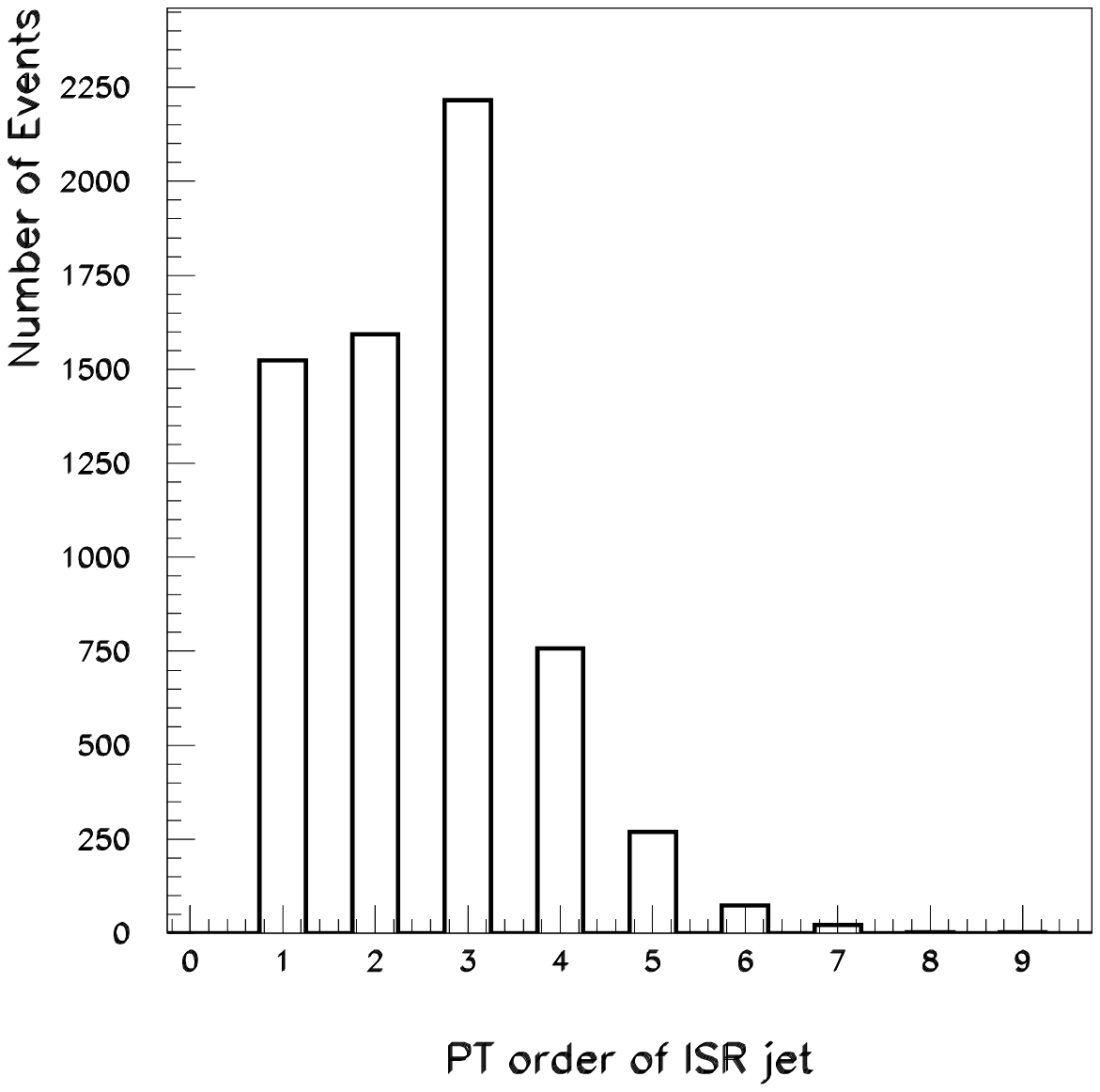}
\includegraphics[width=4.7cm]{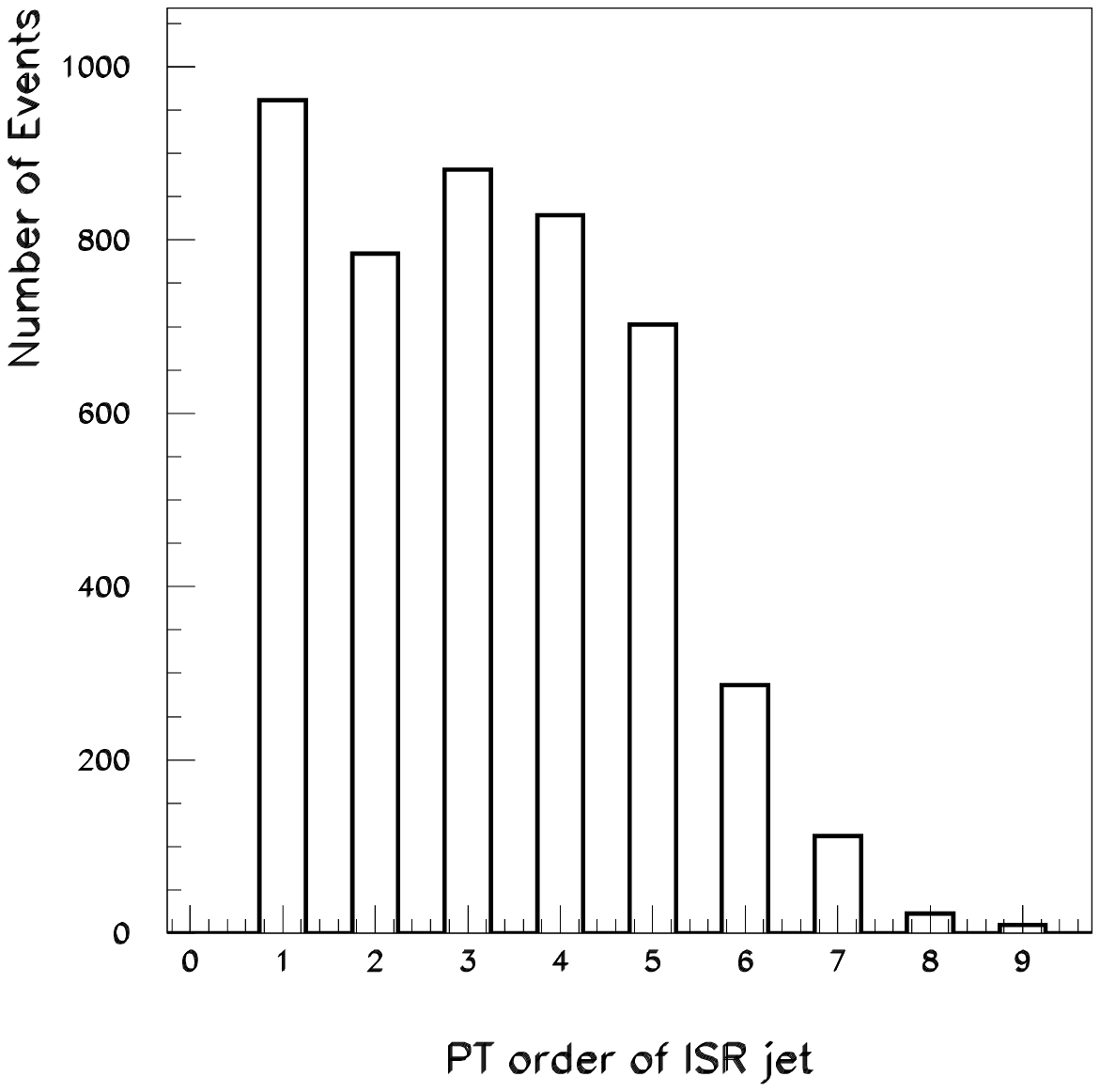}
\includegraphics[width=4.7cm]{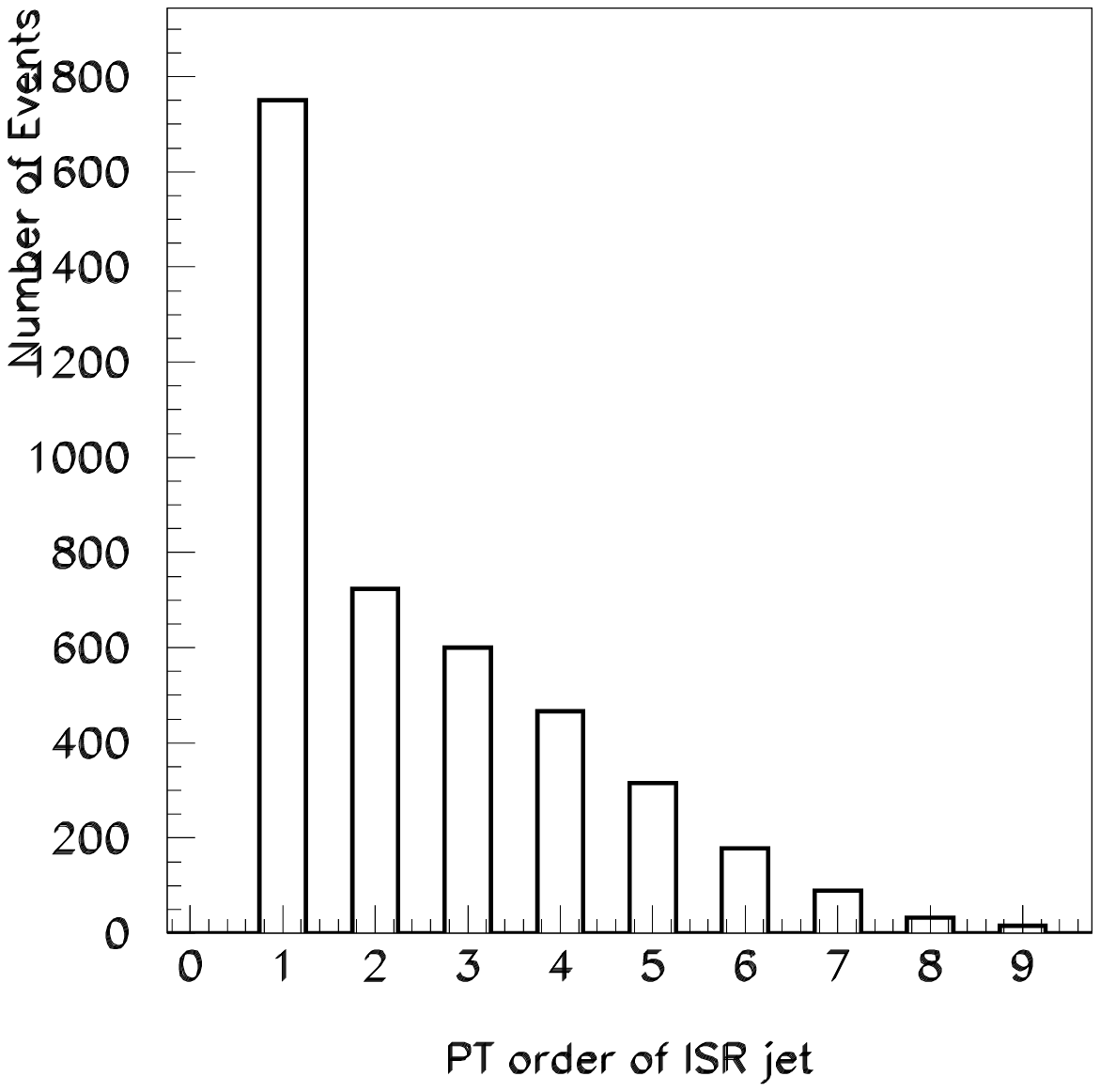}
\caption{The $p_T$ order of the ISR parton among all partons of the inclusive  $\tilde g \tilde g j$
sample.
Left, center and right figures are at  Points 1, 5 and 3' respectively. }
\label{isrorder}
\end{center}
\end{figure}

In Fig.\,\ref{isrorder}, we 
plot the distribution of $i_{\rm ISR}$ where $i_{\rm ISR}$ is the order of an ISR parton among  
all the $p_T$ ordered partons in the $\gl\gl j$ inclusive sample   at Points 1 (left), 5 (middle) and 3' (right).  
Distributions for  $\sq \gl j$ inclusive sample are similar to those for $\gl \gl j$.

At Point 1, the average number of jets in the events are rather 
small,  but about a half of the events has more than four partons with $p_T> 50$\,GeV for $\gl$-$\gl$ production.  
The probablity that the  ISR parton  becomes the 1st or the 2nd hardest parton is smaller than the 
one to be the 3rd hardest parton, because at this point a squark dominantly decays into 
$\tilde \chi_i$ and  a high $p_T$  quark. 
 At Point 5, a gluino decays into three body final state $jj \tilde \chi$. 
 The ISR parton is one among the five highest $p_T$ partons  in that case. 
 Finally, at Point 3', the ISR tends to be the highest $p_T$ jet.
This is because, a gluino dominantly decays into $t \tilde t$ at this point.  
The average number of jets is large, 
and at the same time each of the  jets from squark/gluino 
decay has relatively small $p_T$. 

\begin{figure}[!t]
\begin{center}
\includegraphics[width=6cm]{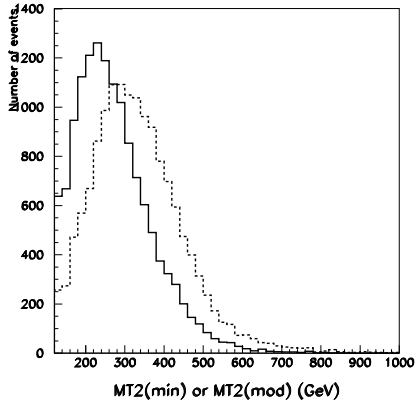}
\caption{Parton level $\mttwmin$(a solid line)  and $\mttwmod$(a dashed line) distributions for the matched $\gl \gl j$ at Point 1. 
The  $\mttwmod$ distribution has more 
events near the endpoint because of the events that ISR jets are correctly 
removed by the minimization. The distributions are based on 40000 matched $\gl$-$\gl$ events 
generated by Madgraph.   }
\label{point1isr}
\end{center}
\end{figure}

We now consider the events that  contribute to the endpoint of the $\mttwmin$.   
For  exclusive samples, the $\mttwmin$ endpoints 
should be smaller than those of the $\mttwo$.  
Especially at Point 1, removing one of  the two highest $p_T$ partons 
 from the system reduce the events near the endpoint too much as 
 discussed previously.  
 Instead of using the $\mttwmin$, 
 we use $\mttwmod$ to obtain the endpoint. 
For the inclusive sample at Point 1, ISR tends 
to be the 3rd hardest parton of the event, and  using 
the $\mttwmod$ is justified also from this point of view.
In Fig.\,\ref{point1isr} 
we show the parton level distributions of $\mttwmod$ and 
$\mttwmin$  for the $\gl\gl j $ matched samples. 
The number of events near the input gluino mass 
 612\,GeV  is quite small for the $\mttwmin$, 
while significant events remain for the $\mttwmod$ in the same region. 
Using the $\mttwmod$ distribution clearly improves the sensitivity to the gluino mass. 

\begin{figure}[t!]
\begin{center}
\includegraphics[width=12cm]{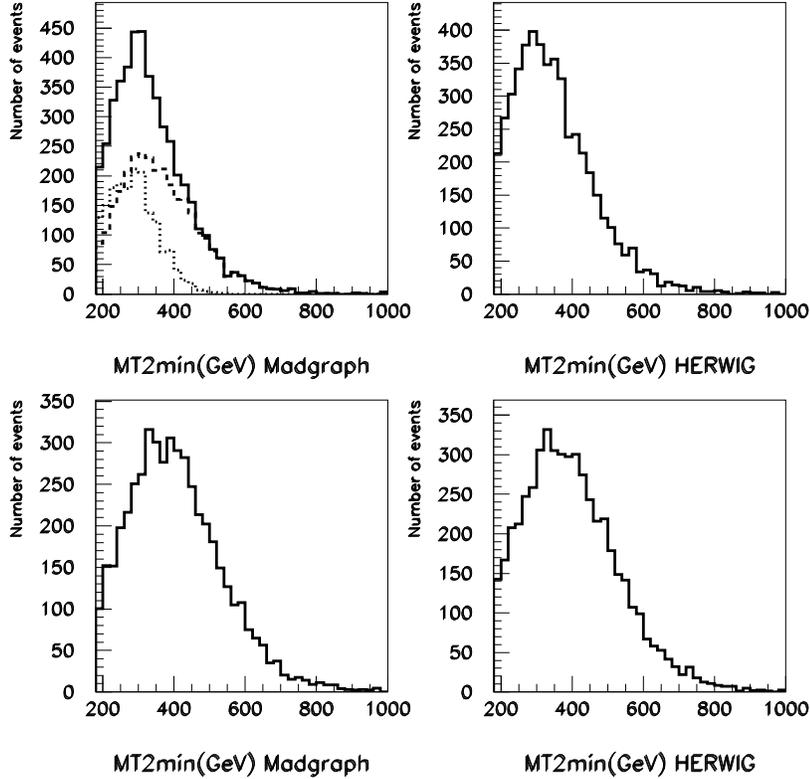} 
\caption{ Parton level $M_{T2}$(min) distributions   
of (top left) matched $\gl \gl j $ sample and (bottom left) 
matched $ \sq \gl j $ sample calculated by Madgraph + 
Pythia.  The corresponding jet level distributions obtained by 
HERWIG  with detector simulation are placed in the right; (top right) $\gl$-$\gl $ production 
and (bottom right) $\sq$-$\gl $ production.} 
\label{fig:isrmt2min}
\end{center}
\end{figure}

For Point 5, the minimization of $M_{T2}(i)$ over the five hardest jets is 
reasonable in removing the ISR jet, because the $i_{\rm ISR}$ distribution
shown in Fig.\,\ref{isrorder} is flat for the five hardest jets.
The endpoint of the $\mttwmin$ distribution for inclusive $\gl\gl j $ production 
would be near the input  gluino mass because  significant events has 
 a hard  ISR jet. 
 The $\mttwmin$ endpoint of the $\sq$-$\gl$ exclusive sample 
 would also be near the gluino mass. 
This is because if we remove the 
parton from $\gl \rightarrow \sq j$ decay, the system becomes $\gl$-$\gl$, 
and the endpoint for such events is given as the input gluino mass. 
Therefore, by minimizing over the five hardest jets, 
a relation $M_{T2}^{\rm mod,max}({\rm min}) \simeq \mgl$ is satisfied. 
The sum of the $\tilde g \tilde g j$ and no-ISR $\tilde q \tilde g$ productions provide the endpoint at $\mgl$.
Additional ISR jets for $\gl$-$\sq$ productions could lead the distribution that ends near the squark mass, 
which may smear the endpoint at $m_{\tilde g}$.  
Fortunately, the fraction of $\gl \sq j$ exclusive sample is less than a half of the total events at Point 5.  

Of course,  the shape of the distributions near the endpoints 
can only be estimated by Monte Carlo simulation.  
In Fig.\,\ref{fig:isrmt2min}, we show the parton level $\mttwmin$ distributions 
of the matched sample (left) at Point 5 generated by Madgraph/Madevent with Pythia, 
and compare it  jet level distributions generated by HERWIG with detector simulation.  Here the jet level event is reconstructed by the AcerDET\cite{acerdet} with jet energy smearing of 
$\delta E/E =50\,\%/\sqrt(E(GeV))$. Jet reconstruction of the 
AcerDET is replaced to the Cambridge-Aachen algorithm with $\Delta R=0.4$ using Fastjet\cite{fastjet}. 
 
  The top figures are the $M_{T2}$(min) distributions for $\gl \gl j$  
production sample and the bottom figures are for $\gl \sq$ production.   
At Point 5, the parton level endpoint  for the $\gl$-$\sq$ sample is around 700\,GeV, 
which is consistent with the input gluino mass, 651\,GeV
and far below the input squark mass, 910\,GeV.  
Note that the left and right distributions are roughly consistent although 
1) Our Madgraph samples include matrix element contribution of the hardest ISR parton, 
2) Our jet level samples obtained by HERWIG include all ISR effects in parton shower approximation.
3) HERWIG takes care spin correlation of matrix elements and all decay processes, 
while Pythia does not for sparticle decays.

\section{The results of jet level simulations }
\label{sec:jet}

\begin{figure}
\begin{center}
\includegraphics[width=9cm]{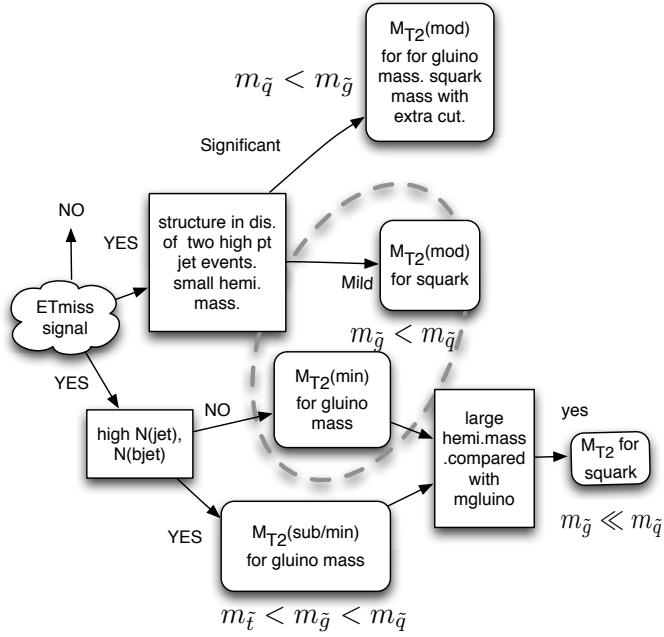}
\end{center}
\caption{Flowchart of the ISR improved $\mttwo$ analysis}
\label{flowchart}
\end{figure}
In this section,  we show  jet level results and discuss mass parameter 
determination by $M_{T2}$(min) and $M_{T2}^{\rm mod}$(min).   
To make our discussion clear,
 we show the 
steps for mass determination in Fig.\,\ref{flowchart}. 
We will see the importance of the  proper choice of improved $\mttwo$ 
depending on the observed event distribution.

\begin{subsection} {$M_{T2}$ distributions and effect of the ISR} \end{subsection}

\begin{figure}[t]
\begin{center}
\includegraphics[width=6cm] {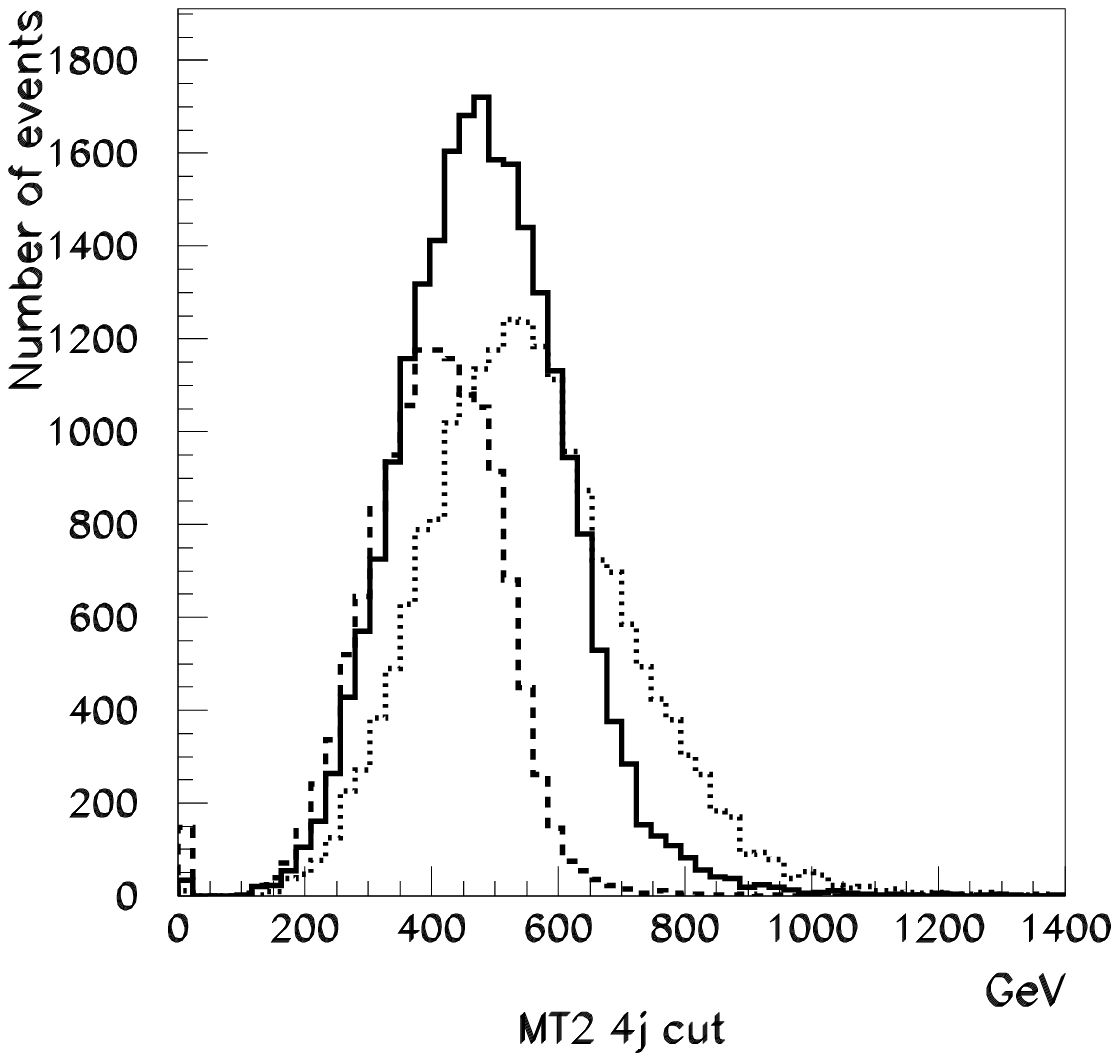}
\includegraphics[width=6cm] {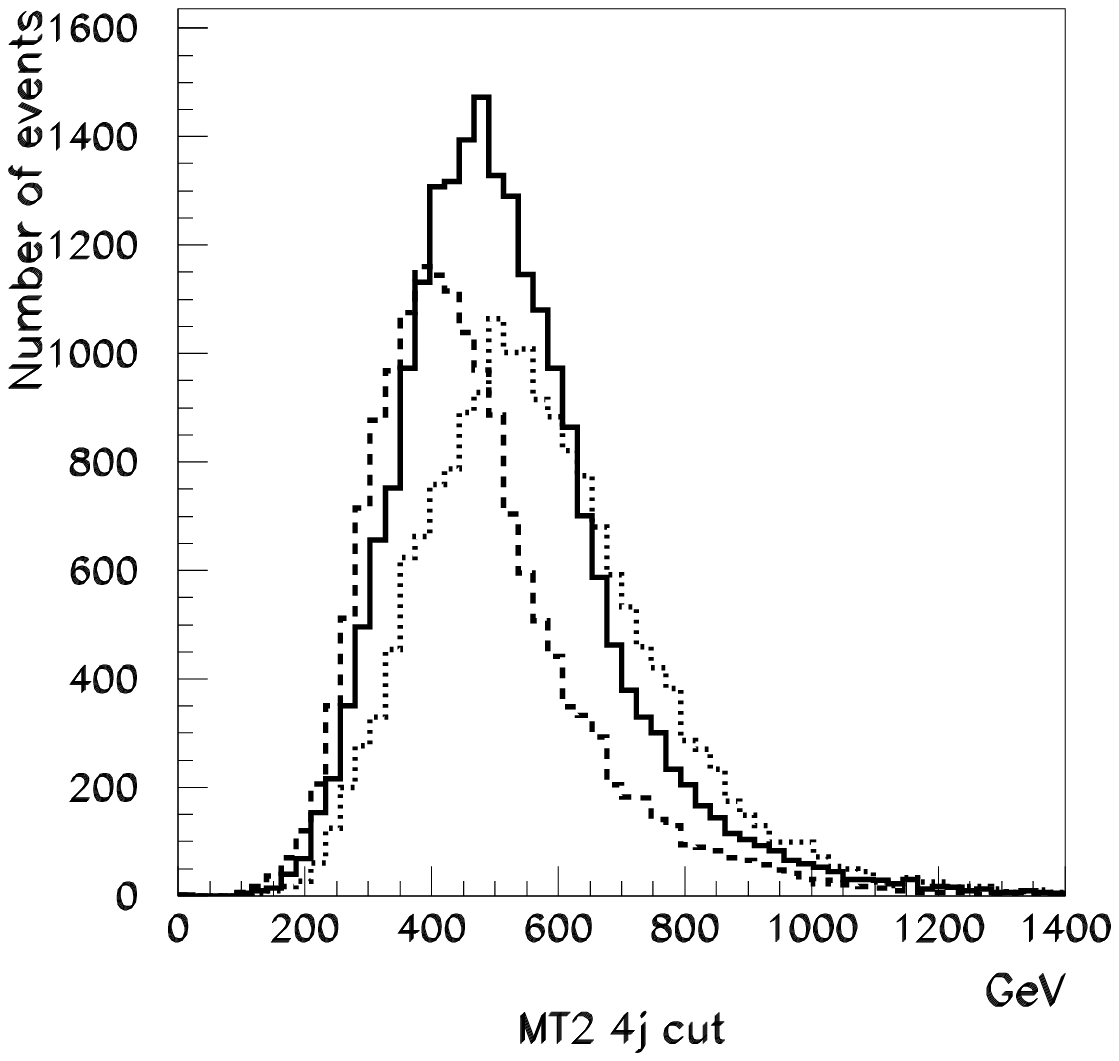}
\caption{$\mttwo$ distributions at Points 1, 3 and 5 in parton level (left) and in jet level (right). 
We generate 60000 SUSY events at 14\,TeV. 
}\label{mttwo135}
\end{center}
\end{figure}

We compare the parton level $\mttwo$ distributions with out ISR 
(left) to jet level ones (right) in Fig.\,\ref{mttwo135}.
We generate the evens by HERWIG+AcerDET+Fastjet as discussed 
in previous secton. 
In the left figure, the distribution reduces rather quickly toward 
$M_{\rm max} \equiv \max( \mgl, \msq)$ and there are 
tails beyond $M_{\rm max}$ due to mis-reconstruction of hemispheres. 
On the other hand, the jet level distributions at Points 1 and 3 do not show structures  
at neither the squark nor gluino masses. 
Especially, the number of events beyond 750\,GeV at Point 5 
is not much different from that of Point 3 in jet level.

In the previous studies\cite{takeuchi},  the endpoint of $\mttwo$ distributions
 are consistent with $M_{\rm max}$.  
The difference comes 
from the choice of the squark and gluino mass scale. 
In the paper, the gluino mass is around 800\,GeV and the squark masses are taken from 881 GeV to 1561 GeV. 
As the squark and gluino masses increase, relative importance of the ISR jet reduces 
as the average $p_T$ of the sparticle decay products increases 
linearly with the sparticle mass, 
while average $p_T$ of the ISR jet does not 
increase so quickly.  
We take the scale of SUSY parameters rather light ($\mgl\lsim 700$\,GeV) in this paper
so that squark and gluino masses are in the discovery region at the early stage of the LHC. 

The ISR effect reduces significantly when $\msq$ is 
increased for fixed $\mgl$.   At Point 5, where the sum $\msq+\mgl$ is 350\,GeV 
higher than that at Point 1,  
the endpoint is roughly at the input  value, 910\,GeV. 
The reduction of the ISR effect may be understood as follows. 
At this point, there are large mass difference 
between squark and gluino and the endpoint is saturated by $M_T$ of the 
squark hemisphere, when hemisphere reconstruction is 
correct. The ISR jet has to be grouped into the ``squark hemisphere" rather than 
the ``gluino hemisphere"  for the evnets contaminating  beyond 
the endpoint, unless the $p_T$ of 
the jet is very high.

\begin{figure}[t]
\begin{center}
\includegraphics[width=6cm] {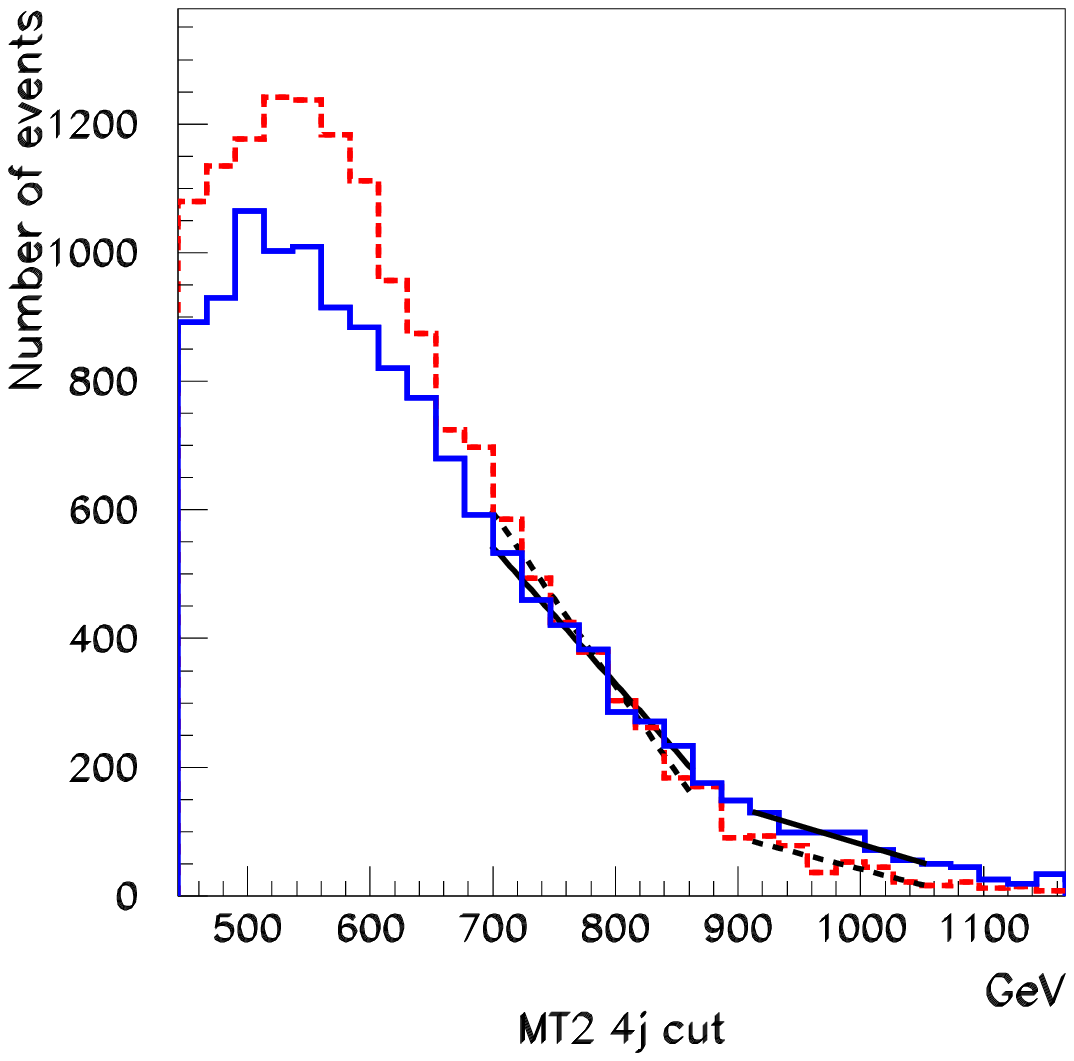}
\includegraphics[width=6cm] {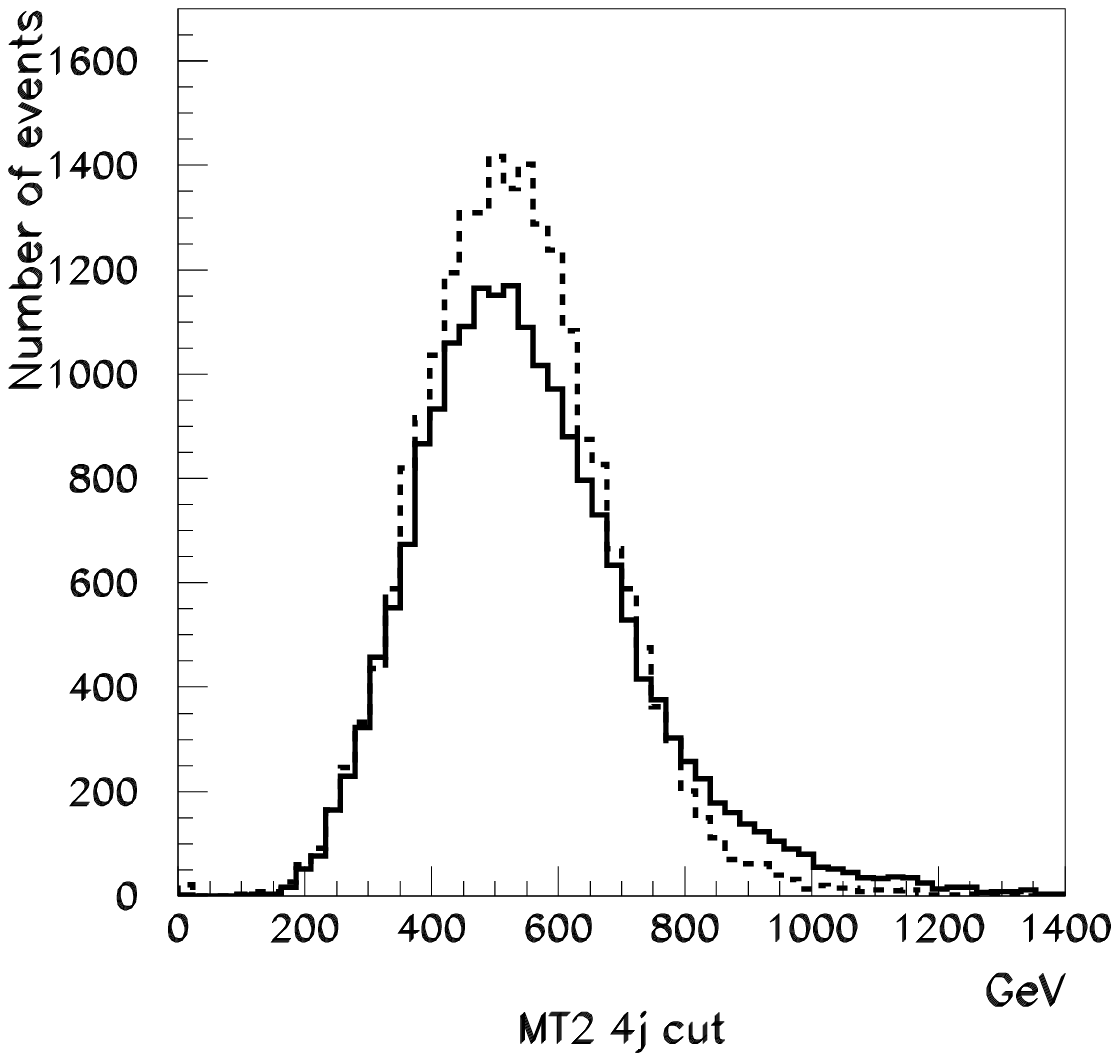}
\end{center}
\caption{Comparison between  parton level $M_{T2}$ (solid lines) 
and jet level $M_{T2}$ (dashed lines) distributions 
at Points 5 (left) and 4 (right).  We generate 60000 events and apply the 4 jet cut.  
In the left figure fitting lines are also shown.  }
\label{mttwo145}
\end{figure}

We compare the parton level distribution and 
the jet level distribution at Point 5 ($m_{\tilde{u}_L}=913$\,GeV) 
more quantitatively in Fig.\,\ref{mttwo145} (left). 
 We found a linearly decreasing  region of the $M_{T2}$  distribution  
 is  close to that of the parton level one.   
Fitting it by a linear function $f(\mttwo) = a + b \mttwo$ 
for both signal and tail regions,  we find ``kink" position at 887\,GeV for jet level  and 
882\,GeV for parton level  with errors around 10\,\% and  6\,\%  respectively, 
which roughly agree with the squark mass.\footnote{$\Delta\chi^2/ {\rm n.o.f }\sim 1$ for all fits. }

In Fig.\,\ref{mttwo145} (right), 
we also show the distributions at  Point 4 ($m_{\tilde{u}_L}=837$\,GeV). 
 The endpoint is more  difficult to see, although slopes of  
 the two distributions agree between 650 $-$ 800\,GeV. 
 Extracting  the mass scale from the distribution 
requires the detailed comparison between Monte Carlo and data, 
and it is not the scope of this paper.  
Therefore we move on to the $\mttwmin$ distributions 
which remove the hardest ISR jet efficiently. 
 
\subsection{Events with Two high $p_T$ jets and proper choice of ISR improved $\mttwo$  for gluino mass determination }

In CMSSM like parameter region, 
event topologies are significantly different between the cases of $m_{\tilde g}>m_{\tilde q}$ (Points 1, 2, 1' and 2') and
$m_{\tilde q} > m_{\tilde g}$ (Points 3, 4, 5, 3', 4' and 5').
As we discussed in section \ref{sec:parton},
in $m_{\tilde q}>m_{\tilde g}$ region, squarks mainly decay to $j \gl$, 
and the $j \chi_i$ mode is subdominant.
On the other hand, in $m_{\tilde g} >m_{\tilde q}$ region,
$\sq \to j \chi_i$ is the main mode and events contain two high $p_T$ jets.
It is likely that 
the  two hard jets are selected as the seeds of hemisphere reconstruction
and thus assigned into different groups.
Because each group has only one hard object,
the masses of both visible systems are expected to be small. 
On the other hand, in $m_{\tilde q}>m_{\tilde g}$ region,
each visible system may have two modest  jets from gluino three body decays,
and the mass scale of visible systems represent  the gluino mass.

\begin{figure}\begin{center}
\includegraphics[width=8cm] {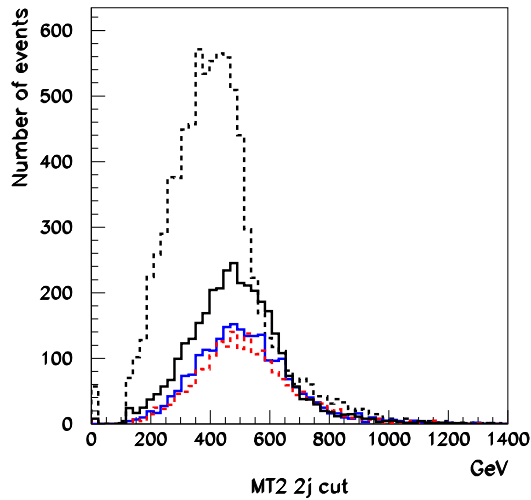}
\end{center}
\caption{The jet level $\mttwo$ distributions 
at  Points 1 (black dotted), 3 (black solid), 4 (blue oslid) and 5 (red dashed) for the events with 
at least two jets with $p_T>200$\,GeV. 
The sharp edge at Point 1 is mostly 
consisted by the events with $n_{50}\le$4.  
The distribution at Point 2 is similar to 
that at Point 1 and is not shown.   }\label{twojet}
\end{figure}

As discussed in the previous sections, 
the number of events beyond $\min(\mgl, \msq)$  will be 
significantly reduced  for  $\mttwmin$ distribution. 
 However, the shape of the distribution depends 
on the mass spectrum,  
and decay pattern. Especially the $\mttwmin$ distribution would be 
flat near the expected  endpoint at Point1 1 and 1', and we should use $\mttwmod$ 
distribution to  obtain the gluino mass.  Another way to say, to obtain the correct gluino mass,  we need  criteria to use $\mttwmod$ instead of $\mttwmin$ from 
experimental data.

\begin{figure}
\begin{center}
\leavevmode
\includegraphics[width=4cm] {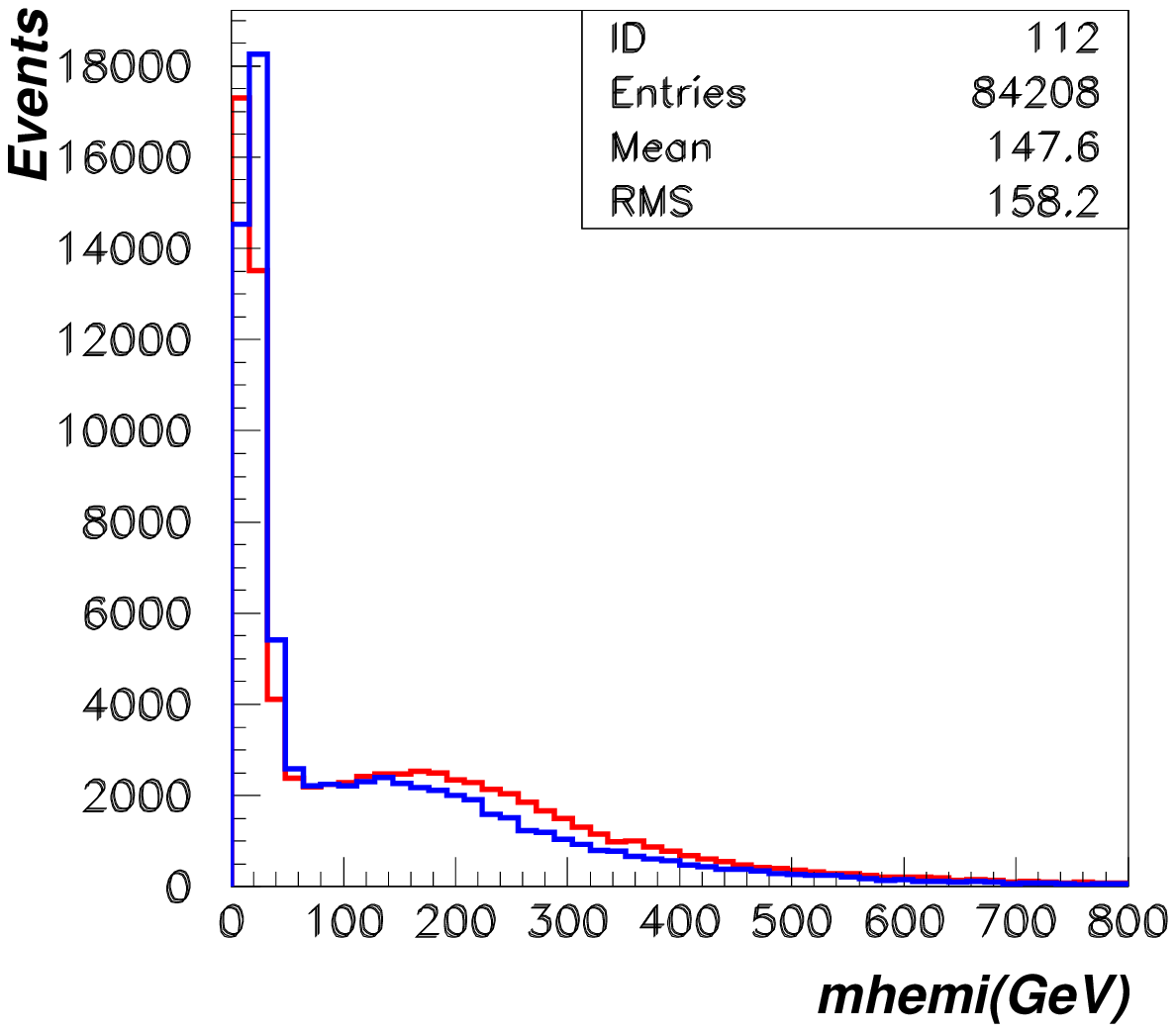}
\hspace{5mm}
\includegraphics[width=4cm] {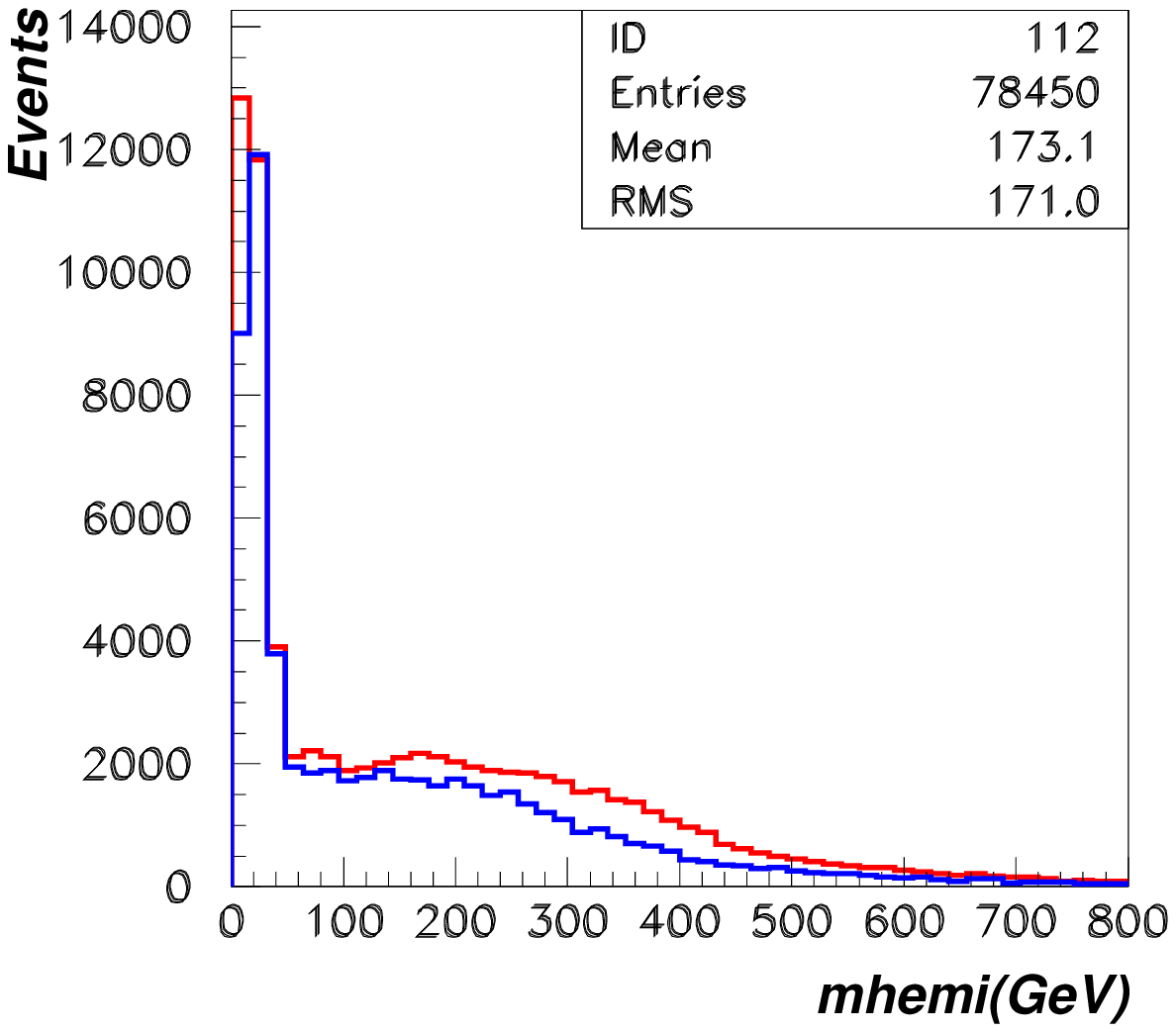}
\put(-190,-10){$(1)$}
\put(-55,-10){$(2)$}
\end{center}
\begin{center}
\includegraphics[width=4cm] {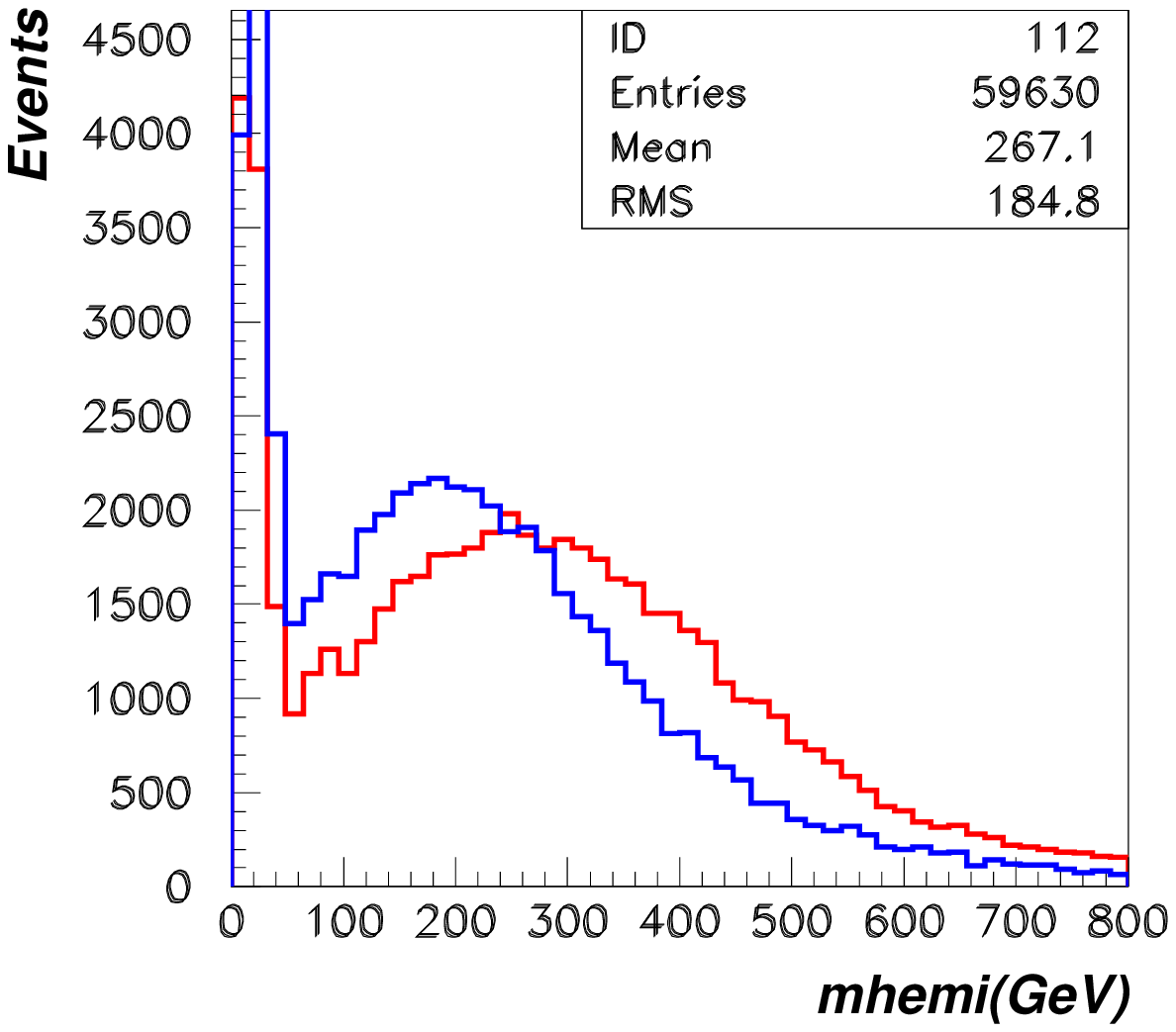}
\includegraphics[width=4cm] {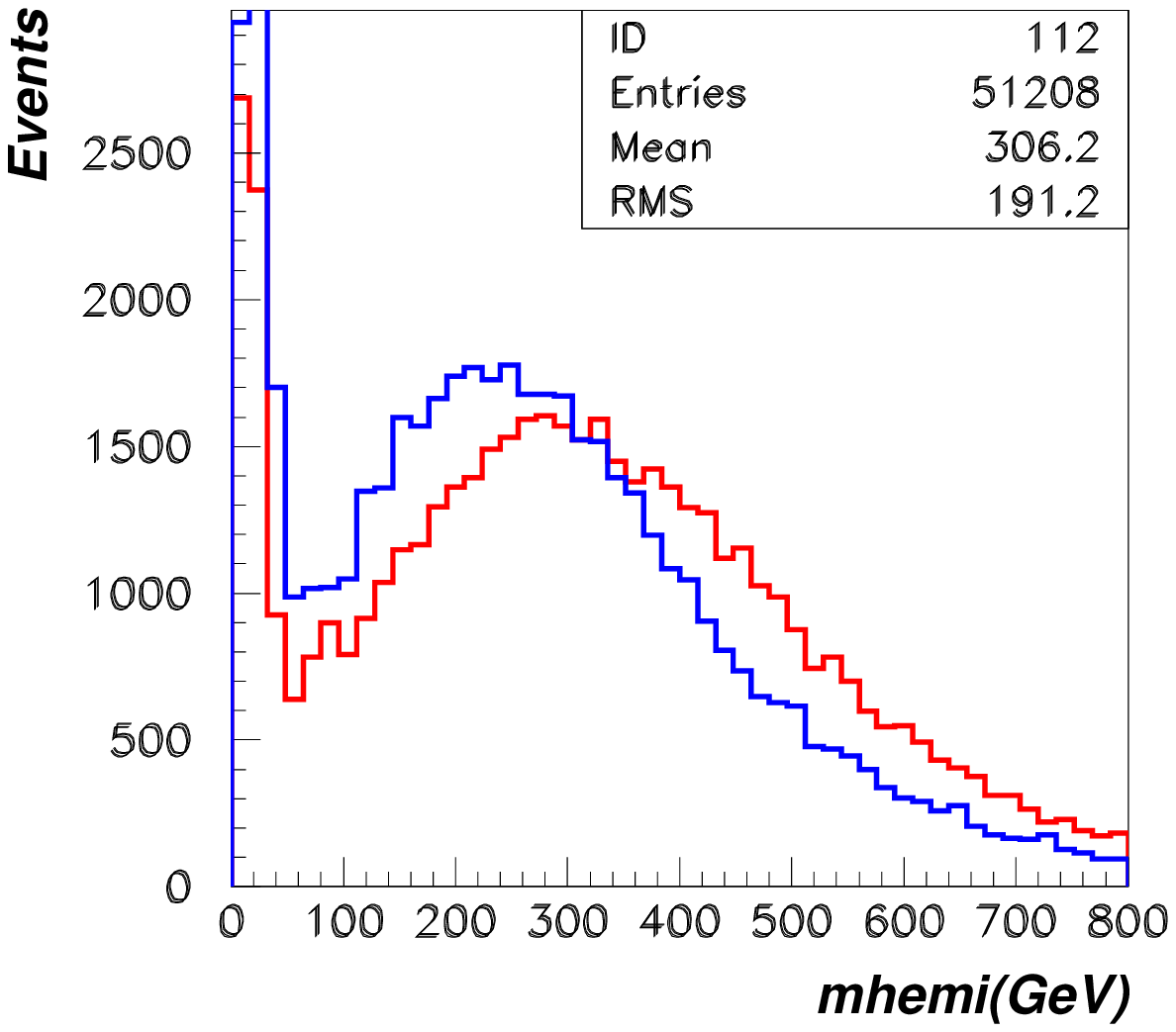}
\includegraphics[width=4cm] {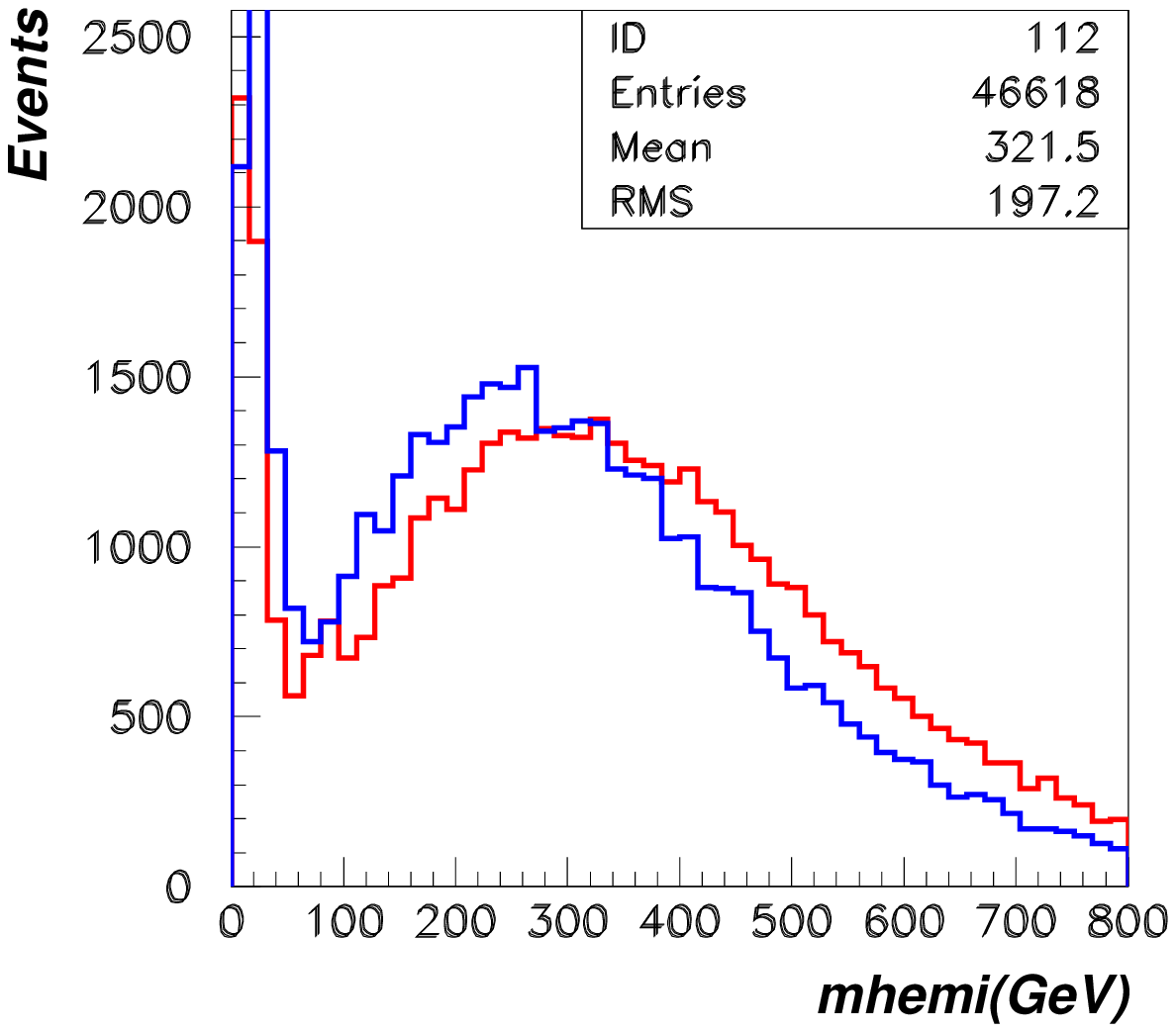}
\leavevmode
\put(-290,-10){$(3)$}
\put(-170,-10){$(4)$}
\put(-55,-10){$(5)$} 
\caption{Jet level $m_{v}$ distributions for the hemisphere reconstruction.
The $m_{v}^{(1)}$ and $m_{v}^{(2)}$ are superposed.
Red (Blue) distributions in figures (1) $-$ (5) correspond to Points 1 $-$ 5 (1' $-$ 5'), respectively. 
}
\label{fig:mhspr}
\end{center}
\end{figure}

 The fraction of  events which survive under the 2 jet cut
 is the one of the key observation to make the choice.  
In Fig.\,\ref{twojet}, we show the 
distribution of the events with at least two jets with $p_T>200$\,GeV. 
The number of events after the cut is large at 
Point 1 (a black dotted line), compared to those at  Points 3 $-$ 5. 
As discussed already, this is because the $\tilde{q} \to j \tilde{\chi}_i $  mode
dominates at Point 1. 
If  we see the excess, we should use $\mttwmod$.  
 Structure in $\mttwo$ distribution under the 2 jet cut 
 might also be useful
for determination of ordering of squark and gluino masses. 
We observe a sharp endpoint at the true squark mass (600\,GeV)  at Point 1.  
Although the number of the events after the 2 jet cuts  is rather small, 
such structure also exists  at Point 3.  This is due to the squark  decay into  electroweak inos 
with significant branching ratio at this point.   
We will discuss about mixed use of $\mttwmod$ and $\mttwmin$ at Points 3 and 3' in the next subsection. 

Alternatively, one can use  hemisphere mass $m_v$ to estimate 
fraction of the events that have gone  though $\tilde{q}\rightarrow j \chi$
decays. 
Red (Blue) distributions in Fig.\,\ref{fig:mhspr} (1) $-$ (5) show
the jet level $m_v$ distributions for Points 1 $-$ 5 (1' $-$ 5'), respectively.
Here $m_v^{(1)}$ and $m_v^{(2)}$ are superposed in the distributions.
The shape of the distributions are very different between $m_{\tilde g}>m_{\tilde q}$ and
$m_{\tilde q}>m_{\tilde g}$ cases.
In $m_{\tilde g}>m_{\tilde q}$ region (See Figs.\,\ref{fig:mhspr} (1) and (2).),
the distributions have a sharp peak at $m_v = 0$, and the fraction 
of the other events are small. 
On the other hand, 
in $m_{\tilde q}>m_{\tilde g}$ (See Figs.\,\ref{fig:mhspr} (3), (4) and (5).),
the distributions have another peak around $m_{\tilde g}/2$, which
is the  contributions from $\tilde g \to \tilde \chi jj$ mode. 
The fraction of the events with $m_v\sim 0 $ becomes small. 
This suggests that the shape can give us information of the 
ordering of gluino and squark masses.  
We can also study the hemisphere mass after removing the jet $i_{\rm min}$, 
$m_v^{(1)}(i_{\rm min})$ and $m_v^{(2)}(i_{\rm min})$, 
where $\mttwmin=M_{T2}(i_{\rm min})$. 
The fractions of the 
events with $\min (m_v^{(1)}(i_{min}), m_v^{(2)}(i_{min}))<50$\,GeV are 73\,\%,  45\,\% and 35\,\%
at Points 1, 3 and 5, respectively.  

\begin{figure}[tbh]
\includegraphics[width=7.5cm] {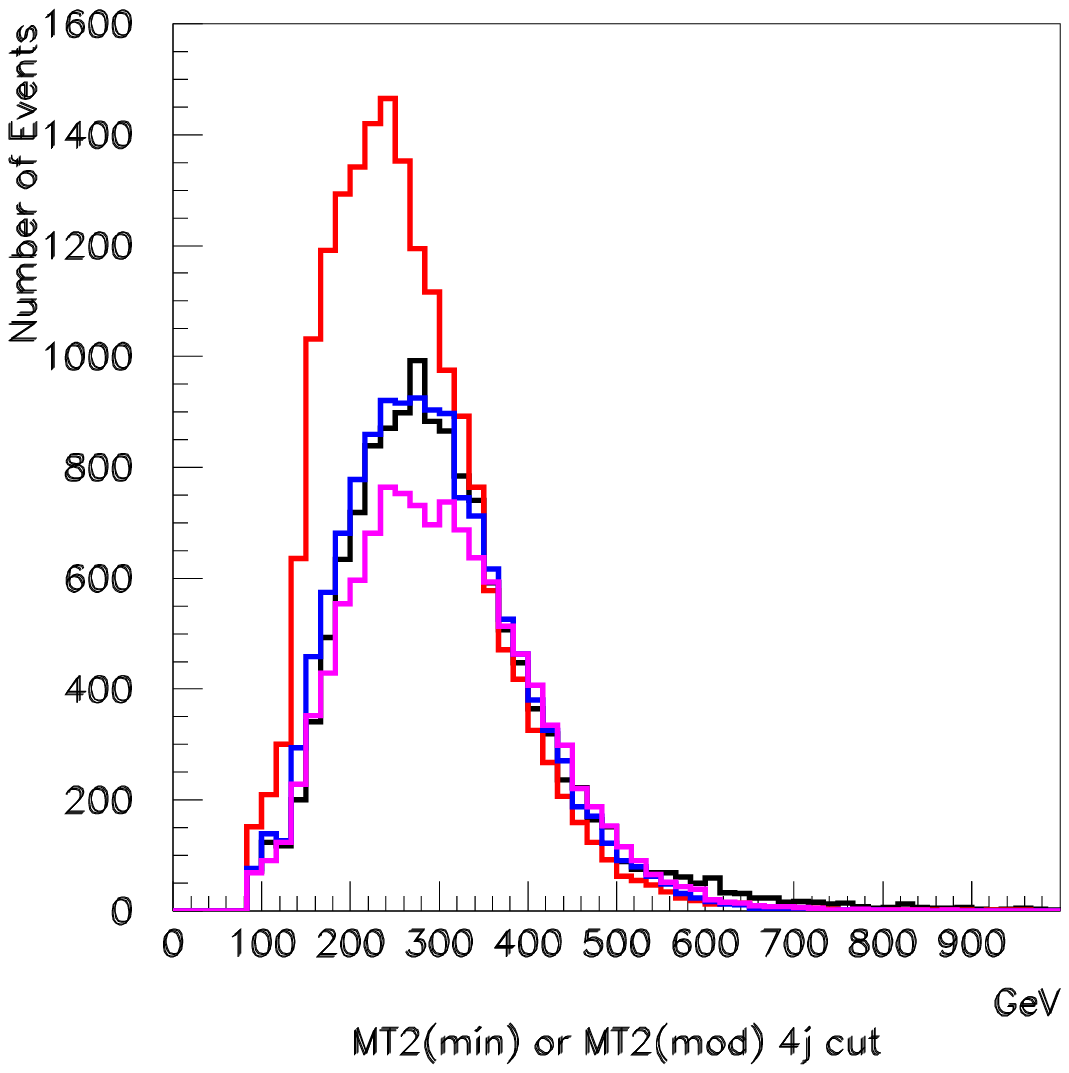}
\includegraphics[width=7.5cm] {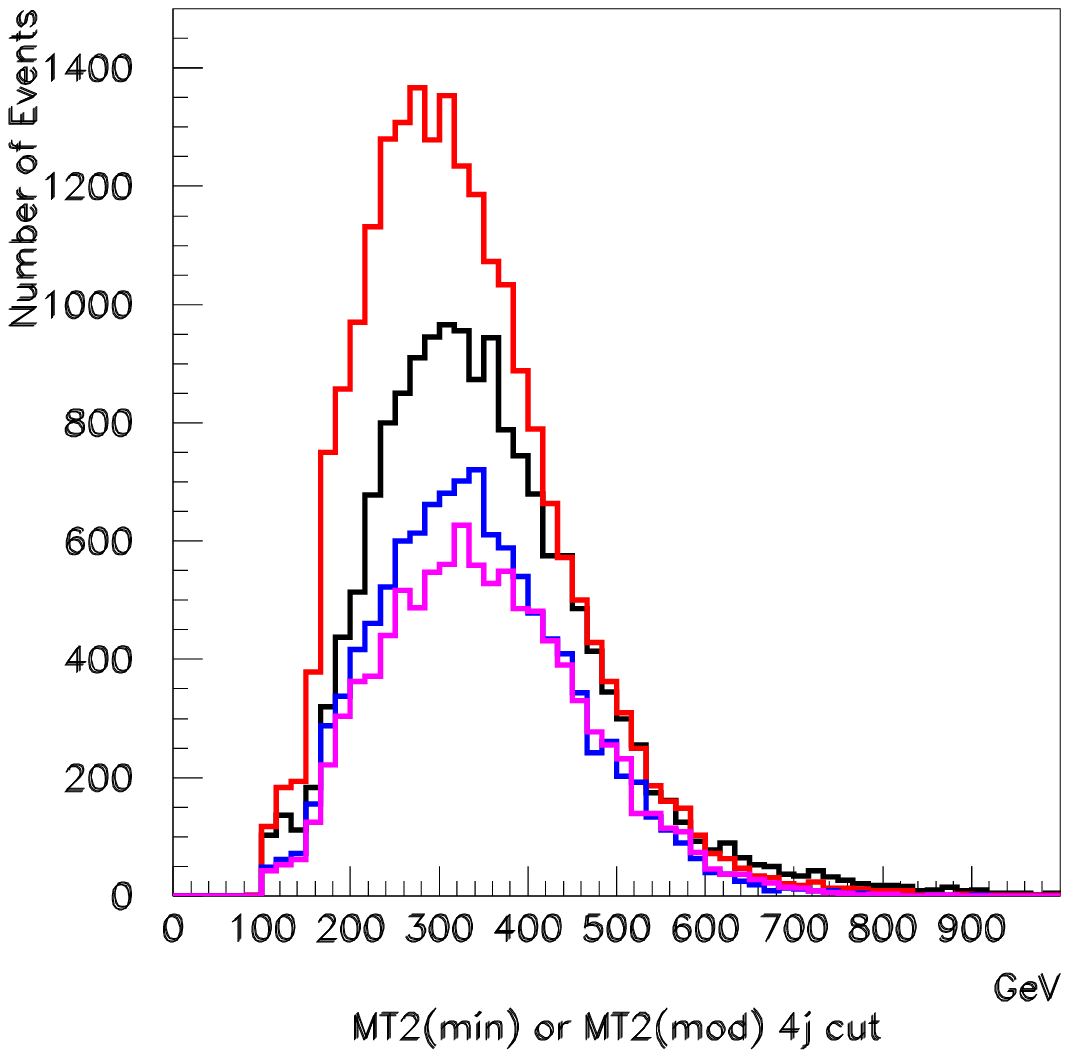}
\caption{Left) $\mttwmod$  (Point 1':black) and $\mttwmin$ distributions 
(Points  3':red, 4':blue, 5':purple).
 Right ) $\mttwmod$ (Point 1: black ) and $\mttwmin$ distributions (Points  3:red, 4:blue, 5:purple).
The mass of gluinos are from  522 $-$ 558\,GeV at Points 1' $-$ 5' and  612 $-$ 651\,GeV at Points 1 $-$ 5,
respectively.  }\label{gluinorecon}
\end{figure}

\begin{table}[tbh]
\begin{center}
\begin{tabular}{|c|c|c|c|c|c|c|}
\hline
point & $a$ & $\delta a$ & $b$& $\delta b$ & $a/b$ & gluino mass \cr
\hline 
1 & 2045 & 84 & 3.42 & 0.16 & 598  &  612\cr
2 & 2103 & 107 & 3.38 & 0.20 & 623 & 620\cr
3 &  2285 & 64 & 3.86  & 0.13 & 592 & 635\cr
4 & 1288 & 57 &  2.10 & 0.10& 612  & 646\cr
5 & 1178 & 82 & 1.90 &   0.13 &620 &  651\cr
\hline  
1' & 1592 & 70 & 2.98 & 0.16 & 534 & 522 \cr
2' & 1571 & 68 & 2.95 & 0.15 & 533 & 528 \cr
3' & 1842& 67 & 3.66 & 0.15 & 502 & 541 \cr
4'  & 1765 & 69 & 3.37 & 0.15 & 524 & 550 \cr
5' & 1514 & 81 & 2.77 & 0.17 & 547 & 558 \cr
\hline 
\end{tabular}
\end{center}
\caption{The results of linear fits of the distributions in Fig. 16 near the endpoint. 
The fitting function is a simple linear function $f=a -b \mttwo$.  For 
Points 1, 2, 1' and 2', $\mttwmod$ distributions are used, and $\mttwmin$ for the 
others.  }\label{fitmt2min}
\end{table}

Based on the discussions above, 
we decide to use $\mttwmod$ for 
Points 1', 2', 1 and 2, 
and $\mttwmin$ for 3' $-$ 5' and 3 $-$ 5 to determine the gluino masses.
The jet level distributions are shown in Fig.\,\ref{gluinorecon}. 
The left figure shows the distributions at Points 1' $-$ 5' 
and the right figure show the distributions at Points 1 $-$ 5'.  The fomar have 
endpoints around 500\,GeV while the latter is around 600\,GeV. 

As we have seen in the parton level distribution,  it is not good idea 
to focus on the events too  close to the endpoint when we use the 
hemisphere algorithm to define $\mttwo$.  Unlike $M_{TGen}$, 
the algorithm tend to preserve the number of events near the 
endpoint, but  there is a tail beyond the true endpoints 
because of the hemisphere misreconstruction.  
Therefore we fit the distributions to linear functions from the bin with the hight 
half of the  maximum  to the bin at the 1/10 $\sim$ 1/6  of the maximum. 
The results of the fitting functions, $f=a-b M_{T2}^{({\rm mod})}({\rm min})$, 
are listed in Table\,\ref{fitmt2min}.  
The errors  are typically 7\,\% and  section of the line  $b/a$ 
 are  consistent with the corresponding gluino masses. 
The fitting function adopted here is too simple,  leading $\chi^2/ n_{\rm d.o.f}$ are around 2. 

\begin{figure}[!t]
\begin{center}
\includegraphics[width=7cm] {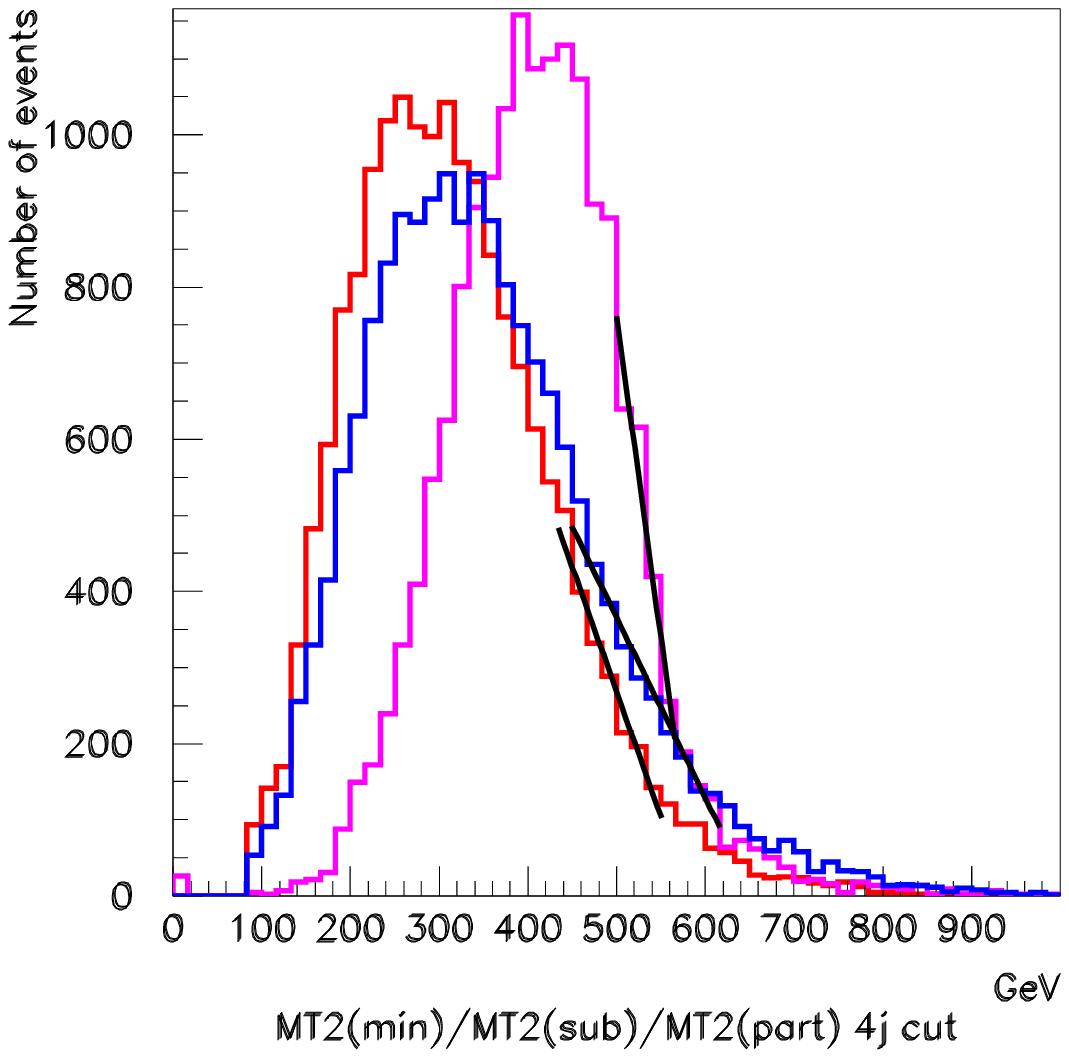}
\includegraphics[width=7cm] {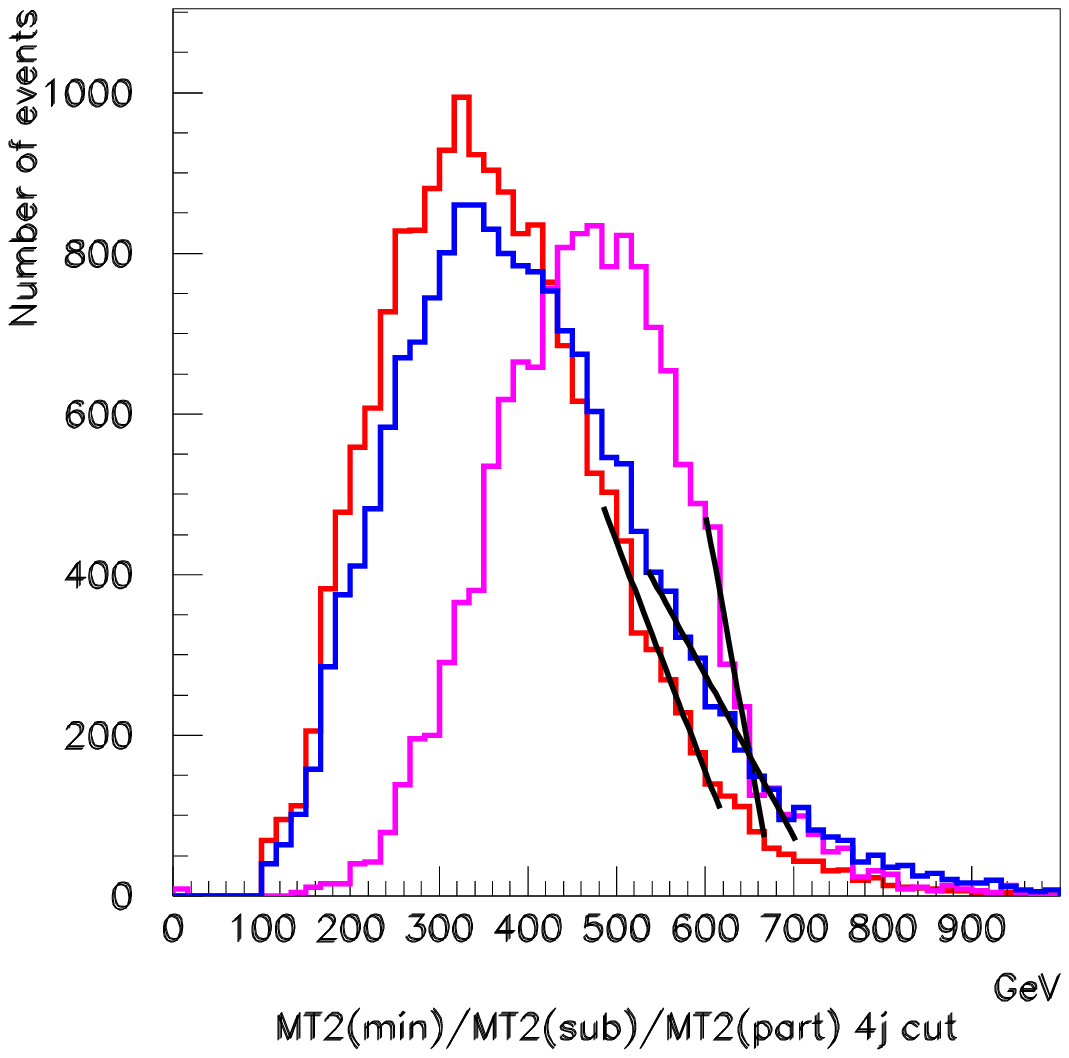}
\end{center}
\caption{$\mttwmin$ distributions (red lines), $\mttwsub$ 
 distributions (blue line)  and parton level $\mttwo$  distributions 
for $\gl$-$\gl$ production (purple lines). 
The left and right figures correspond to Points 5' and 5, respectively. 
Fitting curves are also shown in the figures. 
}\label{compare}
\end{figure}

In the previous paper, we have proposed $\mttwsub$ to determine the 
gluino mass when $\mgl<\msq$. 
The endpoint of $\mttwmin$ distribution  
agrees with the gluino mass better then 
$\mttwsub$. The minimization among the 
possible assignment of jets from $\tilde{q}\rightarrow \tilde{g}q $ decay  
among high $p_T$ jets ensures  $\mttwmin$ distribution in $\sq\gl$ production is below 
$\mgl$ when there are no significant ISR jet. 
 We show $M_{T2}$(min) (red lines) and $M_{T2}$(sub) (blue lines) the distributions 
 at Points 5' (left) and 5 (right) in Fig.\,\ref{compare} 
 with fitting curves near the endpoint. 
 For comparison, parton level $M_{T2}$ distributions for $\gl$-$\gl$ samples without ISR jets
 are also shown by purple lines.  
 The $\mttwsub$ distributions suffer  higher tails and tend to 
 overshoot the true endpoints. 
 Exceptional case is Point 3', which we will discuss separately.

\subsection{Squark mass determination}

\begin{figure}
\begin{center}
\includegraphics[width=6cm]{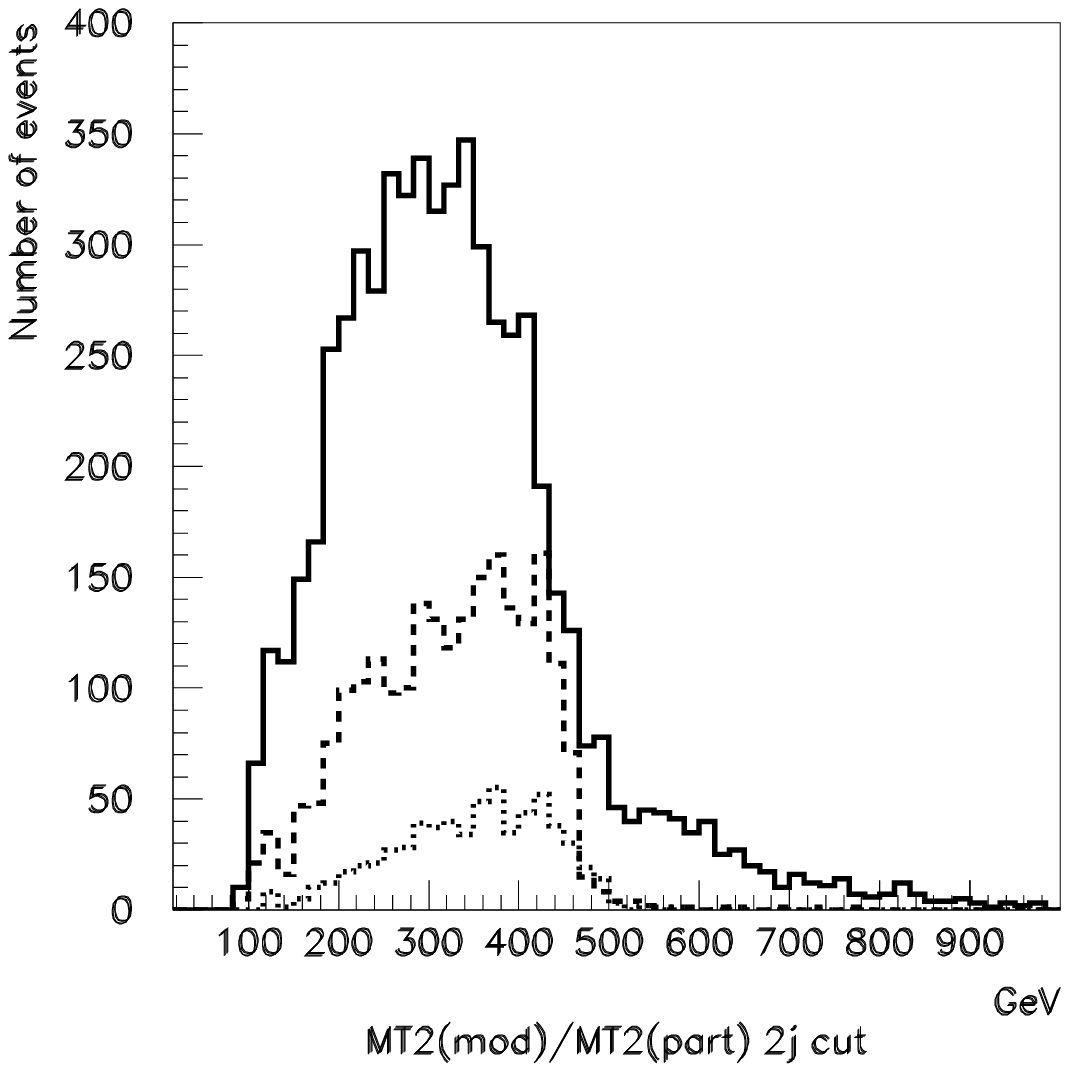}
\includegraphics[width=6cm]{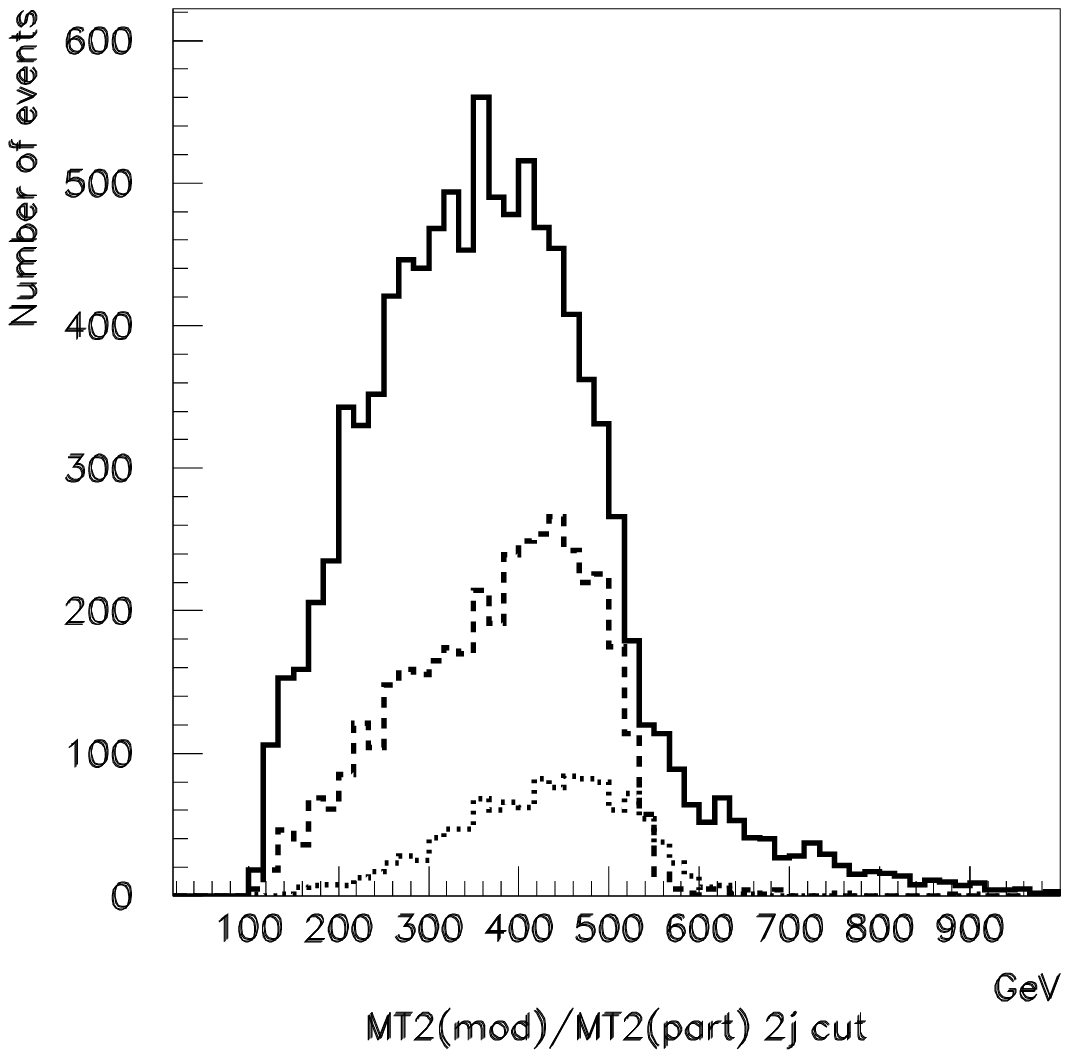}
\end{center}
\caption{  $\mttwmod$ distributions after the two jet cut (solid lines), and parton level $\mttwo$ distributions of $\sq\sq$ (thick dashed lines ) and $\gl\gl$ distributions (thin dashed lines). The left and right figures are for Points 1' and 1, respectively. }
\label{point1jet2}
\end{figure}

In this section we discuss  squark mass determination  
 from the experimental data. 
 In Fig.\,\ref{point1jet2}, 
we show the distributions of $\mttwmod$ at Points 1' (left) and 1 (right) after 2 jet cut. 
Under the cut, the $\tilde q$-$\tilde q$ production events are significant part of the survived events,
because events with $\tilde g \to \tilde t t$ tend to be rejected. 
The structure due to the $\tilde q$-$\tilde q$ production events can be seen in the 
total  distribution. 
The endpoints of  $\sq$-$\sq$ parton level 
distributions (the thick dotted lines)  are consistent with the positions of the edge structure 
in the jet level distributions (the solid lines), 
on the other hand $\gl$-$\gl$ distribution (the thin dashed lines) 
are negligible.  Taking $\mttwmod$ reduces the contamination of 
 $\sq$-$\gl$ production beyond the edge.    

The jet distribution reduces very quickly  near the endpoint. 
At  Point 1', we fit the distribution by two linear functions. 
At point 1 we fit the distribution to a quadratic function bellow the kink 
and  a linear function beyond the kink.  
  The obtained kink position is at 483 (540.0)\,GeV at Point 1' (1). 
  The values are very close to the input  squark masses, and much 
smaller than gluino masses.
Errors in the values are around 7\,\%.

\begin{figure}
\begin{center}
\includegraphics[width=6cm]{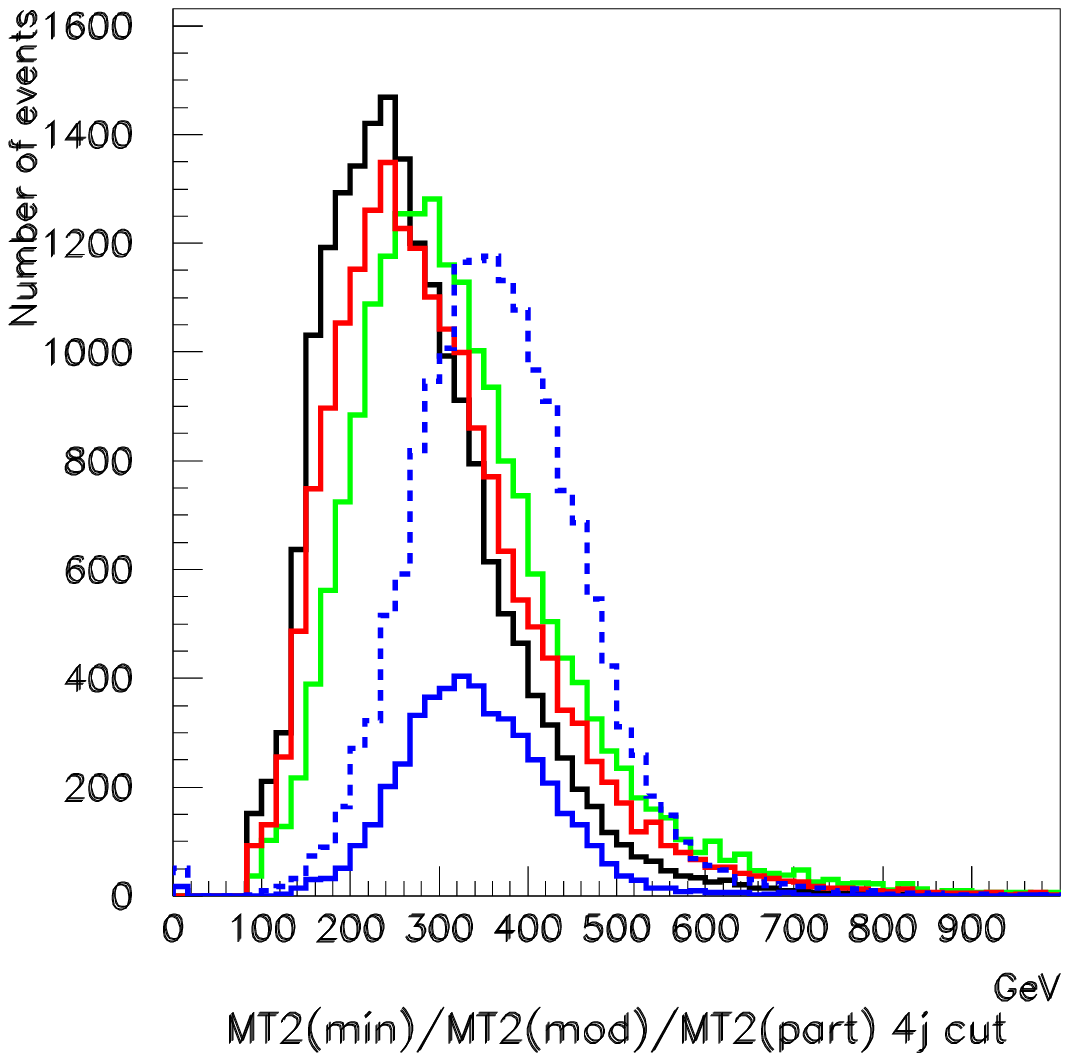}
\includegraphics[width=6cm]{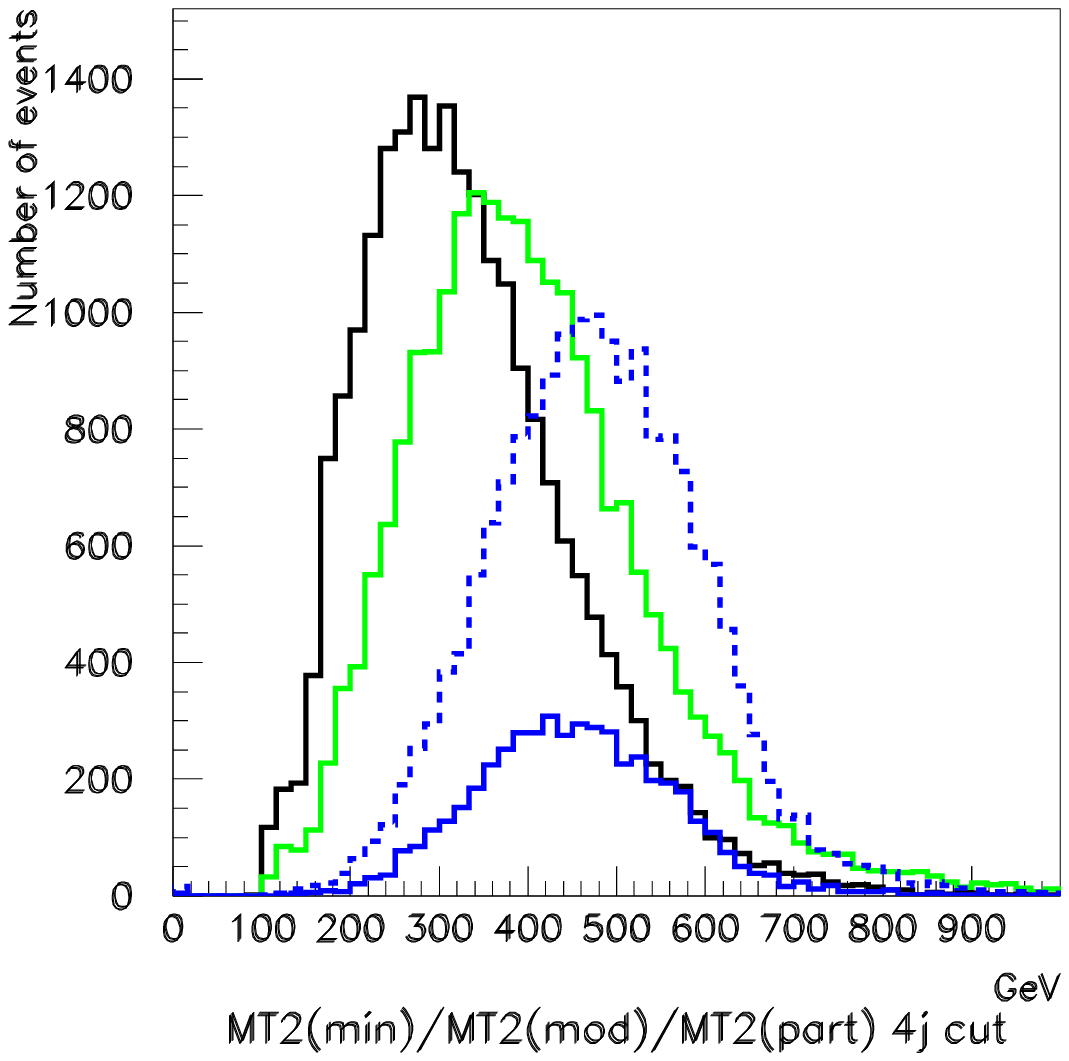}
\end{center}
\caption{$\mttwmin$ (black) and $\mttwmod$ (green) distributions 
at Point  3' (left) and  Point 3 (right). 
The blue sold and blue dotted lines  correspond to parton level
$\gl\gl$ and $\gl\sq$  distributions, respectively.  
At  Point 3 gluino decay into three body final state, while  a gluino decays into a stop and a top  at  Point 3'.  
We also show a distribution of  $\mttwsub$ (red line) at Point 3'  
whose endpoint is slightly lower than that of $\mttwmod$.   }
\label{point13two}
\end{figure}

Now we show distributions at Points 3' and 3. 
The squark masses  is  around 619 (734)\,GeV at Point 3' (3). 
At  these model points, 
the squarks are heavier than the gluino but the decay $\tilde{q}\rightarrow j \chi_i $ 
still has a significant branching ratio (See Table.\,\ref{masstable}). 
Especially the fraction of the events survives under 2 jets cut at Point 3 is larger than 
those at Points 4 and 5.  
Then there is a possibility 
to obtain the squark mass scale by fitting the $\mttwmod$ endpoint 
instead of $\mttwmin$. 
In Fig.\,\ref{point13two} we fit the  events near the endpoint 
to a linear function. 
The  endpoints of $\mttwmod$  distributions 
are 557.4\,GeV and  676.1\,GeV  with about 6\,\% of statistical 
errors at Points 3' and 3, respectively, higher than 
fitted  endpoint of $\mttwmin$($\sim \mgl$)  listed in Table\,\ref{fitmt2min}.  
Fits of the parton level $\mttwo$  distirubtions of $\gl\sq$  production process lead 
601\,GeV and 706\,GeV at Points 3' and 3, respectively.

Point 3' is a model point with special feature.  
At the point, gluino decays into $\tilde t t$ dominantly. 
It is easy to see that
$\gl\rightarrow \tilde{t} t$ is the dominant gluino decay mode 
through the numbers of both jets and $b$-jets in the events. 
 The parton level $\mttwo$ distributions at Point 3' is suppressed near 
squark and gluino masses. This is due to the softness 
of average jet $p_T$.  
We only use the jets with $p_T>50$\,GeV. 
On the other hand, the average $p_T$  of the 
jets are  small  so that  some jets from gluino decays are not taken into 
account in the mass reconstruction.  
We may reduce the $p_T$ cuts but 
in that case the probability to assign jets into the  wrong hemisphere becomes 
so large and  the endpoints are smeared. 
Indeed, the endpoint of the parton and jet level distribution is systematically 
lower than the input squark mass. 





In the previous section, we have seen that the highest $p_T$ jet 
is likely to be the ISR at Point 3' for $\gl$-$\gl$ production.
The main decay modes of squarks are $\sq \to j \gl$,
and $\tilde q \to j \tilde \chi_i$ mode is subdominant.  
Therefore removing the highest $p_T$ jet  
might be equally useful to remove the hard ISR in $\sq\gl j$ production.
To check this, we also plot 
$\mttwo$(sub) distribution in the same plot (red line). 
Both $\mttwo$(sub) and 
$\mttwmod$ have  similar endpoints.

\section{Expectation at $\sqrt{s}=7$~TeV and  $\int {\cal L } dt =1{\rm fb}^{-1}$ }

Currently the LHC is operated at 7\,TeV and will accumulate $\int {\cal L}dt \simeq 1\,{\rm fb^{-1}}$.
Although the aim of this paper is to improve SUSY parameter determination in general 
hadron collider  
by using ISR improved inclusive $M_{T2}$, the event distribution 
at  7\,TeV and $1\,{\rm fb^{-1}}$ of luminosity is also our interest. 

\begin{figure}
\begin{center}
\includegraphics[width=6cm]{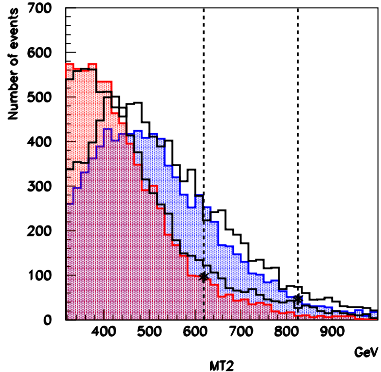}
\end{center}
\caption{ $\mttwo$ distributions under the  4 jet cut  at Points 3' (red filled)  and  5'(blue filled)  and 
$\sqrt{s}=7$~TeV and those at 14~TeV( black solid line) . We use 
30000 events at  each model point. }\label{mt27tev}
\end{figure}

Again, we start with discussion of the $\mttwo$ distribution at 7\,TeV.
Fig.\,\ref{mt27tev} shows the $M_{T2}$ distributions at Points 3' (red filled) and 5' (blue filled). 
The figures are based on 30000 events, but the expected number of events 
by the end of 7\,TeV run is smaller (See Table\,\ref{masstable}.). 

The smearing due to ISR is smaller than those at 14\,TeV shown by the open histogram by black solid lines. 
The distributions beyond the  true masses (shown by $*$ marks and the lines) at  7\,TeV  
are less than corresponding distributions at 14\,TeV. 
 Especially some structure remains around 600\,GeV at Point 3'. 
 At Point 5',  $M_{T2}$ distribution is straight between 600\,GeV to 800\,GeV, 
 and the endpoint obtained by fitting it to a linear function roughly consistent with the squark mass. The 
$M_{T2}$ distribution is correlated with squark mass and may be used for mass determination by
fitting the distribution to the template based on  Monte Carlo simulation data.

\begin{figure}
\includegraphics[width=6cm]{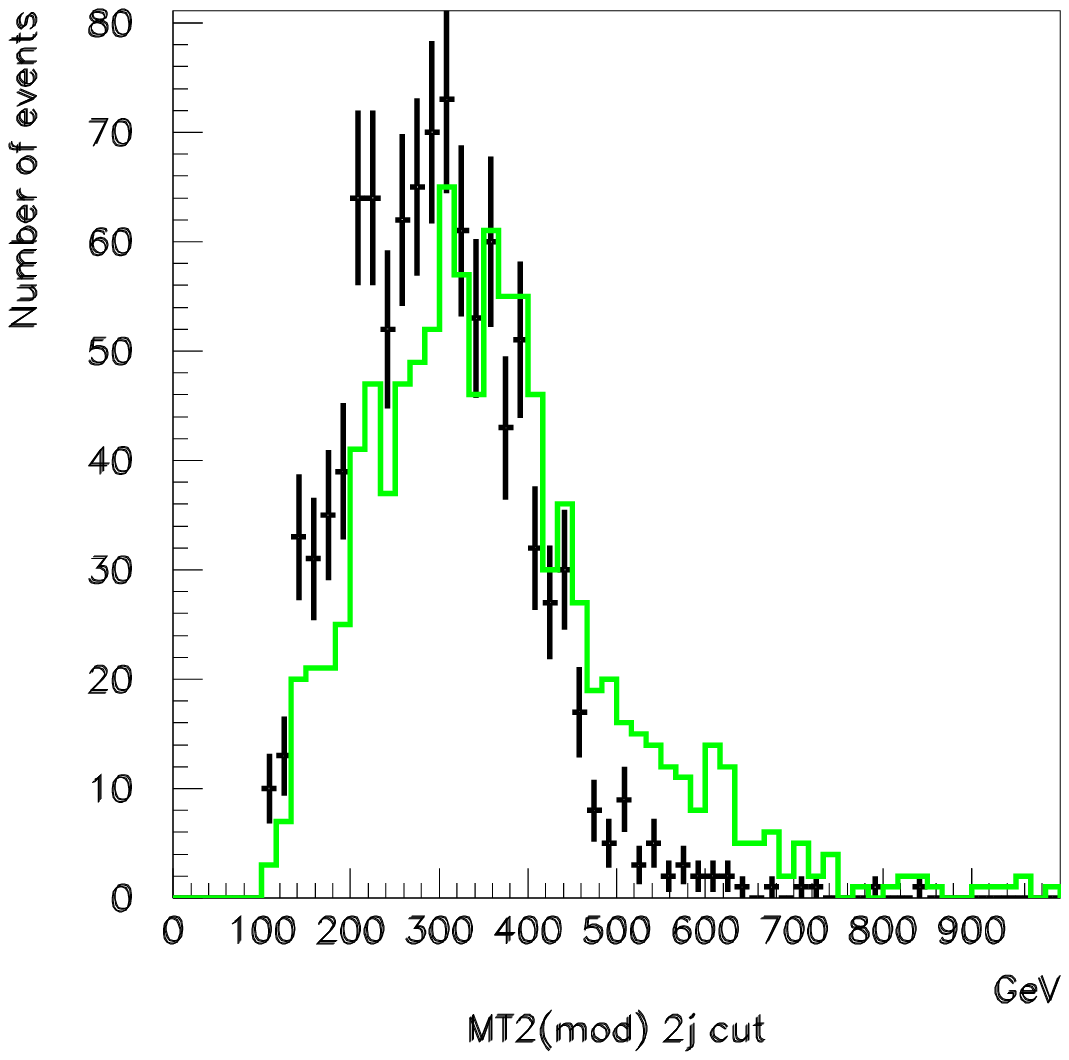}
\includegraphics[width=6cm]{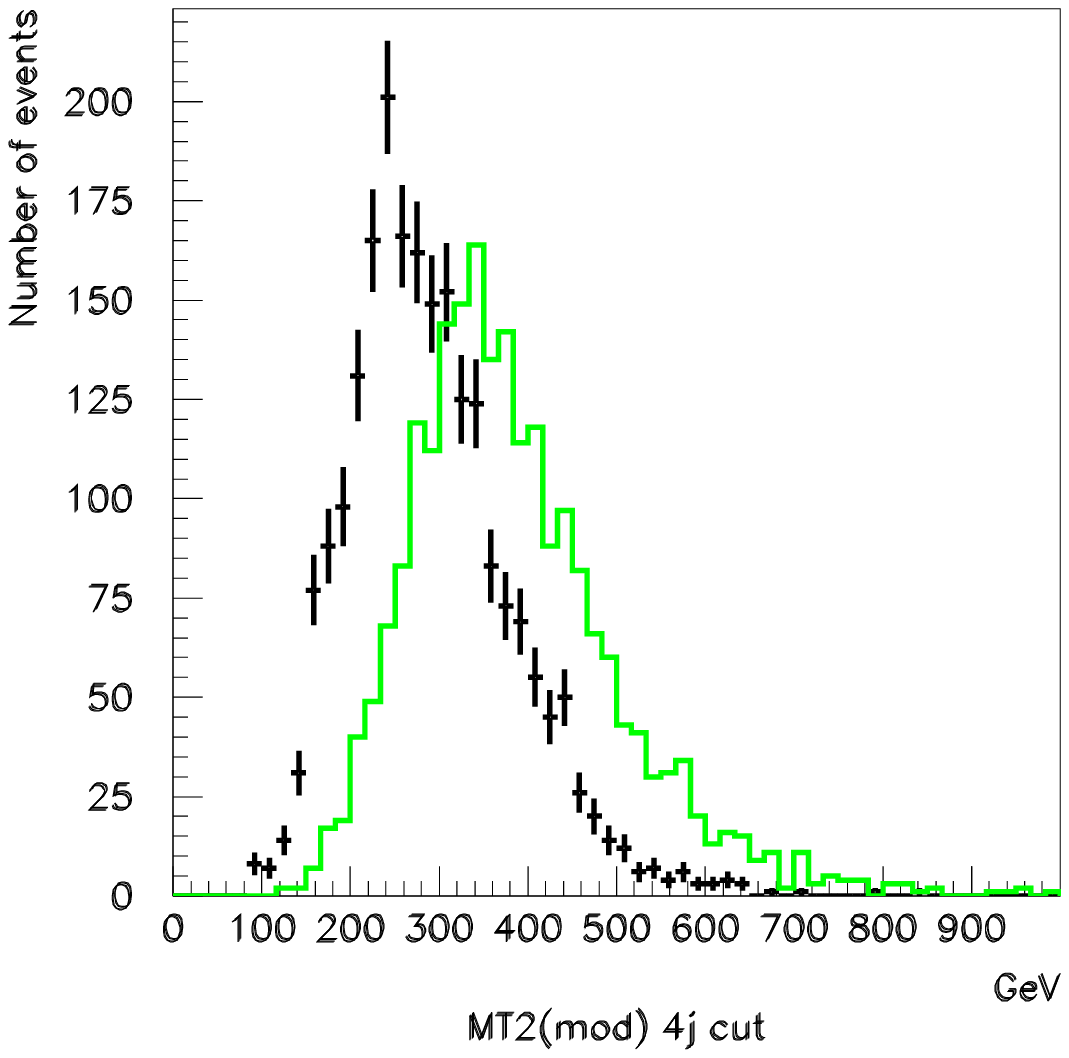}
\caption{ $\mttwo$  distributions at 7\,TeV and 1\,fb$^{-1}$. $\mttwmod$  distirubtions with error bars
at Point 1' for the 2 jet cuts (left), and the 4 jet cut (right). Green lines show $\mttwo$ distribution under the same cut.  } \label{point117tev}
\end{figure}

In Fig.\,\ref{point117tev}, we show the $\mttwmod$ distributions (bars) at Point 1'  
(Type $D$ in Fig.\,\ref{fig:decay_figs}.)
with our 2 jet (left) and 4 jet (right) cuts together with the
$\mttwo$ distribution (green) under the same cut. 
In the figure we generate events corresponding to $\int {\cal L} =1\,{\rm fb^{-1}}$ at 7\,TeV. 
The coincidence of the endpoint with the input squark/gluino masses (478/522\,GeV) are already visible. 

\begin{figure}
\includegraphics[width=4.9cm]{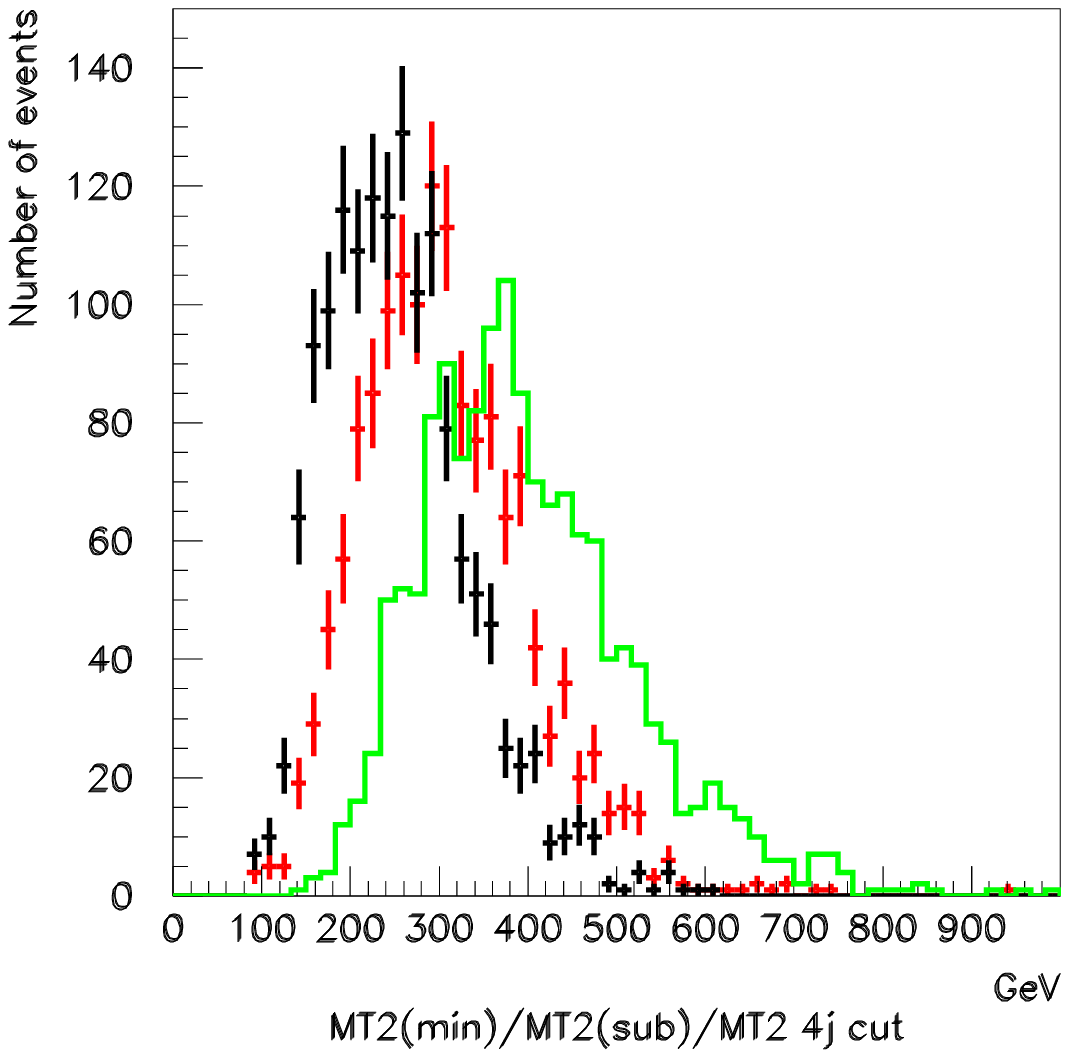}
\includegraphics[width=4.9cm]{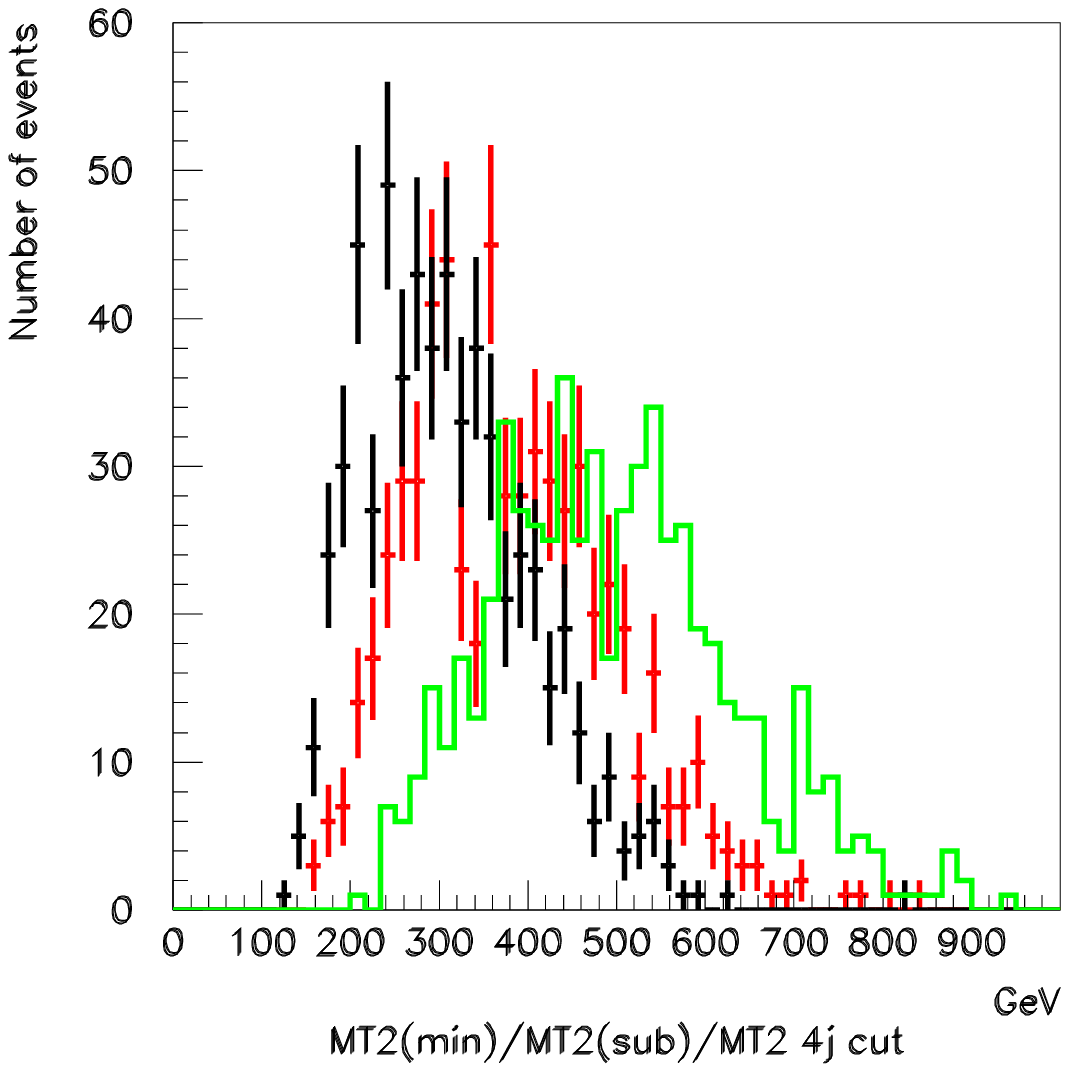}
\includegraphics[width=4.9cm]{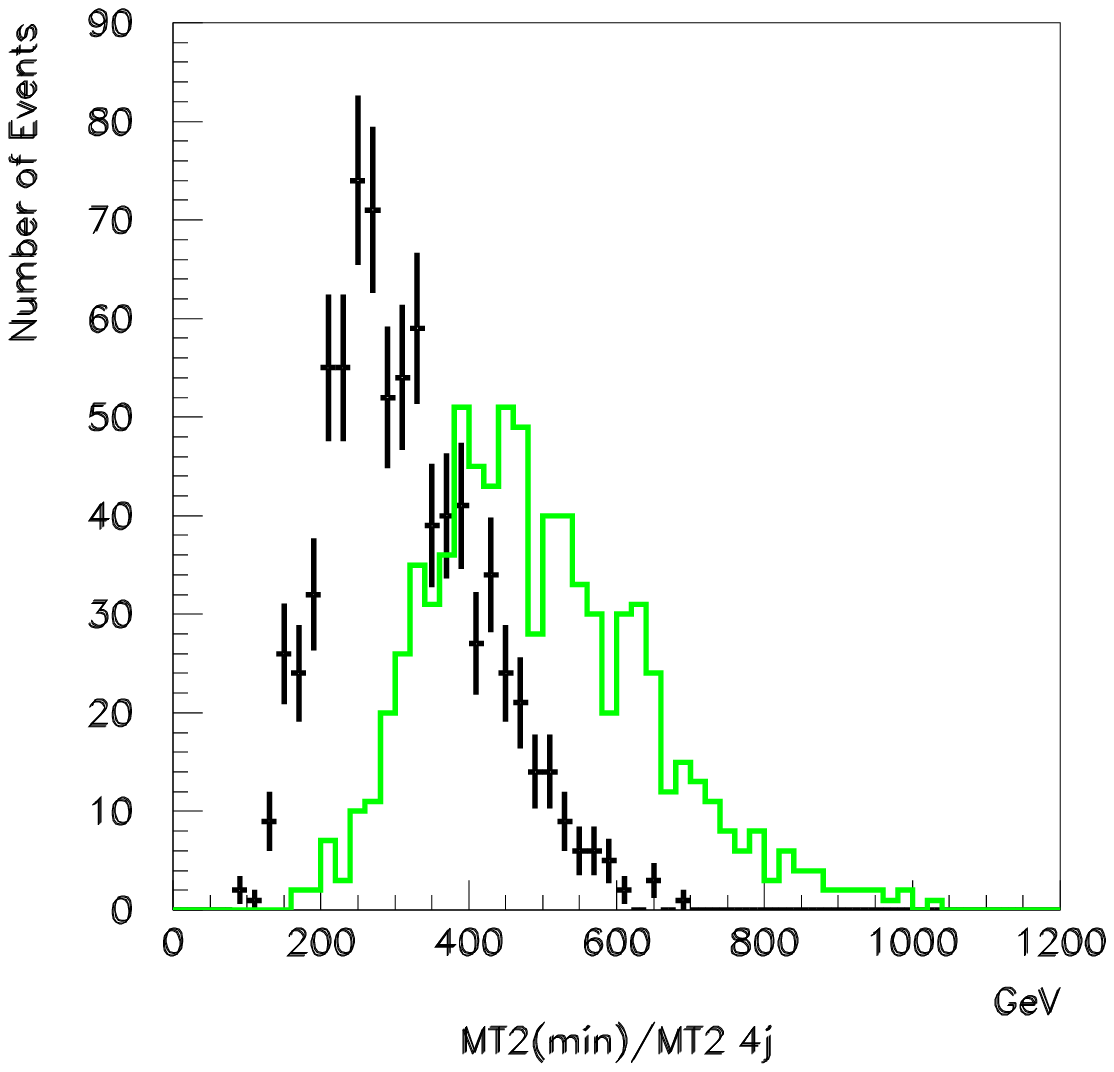}
\caption{ $\mttwo$ distributions at 7\,TeV and 1\,fb$^{-1}$.
 The distributions of $\mttwmin$ (black bars)  $\mttwmod$ (red bars)  and $\mttwo$ (green lines)  
 for  Point 3' (left figure), 3 (central figure) and 5' (right figure).  
}\label{other7tev}
\end{figure}

Fig.\,\ref{other7tev} shows the $\mttwmin$ and $\mttwmod$ 
distributions at  Point 3',  3 and 5' which have mass spectraType $A-C$.   
At Points 3 and 5' production cross sections at 7\,TeV  are  only 
around 1.7\,pb and 2.5\,pb  and the distributions are under significant statistical 
flactuation.\footnote{ 
For example, the distirbution at Point 5' ends around 600\,GeV, 
but an  endpoint structure at 550\,GeV is found for 60000 signal events.} 
The $M_{T2}$ distribution ends around 800\,GeV at Point 5, which 
reflects the input squark mass.  
Note that we have reason not to use $\mttwmod$  distribution for mass determination, 
because  the number of the events which  survives after 2 jet cut is smal. 
Therefore  we do not show the $\mttwmod$  
distribution in the figure. 

\section{Conclusion}

At the early stage of the LHC experiment,  useful discovery 
channels are jets + $E_{\rm Tmiss}$ channel and jets+ 1 lepton + $E_{\rm Tmiss}$ . 
The luminosity is rather low, so we want to measure sparticle nature from inclusive measurement rather than  exclusive and clean modes.  While $\mttwo$ is useful kinematical variables in measuring   parent SUSY partilce masses,  an inclusive definition proposed in\cite{shimizu}  is  not protected from smearing due to ISR.

In this paper we have proposed ISR improved inclusive $\mttwo$ variables  
which might be useful to determine the squark and gluino  masses separately. 
The modified $\mttwo$ variables discussed in this paper are 
 $\mttwmin$ and $\mttwmod$. 
 Endpoints of those distributions represent squark and gluino masses 
as summarized in Table\,\ref{summary}.  
Steps to identify sparticle masses  are shown in Fig.\,\ref{flowchart}. 
 The $\mttwmod$ distribution is used for determination of  gluino and squark masses  for $\mgl>\msq$ case as the algorithm keeps the highest $p_T$ jets in the hemisphere which is likely comes from squark two body decays for  
 $\msq\gg m_{\chi}$ case.  
 Even if  $\mgl \lsim \msq$,   
 $Br(\tilde{q}\rightarrow \chi q') $ 
 is large enough so that $\mttwmod$ is useful to determine $\msq$. 
On the other hand,  $\mttwmin$ is important to determine $\mgl$ when $\mgl<\msq$  
and the decay of $\tilde{g}$ is three body so that there are no dominant jets in the cascade decay. 

Errors in mass determination which we have obtained from the endpoint fits are typically 6 $-$ 10\,\% for 60000  generated SUSY events.  
On the other hand, we expect 4000 (1000)  events at 7\,TeV and $\int {\cal L} dt = 1\,{\rm fb^{-1}}$ 
and 380000 (10000) events at 14\,TeV and $\int {\cal L} dt = 10\,{\rm fb^{-1}}$ at Point 1 (5). 
Therefore, the model points with different gluino mass in Table\,\ref{masstable} can be clearly 
separated using the  $\mttwmin$ or $\mttwmod$ endpoints. 
We may identify an  unique 
point in the squark and gluino mass parameter space up to the LSP mass 
uncertainty.
 This is very important 
meaning, because squark and gluino production crosss section is 
 controlled by the masses up to small chargino and neutralino exchange 
contributions. We can cross check the predicted 
cross section to the observed  number of events. 
In addition, the expected cross section would be 
significantly different if we change spin of the produced particles 
from scalar to fermion as is expected in Little Higgs models with T partity 
for example. Determination of the mass and decay 
pattern allows us to exclude/prove such models as well. 

\begin{table}[!t]
\begin{tabular}{|c|c|c|c|c|c|}
\hline
& 2-jets $\mttwo$ & \begin{tabular}{c}N(peak)\cr
 of $M_{\rm hemi}$\end{tabular}& N(b-jets)& determine S &  determine G \cr
\hline
$G>S$  & $\begin{tabular}{c} significant \cr clear edge \end{tabular}$ & 1 & few &
\begin{tabular}{c} $\mttwmod$ \cr
2-jets cut
\end{tabular}
 &  \begin{tabular}{c}$\mttwmod$ \cr
  4-jets cut
  \end{tabular}\cr
\hline
\begin{tabular}{c}$S>G$\cr
($\gl\rightarrow \tilde{t} t$) \end{tabular}
  & \begin{tabular}{c} mild \cr  edge \end{tabular} & 2 & many &
\begin{tabular}{c} $\mttwmod$ \cr
2-jets cut \cr 
or $\mttwsub$ 
\end{tabular}
 &  \begin{tabular}{c}$\mttwmin$ \cr
  4-jets cut
  \end{tabular}\cr
\hline
\begin{tabular}{c}$S>G$ \cr( $\gl \rightarrow jj\chi$) \end{tabular} 
& \begin{tabular}{c} mild\cr  edge \end{tabular} & 2 & few &
\begin{tabular}{c} $\mttwmod$ \cr
4-jets cut
\end{tabular}
 &  $\mttwmin$ \cr
 \hline
$S\ll G$ & \begin{tabular}{c} no clear \cr  edge \end{tabular} & 2 & few &
 $\mttwo$ &  \begin{tabular}{c} $\mttwmin$ \cr $\mttwsub$\end{tabular}
 \cr
  \hline
\end{tabular}
\caption{Summary of the results obtained in this paper.  G(S) denotes $\mgl$ and 
$\msq$ respectively. }\label{summary}
\end{table}

There exist certain systematical errors  through decay patterns of the SUSY particles. 
In this paper, we study the mass spectrum that can be obtained 
in CMSSM.  The procedure should be modified if 
mass hierarchies are significantly different from those in CMSSM.  
For example when wino or bino mass is much close to the squark mass, 
while the lightest neutralino remains much lighter than squark, 
we my expect moderate   $p_T$ jets from neutralino/chargino 
cascade decays,  even though  $\msq<\mgl$, rather than 
prominent two high $p_T$ jets.  
Some of the squarks still directly decay into the lightest 
neutralino producing a high $p_T$ jet if it is gaugino like.
In such a case, although $\mgl >\msq$, the distribution of the number of jets 
is close to that in $\msq > \mgl$ case.  
Such classification of non-CMSSM models might be useful to reduce the 
systematics further. 

\section*{Aknowledgements}
We thank to J. Alwall and M. Takeuchi for helpful discussions in the early stage of this work.
K.S. is grateful for helpful discussions with other members of
the Cambridge SUSY Working Group.  
This work is supported in part by the World Premier International Center Initiative 
(WPI Program), MEXT, Japan and the Grant-in-Aid for Science Research, Japan 
Society for the Promotion of Science (for MN) and the Monbukagakusho (Japanese 
Government).
K.S. is supported by the U.K. Science and Technology Facilities Council.


\begin{thebibliography}{99}

\bibitem{atlas}
  ATLAS Collaboration,
  ``The ATLAS Experiment at the CERN Large Hadron Collider,''
2008 JINST 3 S08003 

\bibitem{cms}
  CMS Collaboration,
  ``The CMS experiment at the CERN LHC,''
2008 JINST 3 S08004

\bibitem{cdms2}
  Z.~Ahmed {\it et al.}  [The CDMS-II Collaboration],
  ``Results from the Final Exposure of the CDMS II Experiment,''
  arXiv:0912.3592 [astro-ph.CO].

\bibitem{xenon100}
  E.~Aprile {\it et al.}  [XENON100 Collaboration],
  ``First Dark Matter Results from the XENON100 Experiment,''
  arXiv:1005.0380 [astro-ph.CO].

\bibitem{review-lester}
  A.~J.~Barr and C.~G.~Lester,
  ``A Review of the Mass Measurement Techniques proposed for the Large Hadron Collider,''
  arXiv:1004.2732 [hep-ph].


\bibitem{Hinchliffe:1996iu}
  I.~Hinchliffe, F.~E.~Paige, M.~D.~Shapiro, J.~Soderqvist and W.~Yao,
  ``Precision SUSY measurements at CERN LHC,''
  Phys.\ Rev.\  D {\bf 55} (1997) 5520
  [arXiv:hep-ph/9610544].
  
\bibitem{Bachacou:1999zb}
  H.~Bachacou, I.~Hinchliffe and F.~E.~Paige,
  ``Measurements of masses in SUGRA models at CERN LHC,''
  Phys.\ Rev.\  D {\bf 62} (2000) 015009
  [arXiv:hep-ph/9907518].
  
\bibitem{Hinchliffe:1999zc}
  I.~Hinchliffe and F.~E.~Paige,
  ``Measurements in SUGRA models with large tan(beta) at LHC,''
  Phys.\ Rev.\  D {\bf 61} (2000) 095011
  [arXiv:hep-ph/9907519].

\bibitem{Allanach:2000kt}
  B.~C.~Allanach, C.~G.~Lester, M.~A.~Parker and B.~R.~Webber,
  ``Measuring sparticle masses in non-universal string inspired models at  the LHC,''
  JHEP {\bf 0009} (2000) 004
  [arXiv:hep-ph/0007009].

\bibitem{Gjelsten:2004ki}
  B.~K.~Gjelsten, D.~J.~.~Miller and P.~Osland,
  ``Measurement of SUSY masses via cascade decays for SPS 1a,''
  JHEP {\bf 0412} (2004) 003
  [arXiv:hep-ph/0410303].

\bibitem{Gjelsten:2005aw}
  B.~K.~Gjelsten, D.~J.~Miller and P.~Osland,
  ``Measurement of the gluino mass via cascade decays for SPS 1a,''
  JHEP {\bf 0506} (2005) 015
  [arXiv:hep-ph/0501033].

\bibitem{Miller:2005zp}
  D.~J.~Miller, P.~Osland and A.~R.~Raklev,
  ``Invariant mass distributions in cascade decays,''
  JHEP {\bf 0603} (2006) 034
  [arXiv:hep-ph/0510356].

\bibitem{Costanzo:2009mq}
  D.~Costanzo and D.~R.~Tovey,
  ``Supersymmetric particle mass measurement with invariant mass correlations,''
  JHEP {\bf 0904} (2009) 084
  [arXiv:0902.2331 [hep-ph]].

\bibitem{Burns:2009zi}
  M.~Burns, K.~T.~Matchev and M.~Park,
  ``Using kinematic boundary lines for particle mass measurements and
  disambiguation in SUSY-like events with missing energy,''
  JHEP {\bf 0905} (2009) 094
  [arXiv:0903.4371 [hep-ph]].

\bibitem{Matchev:2009iw}
  K.~T.~Matchev, F.~Moortgat, L.~Pape and M.~Park,
  ``Precise reconstruction of sparticle masses without ambiguities,''
  JHEP {\bf 0908} (2009) 104
  [arXiv:0906.2417 [hep-ph]].


\bibitem{lester:1999tx}
  C.~G.~Lester and D.~J.~Summers,
  ``Measuring masses of semiinvisibly decaying particles pair produced at
  hadron colliders,''
  Phys.\ Lett.\  B {\bf 463} (1999) 99
  [arXiv:hep-ph/9906349].

\bibitem{Barr:2003rg}
  A.~Barr, C.~Lester and P.~Stephens,
  ``m(T2) : The Truth behind the glamour,''
  J.\ Phys.\ G {\bf 29} (2003) 2343
  [arXiv:hep-ph/0304226].

\bibitem{Lester:2007fq}
  C.~Lester and A.~Barr,
  ``MTGEN : Mass scale measurements in pair-production at colliders,''
  JHEP {\bf 0712} (2007) 102
  [arXiv:0708.1028 [hep-ph]].

\bibitem{Cho:2007qv}
  W.~S.~Cho, K.~Choi, Y.~G.~Kim and C.~B.~Park,
  ``Gluino Stransverse Mass,''
  Phys.\ Rev.\ Lett.\  {\bf 100} (2008) 171801
  [arXiv:0709.0288 [hep-ph]].

\bibitem{Gripaios:2007is}
  B.~Gripaios,
  ``Transverse Observables and Mass Determination at Hadron Colliders,''
  JHEP {\bf 0802} (2008) 053
  [arXiv:0709.2740 [hep-ph]].

\bibitem{Barr:2007hy}
  A.~J.~Barr, B.~Gripaios and C.~G.~Lester,
  ``Weighing Wimps with Kinks at Colliders: Invisible Particle Mass
  Measurements from Endpoints,''
  JHEP {\bf 0802} (2008) 014
  [arXiv:0711.4008 [hep-ph]].

Cho:2007qv, Barr:2007hy, Cho:2007dh

\bibitem{Cho:2007dh}
  W.~S.~Cho, K.~Choi, Y.~G.~Kim and C.~B.~Park,
  ``Measuring superparticle masses at hadron collider using the transverse mass
  kink,''
  JHEP {\bf 0802} (2008) 035
  [arXiv:0711.4526 [hep-ph]].

\bibitem{shimizu}
  M.~M.~Nojiri, Y.~Shimizu, S.~Okada and K.~Kawagoe,
  ``Inclusive transverse mass analysis for squark and gluino mass
  determination,''
  JHEP {\bf 0806} (2008) 035
  [arXiv:0802.2412 [hep-ph]].

\bibitem{takeuchi}
  M.~M.~Nojiri, K.~Sakurai, Y.~Shimizu and M.~Takeuchi,
  ``Handling jets + missing ET channel using inclusive mT2,''
  JHEP {\bf 0810} (2008) 100
  [arXiv:0808.1094 [hep-ph]].

\bibitem{Cheng:2008hk}
  H.~C.~Cheng and Z.~Han,
  ``Minimal Kinematic Constraints and MT2,''
  JHEP {\bf 0812} (2008) 063
  [arXiv:0810.5178 [hep-ph]].

\bibitem{hiramatsu}
  J.~Alwall, K.~Hiramatsu, M.~M.~Nojiri and Y.~Shimizu,
  ``Novel reconstruction technique for New Physics processes with initial state
  radiation,''
  Phys.\ Rev.\ Lett.\  {\bf 103} (2009) 151802
  [arXiv:0905.1201 [hep-ph]].

\bibitem{Barr:2009jv}
  A.~J.~Barr, B.~Gripaios and C.~G.~Lester,
  ``Transverse masses and kinematic constraints: from the boundary to the
  crease,''
  JHEP {\bf 0911} (2009) 096
  [arXiv:0908.3779 [hep-ph]].

\bibitem{Matchev:2009fh}
  K.~T.~Matchev, F.~Moortgat, L.~Pape and M.~Park,
  ``Precision sparticle spectroscopy in the inclusive same-sign dilepton
  channel at LHC,''
  arXiv:0909.4300 [Unknown].

\bibitem{Konar:2009wn}
  P.~Konar, K.~Kong, K.~T.~Matchev and M.~Park,
  ``Superpartner mass measurements with 1D decomposed MT2,''
  arXiv:0910.3679 [Unknown].

\bibitem{Konar:2009qr}
  P.~Konar, K.~Kong, K.~T.~Matchev and M.~Park,
  ``Dark Matter Particle Spectroscopy at the LHC: Generalizing MT2 to
  Asymmetric Event Topologies,''
  JHEP {\bf 1004} (2010) 086
  [arXiv:0911.4126 [Unknown]].


\bibitem{hemi1}
F.~Moortgat and L.~Pape, {\it CMS Physics TDR, vol. II}, Report No. CERN-LHCC-2006, 
section 13.4, pg.410. 

\bibitem{hemi2}
  S.~Matsumoto, M.~M.~Nojiri and D.~Nomura,
  ``Hunting for the top partner in the littlest Higgs model with T-parity at
  the LHC,''
  Phys.\ Rev.\  D {\bf 75} (2007) 055006
  [arXiv:hep-ph/0612249].


\bibitem{Plehn:2005cq}
  T.~Plehn, D.~Rainwater and P.~Z.~Skands,
  ``Squark and gluino production with jets,''
  Phys.\ Lett.\  B {\bf 645} (2007) 217
  [arXiv:hep-ph/0510144].

\bibitem{Alwall:2008qv}
  J.~Alwall, S.~de Visscher and F.~Maltoni,
  ``QCD radiation in the production of heavy colored particles at the LHC,''
  JHEP {\bf 0902} (2009) 017
  [arXiv:0810.5350 [hep-ph]]. 

\bibitem{Papaefstathiou:2009hp}
  A.~Papaefstathiou and B.~Webber,
  ``Effects of QCD radiation on inclusive variables for determining the scale
  of new physics at hadron colliders,''
  JHEP {\bf 0906} (2009) 069
  [arXiv:0903.2013 [hep-ph]].

\bibitem{Papaefstathiou:2010ru}
  A.~Papaefstathiou and B.~Webber,
  ``Effects of invisible particle emission on global inclusive variables at
  hadron colliders,''
  JHEP {\bf 1007} (2010) 018
  [arXiv:1004.4762 [hep-ph]].  




\bibitem{Baer:1995nq}
  H.~Baer, C.~h.~Chen, F.~Paige and X.~Tata,
  ``Signals for minimal supergravity at the CERN large hadron collider: Multi -
  jet plus missing energy channel,''
  Phys.\ Rev.\  D {\bf 52} (1995) 2746
  [arXiv:hep-ph/9503271].

\bibitem{Baer:1995va}
  H.~Baer, C.~h.~Chen, F.~Paige and X.~Tata,
  ``Signals for Minimal Supergravity at the CERN Large Hadron Collider II:
  Multilepton Channels,''
  Phys.\ Rev.\  D {\bf 53} (1996) 6241
  [arXiv:hep-ph/9512383].

\bibitem{Abdullin:1998pm}
  S.~Abdullin {\it et al.}  [CMS Collaboration],
  ``Discovery potential for supersymmetry in CMS,''
  J.\ Phys.\ G {\bf 28} (2002) 469
  [arXiv:hep-ph/9806366].
  
\bibitem{Baer:1998sz}
  H.~Baer, C.~h.~Chen, M.~Drees, F.~Paige and X.~Tata,
  ``Probing minimal supergravity at the CERN LHC for large tan(beta),''
  Phys.\ Rev.\  D {\bf 59} (1999) 055014
  [arXiv:hep-ph/9809223].

\bibitem{Abdullin:1998nv}
  S.~Abdullin and F.~Charles,
  ``Search for SUSY in (leptons +) jets + E(T)(miss) final states,''
  Nucl.\ Phys.\  B {\bf 547} (1999) 60
  [arXiv:hep-ph/9811402].

\bibitem{Allanach:2000ii}
  B.~C.~Allanach, J.~P.~J.~Hetherington, M.~A.~Parker and B.~R.~Webber,
  ``Naturalness reach of the Large Hadron Collider in minimal supergravity,''
  JHEP {\bf 0008} (2000) 017
  [arXiv:hep-ph/0005186].

\bibitem{Baer:2003wx}
  H.~Baer, C.~Balazs, A.~Belyaev, T.~Krupovnickas and X.~Tata,
  ``Updated reach of the CERN LHC and constraints from relic density, $b \to s
  \gamma$ and a($\mu$) in the mSUGRA model,''
  JHEP {\bf 0306} (2003) 054
  [arXiv:hep-ph/0304303].

\bibitem{Izaguirre:2010nj}
  E.~Izaguirre, M.~Manhart and J.~G.~Wacker,
  ``Bigger, Better, Faster, More at the LHC,''
  arXiv:1003.3886 [hep-ph].

\bibitem{Baer:2010tk}
  H.~Baer, V.~Barger, A.~Lessa and X.~Tata,
  ``Capability of LHC to discover supersymmetry with $\sqrt{s}=7$ TeV and 1 ${\rm fb^{-1}}$,''
  JHEP {\bf 1006} (2010) 102
  [arXiv:1004.3594 [hep-ph]].

\bibitem{cmsnote}
CMS~collaboration,
``The CMS physics reach for searches at 7 TeV,''
CMS NOTE-2010/008 


\bibitem{herwig1}
  G.~Corcella {\it et al.},
  ``HERWIG 6.5: an event generator for Hadron Emission Reactions With
  Interfering Gluons (including supersymmetric processes),''
  JHEP {\bf 0101} (2001) 010
  [arXiv:hep-ph/0011363].

\bibitem{herwig2}
  G.~Corcella {\it et al.},
  ``HERWIG 6.5 release note,''
  arXiv:hep-ph/0210213.

\bibitem{herwig3}
  S.~Moretti, K.~Odagiri, P.~Richardson, M.~H.~Seymour and B.~R.~Webber,
  ``Implementation of supersymmetric processes in the HERWIG event
  generator,''
  JHEP {\bf 0204} (2002) 028
  [arXiv:hep-ph/0204123].





\bibitem{Paige:2003mg}
  F.~E.~Paige, S.~D.~Protopopescu, H.~Baer and X.~Tata,
  ``ISAJET 7.69: A Monte Carlo event generator for p p, anti-p p, and e+ e-
  reactions,''
  arXiv:hep-ph/0312045.

\bibitem{Beenakker:1996ch}
  W.~Beenakker, R.~Hopker, M.~Spira and P.~M.~Zerwas,
  ``Squark and gluino production at hadron colliders,''
  Nucl.\ Phys.\  B {\bf 492} (1997) 51
  [arXiv:hep-ph/9610490].

\bibitem{Datta:1}
  A.~Datta, G.~L.~Kane and M.~Toharia,
  ``Is it SUSY?,''
  arXiv:hep-ph/0510204.

\bibitem{Kane:1}
G.~L.~Kane, A.~A.~Petrov, J.~Shao and L.~T.~Wang,
  ``Initial determination of the spins of the gluino and squarks at LHC,''
  J.\ Phys.\ G {\bf 37} (2010) 045004
  [arXiv:0805.1397 [hep-ph]].


\bibitem{Tovey:meff}
  D.~R.~Tovey,
  ``Measuring the SUSY mass scale at the LHC,''
  Phys.\ Lett.\  B {\bf 498} (2001) 1
  [arXiv:hep-ph/0006276].




\bibitem{madgraph}
  J.~Alwall {\it et al.},
  ``MadGraph/MadEvent v4: The New Web Generation,''
  JHEP {\bf 0709} (2007) 028
  [arXiv:0706.2334 [hep-ph]].

\bibitem{phythia}
  T.~Sjostrand, S.~Mrenna and P.~Z.~Skands,
  ``PYTHIA 6.4 Physics and Manual,''
  JHEP {\bf 0605} (2006) 026
  [arXiv:hep-ph/0603175].


\bibitem{acerdet}
  E.~Richter-Was,
  ``AcerDET: A particle level fast simulation and reconstruction package  for
  phenomenological studies on high p(T) physics at LHC,''
  arXiv:hep-ph/0207355, http://erichter.home.cern.ch/erichter/AcerDET.html

\bibitem{fastjet}
  M.~Cacciari and G.~P.~Salam,
  Phys.\ Lett.\  B {\bf 641} (2006) 57
  [arXiv:hep-ph/0512210].




\end{thebibliography}
\end{document}